\documentclass[journal]{IEEEtran}
\usepackage{amsmath,amsfonts}
\usepackage{mathrsfs}
\usepackage{pifont}
\usepackage{algorithmic}
\usepackage{algorithm}
\usepackage{array}
\usepackage{textcomp}
\usepackage{stfloats}
\usepackage{url}
\usepackage{verbatim}
\usepackage{graphicx}
\usepackage{cite}
\usepackage{subcaption}  
\usepackage{amsthm}
\newtheoremstyle{myremark}
  {1pt} 
  {1pt} 
  {\itshape} 
  {} 
  {\bfseries} 
  {:} 
  {0.5em} 
  {} 

\theoremstyle{myremark}
\newtheorem*{remark}{\textbf{Remark}}

\usepackage{acronym}

\acrodef{cc}[CC]{channel charting}
\acrodef{gospa}[GOSPA]{generalized optimal sub-pattern assignment}
\acrodef{kde}[KDE]{kernel density estimation}
\acrodef{kde}[KDE]{kernel density estimation}
\acrodef{gmm}[GMM]{Gaussian mixture model}
\acrodef{em}[EM]{expectation-maximization}

\acrodef{ot}[OT]{optimal transport}
\acrodef{csi}[CSI]{channel state information}
\acrodef{cir}[CIR]{channel impulse response}

\acrodef{5g}[5G]{the fifth generation}
\acrodef{6g}[6G]{the sixth generation}
\acrodef{iot}[IoT]{internet of things}

\acrodef{aoa}[AoA]{angle-of-arrival}
\acrodef{toa}[ToA]{time-of-arrival}
\acrodef{tdoa}[TDoA]{time-difference-of-arrival}
\acrodef{rtt}[RTT]{round-trip time}

\acrodef{ap}[AP]{access point}
\acrodef{bs}[BS]{base station}
\acrodef{trp}[TRP]{transmission reception point}

\acrodef{cdf}[CDF]{cumulative distribution function}
\acrodef{pdf}[PDF]{probability density function}
\acrodef{crb}[CRB]{Cram\'er-Rao bound}
\acrodef{gnss}[GNSS]{global navigation satellite system}
\acrodef{los}[LoS]{line-of-sight}
\acrodef{nlos}[NLoS]{non-line-of-sight}
\acrodef{mae}[MAE]{mean absolute error}
\acrodef{rmse}[RMSE]{root mean square error}
\acrodef{ls}[LS]{least squares}
\acrodef{music}[MUSIC]{multiple signal classification}
\acrodef{omp}[OMP]{orthogonal matching pursuit}
\acrodef{ofdm}[OFDM]{orthogonal frequency division multiplexing}
\acrodef{snr}[SNR]{signal-to-noise ratio}

\acrodef{siso}[SISO]{single-input single-output}
\acrodef{mimo}[MIMO]{multiple-input-multiple-output}

\acrodef{ue}[UE]{user equipment}
\acrodef{upa}[UPA]{uniform planar array}
\acrodef{ula}[ULA]{uniform linear array}

\acrodef{ml}[ML]{machine learning}
\acrodef{nn}[NN]{neural network}
\acrodef{cnn}[CNN]{convolutional neural network}
\acrodef{mlp}[MLP]{multi-layer perceptron}

\acrodef{cdf}[CDF]{cumulative distribution function}
\acrodef{kkt}[KKT]{Karush-Kuhn-Tucker}

\usepackage{pgfplots}
\usepackage{tikz}
\usetikzlibrary{calc}
\makeatletter
\newcommand{\gettikzxy}[3]{%
  \tikz@scan@one@point\pgfutil@firstofone#1\relax
  \edef#2{\the\pgf@x}%
  \edef#3{\the\pgf@y}%
}
\usetikzlibrary{spy,backgrounds}
\usepackage{pgfplotstable}

\usepackage{microtype} 

\usepackage{titlesec}
\titlespacing\section{0pt}{4pt}{2pt}
\titlespacing\subsection{0pt}{1pt}{1pt}
\titlespacing\subsubsection{0pt}{1pt}{1pt}

\begin{document}

\title{UNILocPro: Unified Localization Integrating Model-Based \\ Geometry and Channel Charting}

\author{Yuhao~Zhang, 
Guangjin~Pan,~\IEEEmembership{Member,~IEEE}, 
Musa~Furkan~Keskin,~\IEEEmembership{Senior Member,~IEEE}, Ossi~Kaltiokallio,~\IEEEmembership{Member,~IEEE}, Mikko~Valkama,~\IEEEmembership{Fellow,~IEEE}, and Henk~Wymeersch,~\IEEEmembership{Fellow,~IEEE} \vspace{-0.5cm}
\thanks{This work was supported in part by the European Union through the project ISLANDS (grant agreement no. 101120544), the Swedish Research Council (VR) through the project 6G-PERCEF under grant 2024-04390, the Swedish Foundation for Strategic Research (SSF) funded by Software Artificial Intelligence for Communications (SAICOM) under Grant FUS21-0004, the Chalmers Areas of Advance in Information and Communication Technology (ICT) and Transport, and Business Finland under the 6G-ISAC project.
\emph{(Corresponding author: Guangjin~Pan.)}}
\thanks{Yuhao~Zhang, Guangjin~Pan, Musa~Furkan~Keskin, and Henk~Wymeersch are with the Department of Electrical Engineering, Chalmers University of Technology,
41296 Gothenburg, Sweden (e-mail: \{yuhaozh, guangjin.pan, furkan, henkw\}@chalmers.se).}
\thanks{Ossi~Kaltiokallio and Mikko~Valkama are with the Tampere Wireless Research Center, Unit of Electrical Engineering, Tampere University, 33100 Tampere, Finland (e-mail: \{ossi.kaltiokallio, mikko.valkama\}@tuni.fi).}
\thanks{Part of this work has been presented in our previous conference paper~\cite{Unif_Loca_Yuha_2025}, which is a special case of the proposed localization framework.}}

\markboth{DRAFT}%
{Shell \MakeLowercase{\textit{et al.}}: A Sample Article Using IEEEtran.cls for IEEE Journals}


\maketitle

\begin{abstract}
In this paper, we propose a unified framework (called UNILocPro) that integrates model-based geometry and \ac{cc} for localization via a single massive \ac{mimo} \ac{bs} in mixed \ac{los}/\ac{nlos} scenarios. Specifically, based on \ac{los}/\ac{nlos} identification, an adaptive activation between the model-based and \ac{cc}-based methods is conducted. Aiming for unsupervised learning, information obtained from the model-based method is utilized to train the \ac{cc} model, where a pairwise distance loss (involving a new dissimilarity metric design), a triplet loss (if timestamps are available), a \ac{los}-based loss, and an \ac{ot}-based loss are jointly employed such that the global geometry can be well preserved. To reduce the training complexity, we propose a low-complexity implementation (called UNILoc), where the \ac{cc} model is trained with self-generated labels produced by a single pre-training \ac{ot} transformation, which avoids iterative Sinkhorn updates involved in the \ac{ot}-based loss computation. Extensive numerical experiments demonstrate that the proposed unified frameworks achieve significantly improved positioning accuracy compared to both model-based and \ac{cc}-based methods. Notably, UNILocPro with timestamps attains performance on par with fully-supervised fingerprinting despite operating without labelled training data. It is also shown that the low-complexity UNILoc can substantially reduce training complexity with marginal performance degradation.
\end{abstract}


\begin{IEEEkeywords}
localization, model-based localization, channel charting, machine learning, unsupervised learning, optimal transport.
\end{IEEEkeywords}

\acresetall

\section{Introduction}\label{sec:introduction}
\begin{figure}[t]
    \centering
    \begin{tikzpicture}[every node/.style={font=\footnotesize}]
    \node (image) [anchor=south west]{\includegraphics[width=0.75\linewidth]{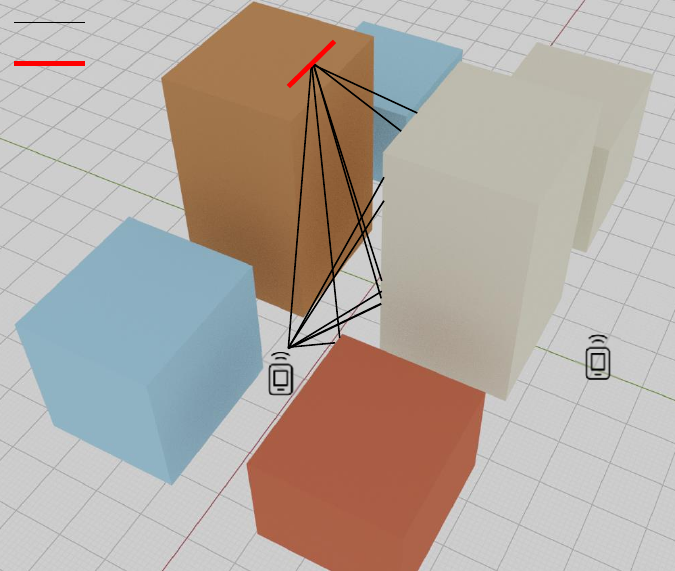}};
    \gettikzxy{(image.north east)}{\ix}{\iy};
    
    \node at (0.28*\ix,0.98*\iy)[rotate=0,anchor=north]{{channel paths}};
    \node at (0.28*\ix,0.91*\iy)[rotate=0,anchor=north]{{antenna array}};
    \node at (0.43*\ix,0.33*\iy)[rotate=0,anchor=north]{{user 1}};
    \node at (0.88*\ix,0.36*\iy)[rotate=0,anchor=north]{{user 2}};
    
    \end{tikzpicture}
    \caption{A street canyon scenario, where channel paths are shown for user 1 for example. Note that user 1 is a \ac{los} user, while user 2 is a \ac{nlos} user as its \ac{los} path is blocked by a building.}
    \label{fig:system_setting}
    \vspace{-3mm}
\end{figure}

Wireless communication systems (e.g., 5G and 6G) provide opportunities for high-precision localization, even in \ac{gnss}-challenged environments~\cite{High_Loca_Witr_2016,A_Tuto_Chen_2022,Near_Loca_Wang_2025}. In general, wireless localization methods fall into two main categories: (1) model-based methods relying on channel parameter estimation and geometric relationships~\cite{An_Intr_Zeka_2012,Harn_NLOS_Mend_2019,A_Surv_Zafa_2019,Robu_Snap_Kalt_2024}, and (2) data-driven methods leveraging \ac{ml} to infer user position from \ac{csi}~\cite{Fing_Loca_Sun_2019,AI_Wire_Pan_2025,Lear_Loca_Wu_2021,Fing_Loca_Uyka_2026,Larg_Wire_Pan_2025}. Model-based methods typically exploit the \ac{los} path (which provides the most direct and reliable information about the user location), whose parameters (e.g., \ac{toa}, \ac{aoa}, \ac{tdoa}) are estimated using algorithms such as \ac{music} or \ac{omp}~\cite{Sign_Reco_Trop_2007}, and then mapped to the user location via geometric relationships. These methods perform well in \ac{los} scenarios (e.g., user 1 in Fig.~\ref{fig:system_setting}) but degrade under \ac{nlos} conditions due to \ac{los} blockage (e.g., user 2 in Fig.~\ref{fig:system_setting}).\footnote{In this paper, \ac{nlos} scenarios refer to scenarios where the \ac{los} path is blocked. Conversely, \ac{los} scenarios describe cases where the \ac{los} path exists, regardless of whether \ac{nlos} paths are present or not.} Although advanced techniques can exploit \ac{nlos} paths, they are often complex and scenario-specific (e.g., relying on specular reflections)~\cite{Harn_NLOS_Mend_2019,An_Intr_Zeka_2012,Robu_Snap_Kalt_2024}. On the contrary, data-driven methods hold some promise for \ac{nlos} scenarios, but suffer from certain drawbacks. For example, supervised learning, e.g., fingerprinting~\cite{Fing_Loca_Sun_2019,Lear_Loca_Wu_2021,Fing_Loca_Uyka_2026,AI_Wire_Pan_2025}, requires extensive data measurement and collection for each specific setting, which may not be practical especially for dynamic environments, where fingerprints can become outdated with time and thus frequent data re-measurement and retraining are required. As an unsupervised learning technique that aims to generate a mapping from high-dimensional \ac{csi} to a low-dimensional space (called channel chart), \ac{cc} can learn the relative geometry based on \ac{csi} without the need for labeled data, while it would usually yield position estimates that are highly distorted~\cite{Chan_Char_Stud_2018,Trip_Wire_Ferr_2021}. In order to preserve the global geometry, side information and/or model-based methods can be utilized in \ac{cc} for absolute localization~\cite{Augm_Chan_Euch_2023,Angl_Prof_Step_2024,Glob_Scal_Omid_2024}.

Particularly interesting is the use of \ac{cc} for absolute localization, which requires auxiliary information, e.g., anchor locations or map information. With a set of labeled \ac{csi} samples, a localization penalty similar to fingerprinting can be incorporated into the loss function to optimize the \ac{cc} model~\cite{Trip_Wire_Ferr_2021,Leve_Chan_Xu_2026,A_Chan_Pan_2026}. In~\cite{Abso_Posi_Pihl_2020,Indo_Loca_Stah_2023,A_Sign_Zhao_2024,Angl_Prof_Step_2024,Leve_The_Euch_2024}, anchor points with known positions are used to find an affine transformation matching the channel chart with the global coordinates. All the above methods are semi-supervised, requiring pre-measured anchor locations or labeled data, and thus share the drawbacks of fingerprinting. To mitigate these issues, map-assisted approaches can exploit, for example, building floor plans or street maps to align the channel chart with the global coordinates. In~\cite{Moda_Topo_Farh_2021}, a \ac{cc} method is proposed for indoor localization, where a Siamese \ac{nn} is trained to coherently optimize a pairwise distance loss and a Sinkhorn distance between the channel chart and the topological map. In~\cite{Velo_Chan_Stah_2024}, a velocity-assisted \ac{cc} is proposed for indoor scenarios where the distance between two \ac{csi} samples is calculated based on velocity information (measured by deploying pedestrian dead reckoning or odometry systems) and an affine transformation is learned to match the generated channel chart and building floor plans.

To improve robustness, interpretability, and accuracy of localization, model-based and data-driven methods can be integrated to exploit their complementary strengths. For \ac{cc}-based localization, it is crucial to select or design an appropriate dissimilarity metric that can adequately achieve accurate global positioning. Since the dissimilarity metric computed solely from \ac{csi} is highly related to channel characteristics, its positioning performance is sensitive to channel conditions, e.g., \ac{los}/\ac{nlos} conditions and the configuration of antenna arrays~\cite{Abso_Posi_Pihl_2020,Chan_Char_Stud_2018,Indo_Loca_Stah_2023,Angl_Prof_Step_2024,Chan_Char_Tous_2022}. Instead of computing the dissimilarity metric directly from \ac{csi}, model-based estimation methods could be employed first to extract high-resolution geometric features, e.g., \ac{aoa} and \ac{toa}, from which the dissimilarity metric can then be computed~\cite{Chan_Char_Tous_2022,Angl_Feat_Uyka_2024,Mult_Chan_Deng_2018}. For example, as proposed in~\cite{Chan_Char_Tous_2022}, the earth mover's distance is used to compute dissimilarity from the model-based location estimates of multipath components, which demonstrates superior \ac{cc} performance compared to some conventional linear algebraic distances on \ac{csi}, e.g., correlation matrix distance.

Moreover, model-based localization has also been used to assist \ac{cc} in achieving unsupervised absolute localization. For example, a \ac{cc} technique is proposed in~\cite{Glob_Scal_Omid_2024} for an indoor scenario by incorporating \ac{toa} measurements and laser scanner data into the loss function. In~\cite{Augm_Chan_Euch_2023}, an augmented \ac{cc} approach is proposed for an indoor scenario, where the joint \ac{tdoa} and \ac{aoa} likelihood is included in the \ac{nn} training loss. In~\cite{TDoA_Self_Mohs_2025}, a self-supervised \ac{cc}-based localization is proposed for distributed systems by leveraging \ac{tdoa} measurements and multiple \ac{trp} locations. However, in~\cite{Glob_Scal_Omid_2024}, an additional laser scanner system is required to calculate the distance between two \ac{csi} samples, which induces extra cost and complexity; \cite{TDoA_Self_Mohs_2025} classifies the \ac{nlos} paths as foes and consequently filters out the \ac{nlos} information, making it unavailable for localization; \cite{Augm_Chan_Euch_2023} relies on distributed settings, and the cosine-based dissimilarity performs inadequately in \ac{nlos} scenarios~\cite{Chan_Char_Tous_2022,Angl_Feat_Uyka_2024}. In the literature, \ac{cc}-based localization is studied mainly for \ac{los} dominant environments~\cite{Indo_Loca_Stah_2023,Angl_Prof_Step_2024,Velo_Chan_Stah_2024,Leve_The_Euch_2024}, and the above model-assisted works~\cite{Augm_Chan_Euch_2023,Glob_Scal_Omid_2024,TDoA_Self_Mohs_2025} require \ac{los} conditions for absolute positioning.\footnote{In this paper, we use ``\ac{los} dominant'' to represent the scenario where most of the user positions in the system have a \ac{los} path to the \ac{bs}.} In~\cite{Chan_Char_Sued_2025}, by distinguishing users through received signal strength, a dedicated loss based on the prior of \ac{los} regions is introduced for \ac{cc} to encourage \ac{los} users to lie within a predefined area. However, this approach is mainly designed for distributed \ac{mimo} systems, where \ac{csi} samples from multiple access points are used. In summary, despite some initial efforts to explore \ac{cc} in mixed \ac{los}/\ac{nlos} scenarios, the in-depth integration of model-based methods and \ac{cc}, i.e., a general and robust solution, for such environments remains an open challenge and has not been addressed to date, especially for single-\ac{bs} localization.

For localization via single massive \ac{mimo} \ac{bs}, we aim to design unified frameworks that integrate model-based and fully-unsupervised \ac{cc}-based methods in challenging mixed \ac{los}/\ac{nlos} scenarios by exploiting available map information (only a coarse map is required, e.g., a street or floor plan, which is usually available for many cities and buildings). Extending our prior research~\cite{Unif_Loca_Yuha_2025}, where model-based estimates and map information are used to generate pseudo-labels for \ac{nn} regression through a one-shot \ac{ot} transformation, we integrate \ac{cc} to develop a more comprehensive unified framework that explicitly learns a model by introducing geometry-aware dissimilarity metrics and a multi-loss training objective. Furthermore, we conduct more systematic numerical experiments to provide a thorough comparative evaluation and analysis. The main contributions of this work are as follows:

\begin{itemize}
    \item \textbf{Unified model- and \ac{cc}-based localization framework for mixed \ac{los}/\ac{nlos} scenarios (Sec.~\ref{sec:UNILocPro})}: We propose a unified localization framework (UNILocPro) that integrates model-based and \ac{cc}-based methods for mixed \ac{los}/\ac{nlos} environments. Based on \ac{los}/\ac{nlos} identification, we apply the model-based method to the identified \ac{los} users and the \ac{cc}-based method to the identified \ac{nlos} users. To enable unsupervised learning, channel parameters estimated from \ac{csi} are used to train the \ac{cc} model, preserving the global spatial geometry. UNILocPro is unified in: (1) separate handling of \ac{los}/\ac{nlos} users; (2) unsupervised learning driven by both model and data; and (3) a multi-component \ac{cc} loss that exploits model and map information.
    \item \textbf{Novel dissimilarity metrics and multi-component loss design (Sec.~\ref{sec:cc_loss})}: Using channel parameters extracted from the model-based method, we design effective dissimilarity metrics, including a \ac{gospa}-based and its fusion with the Wasserstein-based dissimilarity, to preserve local geometry in \ac{cc}. The global geometry is maintained by using model-based position estimates of identified \ac{los} users as anchors and incorporating an \ac{ot}-based loss during training to align with the map. Additionally, when timestamps are available, a timestamp-based triplet loss can also be applied to further enhance the \ac{cc} model.
    \item \textbf{Low-complexity implementation with pre-trained \ac{ot} transformation (Sec.~\ref{sec:UNILoc})}: To reduce the high computational cost of the \ac{ot}-based loss in the \ac{cc} model, we propose a low-complexity approach (UNILoc) that replaces iterative \ac{ot} execution during training with a one-shot pre-trained \ac{ot} transformation. Specifically, the \ac{cc} model is trained using self-generated labels obtained from model-based position estimates and map information through \ac{ot} alignment. This strategy greatly decreases training overhead and complexity while maintaining high positioning accuracy. In our experiments, UNILoc achieves a training complexity reduction of about $\mathcal{O}\left( N_u^2 \right)$, where $N_u$ is the number of data samples, and runs $4 \text{-} 5$ times faster than UNILocPro.
\end{itemize}

\textit{Notations}: We denote column vectors as bold-faced lower-case letters, $\mathbf{a}$, and matrices as bold-faced upper-case letters, $\mathbf{A}$. A column vector with a size of $N$ whose entries are all equal to $1$ is denoted as $\mathbf{1}_N$. The identity matrix of size $N \times N$ is denoted as $\mathbf{I}_N$. The transpose and conjugate transpose operations are denoted by $(\cdot)^\mathsf{T}$ and $(\cdot)^\mathsf{H}$, respectively. The $i$-th element of a vector and the $(i,j)$-th element of a matrix are denoted by $[\mathbf{a}]_i$ and $[\mathbf{A}]_{i,j}$, respectively. $\otimes$ represents the Kronecker product and $\oslash$ represents the element-wise division. $\operatorname{vec}(\cdot)$ denotes the matrix vectorization operator and $\mathrm{diag}(\cdot)$ denotes the diagonalization operator. $\| \cdot \|_2$ and $\| \cdot \|_F$ are the Euclidean norm and the Frobenius norm, respectively. $\langle \cdot, \cdot\rangle_F$ represents the Frobenius dot product.

\section{System Model}\label{sec:system_model}
We consider a single-cell \ac{ofdm} system, where a \ac{bs} equipped with $M$ antennas serves multiple single-antenna users over $N_c$ subcarriers (with $\Delta_f$ being the subcarrier spacing). A dedicated sequence of pilots is transmitted by each user such that its associated \ac{csi} can be estimated by the \ac{bs}, where the estimated \ac{csi} from user $i$ at all subcarriers is denoted by $\mathbf{H}_i \in \mathbb{C}^{M \times N_c}$. In this work, we assume that the \ac{bs} and users are tightly synchronized such that the residual clock bias between them is negligible.\footnote{This assumption limits the applicability of the proposed framework to scenarios in which such time synchronization is available. When the user is non-synchronized, its clock bias introduces an unknown offset in delay measurements and must be accounted for when estimating position. We leave the joint estimation of user clock bias and position to future work.} Our goal is to estimate the position of each user, denoted by $\mathbf{p}_i = [ x_i, y_i, z_i]^\mathsf{T} $, based solely on its associated \ac{csi}, i.e., $\hat{\mathbf{p}}_i = f_e(\mathbf{H}_i)$, $\forall i$, where $\hat{\mathbf{p}}_i$ is the estimate of $\mathbf{p}_i$ and $f_e(\cdot)$ is a position estimation function.

\subsection{Channel Model}\label{sec:channel_model}
Considering the multi-path effect, the estimated frequency domain channel matrix for user $i$ can be modeled as $\mathbf{H}_i = \sum_{l = 1}^{L_i} \beta_l^i \mathbf{a}_{\mathbf{t}}(\theta_l^i) \mathbf{b}^\mathsf{T}(\tau_l^i) + \mathbf{N}_i$~\cite{Mass_MIMO_Emil_2017,Mult_ISAC_Visa_2024,Mode_End_Jose_2025,Fund_Trad_Kesk_2025}, where there are $L_i$ channel paths for user $i$; $\beta_l^i$, $\theta_l^i$, and $\tau_l^i$ are the complex channel gain, \ac{aoa}, and \ac{toa} of the $l$-th path, respectively; $\mathbf{b}(\tau) \in \mathbb{C}^{N_c \times 1}$ is the frequency-domain steering vector, as given by $[\mathbf{b}(\tau)]_n = \exp [-\jmath 2 \pi (n-1) \Delta_f \tau]$, $n = 1,2, \cdots,N_c$; $\mathbf{a}_{\mathbf{t}}(\theta) \in \mathbb{C}^{M \times 1}$ is the array steering vector, which depends on the antenna array geometry. For example, when a \ac{ula} with an inter-antenna separation of $d$ is adopted, the steering vector is $[\mathbf{a}_{\mathbf{t}}(\theta)]_m = \exp [\jmath 2 \pi (m-1) \frac{d}{\lambda} \sin (\theta)]$, $m = 1,2, \cdots,M$, where $\lambda = {c}/{f_c}$ with $c$ being the speed of light and $f_c$ being the carrier frequency. The steering vector $\mathbf{a}_{\mathbf{t}}(\theta)$ for a \ac{upa} is referred to~\cite{Lear_Loca_Wu_2021,Mass_MIMO_Emil_2017}. $\mathbf{N}_i$ is the noise matrix that results from imperfect channel estimation at the \ac{bs}, where $\operatorname{vec}(\mathbf{N}_i) \sim \mathcal{CN}(0, \sigma^2 \mathbf{I}_{M N_c})$ is assumed with $\sigma^2$ being the noise power~\cite{Mass_MIMO_Emil_2017}.

With \ac{ofdm} transmission and multi-antenna \ac{bs}, channel parameters, including channel gain, \ac{aoa}, and \ac{toa} for different channel paths, can be estimated from the \ac{csi} $\mathbf{H}_i$ by channel estimation algorithms, e.g., the \ac{omp} algorithm~\cite{Sign_Reco_Trop_2007,Mode_End_Jose_2025}. Specifically, with \ac{omp}, $\hat{L}_i$ channel paths can be recovered with channel parameters $\{ \hat{\beta}_l^i, \hat{\theta}_l^i, \hat{\tau}_l^i \}_{l = 1,2,\ldots,\hat{L}_i} = f_{\rm ce}(\mathbf{H}_i)$ based on the \ac{csi} $\mathbf{H}_i$ for user $i$, where $f_{\rm ce}(\cdot)$ is the channel parameter estimation function.

\subsection{Map Information}
We assume that a map of the environment is accessible, from which the region for all possible user positions can be extracted as $\mathcal{R} \subseteq \mathbb{R}^{3 \times 1}$ such that we have $\mathbf{p}_i \in \mathcal{R}$, $\forall i$. Moreover, the map enables us to determine whether a given point in $\mathcal{R}$ has a \ac{los} connection to the \ac{bs} or not, based on which $\mathcal{R}$ can be divided into $\mathcal{R}_{\rm LoS} \subseteq \mathbb{R}^{3 \times 1}$ (the region for \ac{los} user) and $\mathcal{R}_{\rm NLoS} \subseteq \mathbb{R}^{3 \times 1}$ (the region for \ac{nlos} users), respectively, where $\mathcal{R}_{\rm LoS} \cap \mathcal{R}_{\rm NLoS} = \emptyset$ and $\mathcal{R}_{\rm LoS} \cup \mathcal{R}_{\rm NLoS} = \mathcal{R}$. In this work, the map serves only to identify possible user locations and \ac{los}/\ac{nlos} regions, which necessitate a coarse map, e.g., a street or floor plan.

Without loss of generality, it is assumed that the prior information of the user distribution in $\mathcal{R}$ is available, which can be characterized as a \ac{pdf}, denoted by $f_{\rm D}(\mathbf{p})$, $\mathbf{p} \in \mathcal{R}$. Then, based on $f_{\rm D}(\cdot)$, the \acp{pdf} for user distribution in \ac{los} and \ac{nlos} regions, denoted by $f_{\rm D}^{\rm LoS}(\cdot)$ and $f_{\rm D}^{\rm NLoS}(\cdot)$ respectively, can also be obtained. Note that if the real user distribution is not precisely known, an estimate or prior belief of the distribution can be adopted for $f_{\rm D}(\cdot)$. Moreover, if no prior information is available, uniform distribution in $\mathcal{R}$ can be assumed by default. However, localization performance degradation may occur for the proposed framework due to prior information mismatch, as will be discussed in Sec.~\ref{sec:localization_results}.

\section{UNILocPro: Unified Localization Framework}\label{sec:UNILocPro}

\begin{figure}[t]
    \centering
    \begin{tikzpicture}[every node/.style={font=\footnotesize}]
    \node (image) [anchor=south west]{\includegraphics[width=0.86\linewidth]{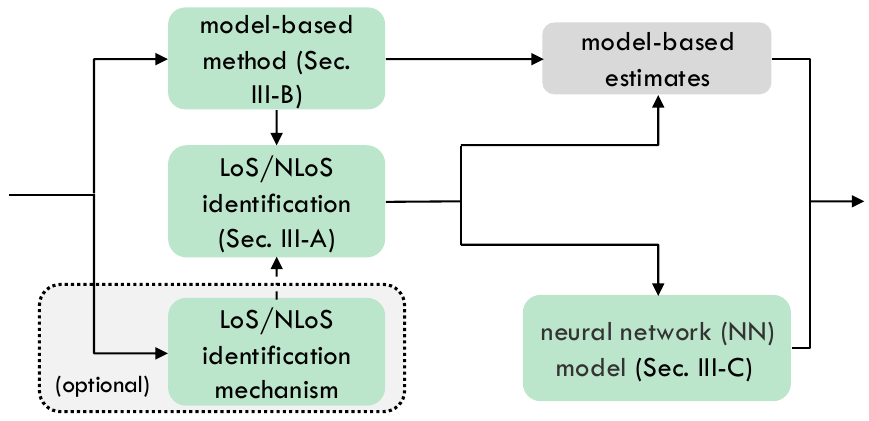}};
    \gettikzxy{(image.north east)}{\ix}{\iy};
    
    \node at (0.07*\ix,0.65*\iy)[rotate=0,anchor=north]{{$\mathbf{H}_i$}};
    \node at (0.625*\ix,0.765*\iy)[rotate=0,anchor=north]{{$I'(\mathbf{H}_i) = 1$}};
    \node at (0.625*\ix,0.44*\iy)[rotate=0,anchor=north]{{$I'(\mathbf{H}_i) = 0$}};
    \node at (0.95*\ix,0.65*\iy)[rotate=0,anchor=north]{{$\hat{\mathbf{p}}_i$}};
    
    \end{tikzpicture}
    \caption{The framework of UNILocPro (and its low-complexity variant UNILoc).}
    \label{fig:framework_UNILocPro}
    \vspace{-1mm}
\end{figure}

In this section, we introduce the proposed UNILocPro framework, which is illustrated in Fig.~\ref{fig:framework_UNILocPro}. First, \ac{los}/\ac{nlos} identification is conducted based on \ac{csi} $\mathbf{H}_i$; then, if user $i$ is identified as a \ac{los} user, the model-based method would be used to estimate its position; otherwise, a \ac{nn} model is adopted to generate $\hat{\mathbf{p}}_i$ based on $\mathbf{H}_i$. The model-based and \ac{nn}-based methods are integrated as
\begin{equation}\label{eq:unified_localization}
    \hat{\mathbf{p}}_i = f_e(\mathbf{H}_i) = \left\{ \begin{array}{ll}
        f_{\rm pe}\left[ f_{\rm ce}^{\rm sh}(\mathbf{H}_i) \right], & {\rm if} \, I'(\mathbf{H}_i) = 1;  \\
        f_{\Theta}\left[f_{\rm extr}(\mathbf{H}_i)\right], & {\rm if} \, I'(\mathbf{H}_i) = 0,
    \end{array} \right.
\end{equation}
where $f_{\rm ce}^{\rm sh}(\cdot)$ is a channel parameter estimation function; $f_{\rm pe}(\cdot)$ is a function that maps channel parameters to position based on geometric relationships; $f_{\rm extr}(\cdot)$ is a feature extraction function that transforms \ac{csi} $\mathbf{H}_i$ into the input of the \ac{nn} model; $f_{\Theta}(\cdot)$ is the forward function of the \ac{nn} model with $\Theta$ as the parameters; and $I'(\cdot)$ is a \ac{csi}-based \ac{los}/\ac{nlos} identification function.

\subsection{LoS/NLoS Identification}\label{sec:LoS_NLoS_Identification}
\ac{csi}-based \ac{los}/\ac{nlos} identification has been extensively studied, achieving accuracies of $93\% \text{-} 100\%$ under various settings~\cite{An_Intr_Zeka_2012,TDoA_Self_Mohs_2025}. We adopt a general approach in which \ac{los}/\ac{nlos} identification is represented by a function $I(\cdot)$ with an identification accuracy $p_I \in [0.5,1]$ that can distinguish \ac{los}/\ac{nlos} users based on the \ac{csi} (user $i$ is identified as a \ac{los} user if $I(\mathbf{H}_i)=1$ and as a \ac{nlos} user if $I(\mathbf{H}_i)=0$). The specific design of $I(\cdot)$ can be flexible (a lightweight example based on \ac{gmm} is provided in Appendix~\ref{app:LoS_NLoS_identification}), which allows our proposed framework to be applicable to various scenarios regardless of the specific method used for $I(\cdot)$. Furthermore, with a \ac{los} path, the mapping function $f_{\rm pe}(\cdot)$, e.g., solving geometric relationships based on \ac{aoa} and \ac{toa}, would result in a position in the \ac{los} region. This allows us to conclude that all \ac{los} users would be localized in the \ac{los} region, i.e., if a user is localized in the \ac{nlos} region, it must be a \ac{nlos} user.\footnote{However, \ac{nlos} users could be localized in \ac{los} region due to reflection, diffraction, or scattering of the physical environment.} Therefore, given the results from $I(\mathbf{H}_i)$, we can further improve the \ac{los}/\ac{nlos} identification by
\begin{equation}\label{eq:LoS_identification}
    I'(\mathbf{H}_i) = \left\{ \begin{array}{ll}
        0, & {\rm if} \, f_{\rm pe}\left[ f_{\rm ce}^{\rm sh}(\mathbf{H}_i) \right] \in \mathcal{R}_{\rm NLoS};  \\
        I(\mathbf{H}_i), & {\rm if} \, f_{\rm pe}\left[ f_{\rm ce}^{\rm sh}(\mathbf{H}_i) \right] \in \mathcal{R}_{\rm LoS},
    \end{array} \right.
\end{equation}
where $I'(\cdot)$ is the final \ac{los}/\ac{nlos} identification function.

\begin{remark}
    When the mechanism $I(\cdot)$ is not accessible or its identification accuracy $p_I$ drops below a certain threshold, a conservative method that does not rely on $I(\cdot)$ can be applied. In particular, without $I(\cdot)$, the model-based results could be used for \ac{los}/\ac{nlos} identification, i.e., if a user is localized in the \ac{los} (resp. \ac{nlos}) region, it would be identified as a \ac{los} (resp. \ac{nlos}) user. Therefore, the \ac{los}/\ac{nlos} identification~\eqref{eq:LoS_identification} can be reformulated as
\begin{equation}\label{eq:LoS_identification_conservative}
    I_{\rm conserv}'(\mathbf{H}_i) = \left\{ \begin{array}{ll}
        0, & {\rm if} \, f_{\rm pe}\left[ f_{\rm ce}^{\rm sh}(\mathbf{H}_i) \right] \in \mathcal{R}_{\rm NLoS};  \\
        1, & {\rm if} \, f_{\rm pe}\left[ f_{\rm ce}^{\rm sh}(\mathbf{H}_i) \right] \in \mathcal{R}_{\rm LoS},
    \end{array} \right.
\end{equation}
based on which \ac{los} and \ac{nlos} users can be identified completely by $f_{\rm pe}\left[ f_{\rm ce}^{\rm sh}(\mathbf{H}_i) \right]$.
\end{remark}

\subsection{Model-based Localization Method}\label{sec:model_based_method}
As discussed in Sec.~\ref{sec:channel_model}, the channel estimation algorithm estimates the channel parameters based on $\mathbf{H}_i$ by $\{ \hat{\theta}_l^i, \hat{\tau}_l^i, \hat{\beta}_l^i \}_{l = 1,2,\ldots,\hat{L}_i} = f_{\rm ce}(\mathbf{H}_i)$. For a low-complexity implementation, the shortest channel path among the recovered $\hat{L}_i$ paths is identified as the \ac{los} path and used to estimate the user position. Thus, the channel parameter estimation function $f_{\rm ce}^{\rm sh}(\cdot)$ that maps the \ac{csi} to \ac{aoa} and \ac{toa} can be written as $\{ \hat{\theta}_s^i, \hat{\tau}_s^i \} = f_{\rm ce}^{\rm sh}(\mathbf{H}_i)$, $\forall i$, where $s = \arg \min_l \hat{\tau}_l^i$, $l = 1,2,\ldots,\hat{L}_i$. Then, with \ac{aoa} and \ac{toa}, the user position estimate can be derived based on geometric relationships~\cite{A_Tuto_Chen_2022,An_Intr_Zeka_2012,Harn_NLOS_Mend_2019,Robu_Snap_Kalt_2024}. Specifically, when a \ac{ula} is adopted, given $\{ \hat{\theta}_s^i, \hat{\tau}_s^i \}$ and regarding this channel path as \ac{los} (yielding sufficient information to estimate the x-coordinate and y-coordinate for a given z-coordinate), the user position estimate can be derived by solving the following geometric relationships:
\begin{equation}\label{eq:geometric_relationships_ULA}
    c \cdot \hat{\tau}_s^i = \| \mathbf{p}_{\rm BS} - \hat{\mathbf{p}}_i \|_2, \quad \sin(\hat{\theta}_s^i) = \frac{(\mathbf{p}_{\rm BS} - \hat{\mathbf{p}}_i) \cdot \mathbf{n}_{\rm ULA}}{\|\mathbf{p}_{\rm BS} - \hat{\mathbf{p}}_i \|_2},
\end{equation}
where $\mathbf{p}_{\rm BS}$ is the position of the \ac{bs} and $\mathbf{n}_{\rm ULA}$ is the normalized vector representing the direction of the \ac{ula}. Similarly, the geometric relationships for a \ac{upa} can be derived based on the \ac{toa} and the \ac{aoa} in both azimuth and elevation angles. Therefore, the model-based estimate can be written as $\hat{\mathbf{p}}_i^{\rm mb} = f_{\rm pe}\left[ f_{\rm ce}^{\rm sh}(\mathbf{H}_i) \right]$, $\forall i$, where $f_{\rm pe}(\cdot)$ is obtained by solving the relationships between the user position and the channel parameters~\cite{Unif_Loca_Yuha_2025,A_Tuto_Chen_2022,An_Intr_Zeka_2012}.

\subsection{CC-based Localization Method}\label{sec:CC_method}
In order to obtain a parametric \ac{cc} model, \ac{nn} is adopted to map \ac{csi} to position.\footnote{In our implementation, an $L_{\rm MLP}$-layer \ac{mlp} with layer widths $\{n_l\}_{l=1, \ldots,L_{\rm MLP}}$ is adopted for the \ac{nn} model (we also use $n_0$ to denote the dimensionality of $\mathbf{s}_i$).} As for \ac{csi} preprocessing, we first convert the antenna-frequency domain \ac{csi} $\mathbf{H}_i$ to the angle-delay domain, as denoted by $\bar{\mathbf{H}}_i$~\cite{Angl_Prof_Step_2024}. Then, $\bar{\mathbf{H}}_i$ can be truncated by taking the columns that have delays within the minimum \ac{toa} $\tau_{\min}$ and the maximum \ac{toa} $\tau_{\max}$, i.e., $\bar{\mathbf{H}}_i^{\rm t} = [\bar{\mathbf{H}}_i]_{:,\left \lfloor{\tau_{\min} \Delta_f N_c}\right \rfloor :\left \lceil \tau_{\max} \Delta_f N_c\right \rceil }$, $\forall i$, whose element-wise amplitude and phase are used to generate the input to the \ac{nn}~\cite{Chan_Char_Stud_2018,Indo_Loca_Stah_2023,Larg_Wire_Pan_2025}, as denoted by $\mathbf{s}_i = {\rm vec}\left( \left[\log (|\bar{\mathbf{H}}_i^{\rm t}|), \angle \bar{\mathbf{H}}_i^{\rm t} \right] \right)$, $\forall i$. Thus, implementing \ac{csi} domain conversion, truncation, and element-wise extraction, $f_{\rm extr}(\cdot)$ transforms $\mathbf{H}_i$ to the input of the \ac{nn} model, i.e., $\mathbf{s}_i = f_{\rm extr}(\mathbf{H}_i)$, based on which the output of the \ac{nn} model, i.e., the user position estimate $\hat{\mathbf{p}}_i^{\rm nb}$, can be expressed by $\hat{\mathbf{p}}_i^{\rm nb} = f_{\Theta}[f_{\rm extr}(\mathbf{H}_i)]$, $\forall i$.

\subsubsection{CC Loss Function}
\begin{figure}[t]
    \centering
    \begin{tikzpicture}[every node/.style={font=\footnotesize}]
    \node (image) [anchor=south west]{\includegraphics[width=0.86\linewidth]{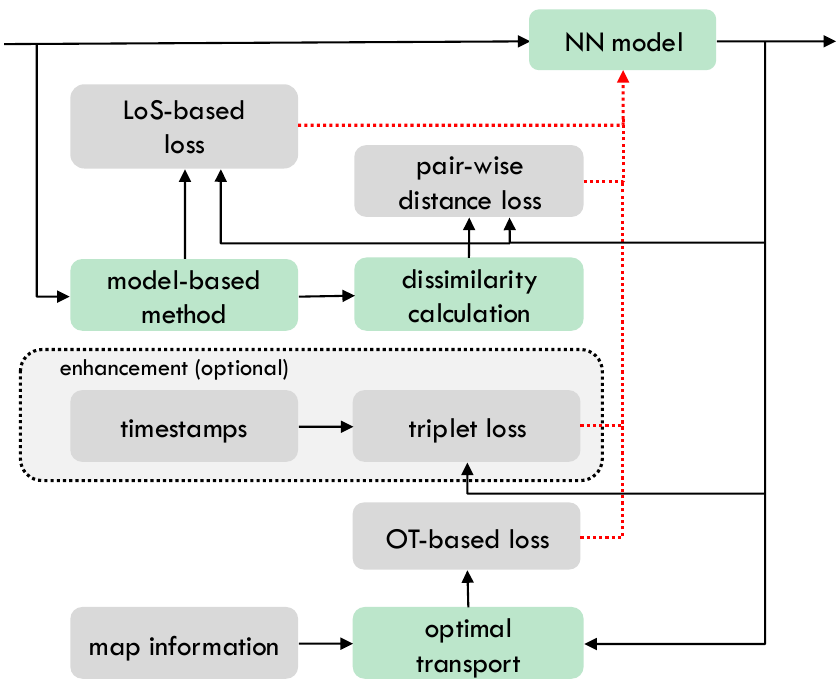}};
    \gettikzxy{(image.north east)}{\ix}{\iy};

    \node at (0.07*\ix,0.99*\iy)[rotate=0,anchor=north]{{$\{\mathbf{H}_i\}$}};
    \node at (0.19*\ix,0.72*\iy)[rotate=0,anchor=north]{{LoS}};
    \node at (0.91*\ix,1*\iy)[rotate=0,anchor=north]{{$\{\hat{\mathbf{p}}_i^{\rm nb}\}$}};
    
    \end{tikzpicture}
    \caption{The \ac{cc} training stage of the \ac{nn} model in UNILocPro (the red dashed arrow line indicates the path of gradient backpropagation).}
    \label{fig:UNILocPro_training}
    \vspace{-1mm}
\end{figure}

Given a dataset $\mathcal{S} = \{(\mathbf{H}_i, \mathbf{p}_i)\}_{i = 1,2,\ldots,N_u}$, where $N_u$ is the number of user positions, by executing $\mathbf{s}_i = f_{\rm extr}(\mathbf{H}_i)$ and $I'(\mathbf{H}_i)$, the dataset $\mathcal{S}' = \{(\mathbf{s}_i, I'(\mathbf{H}_i))\}_{i = 1,2,\ldots,N_u}$ can be constructed (the ground-truth positions $\{\mathbf{p}_i\}$ are not included), which can be used to update the parameters $\Theta$ by minimizing an overall \ac{cc} loss function (as shown in Fig.~\ref{fig:UNILocPro_training}):
\begin{equation}\label{eq:loss_func_UNILocPro}
    \mathcal{L}_{\rm CC} = \omega_{\rm B-CC} \cdot \mathcal{L}_{\rm B-CC} + \omega_{\rm LoS} \cdot \mathcal{L}_{\rm LoS} + \omega_{\rm OT} \cdot \mathcal{L}_{\rm OT},
\end{equation}
where $\mathcal{L}_{\rm B-CC}$ is the baseline \ac{cc} loss~\cite{Chan_Char_Stud_2018,Trip_Wire_Ferr_2021}, which generates a channel chart that preserves local geometry; $\mathcal{L}_{\rm LoS}$ is the \ac{los}-based loss, which aligns the channel chart with the global coordinates by using identified \ac{los} users as anchors and also improves their localization accuracy; $\mathcal{L}_{\rm OT}$ is the map-assisted loss (computed based on \ac{ot}), which constrains the estimated positions of identified \ac{nlos} users in the \ac{nlos} region (these losses will be detailed in Sec.~\ref{sec:cc_loss}). These losses are weighted by positive coefficients $\omega_{\rm B-CC}$, $\omega_{\rm LoS}$, and $\omega_{\rm OT}$, respectively.

\subsubsection{Data Augmentation and Training Strategy}
Due to imperfect channel estimation, \ac{csi} samples collected from the same physical location may exhibit varying input features, potentially degrading the generalization performance of the \ac{nn} model. To mitigate this issue, we adopt a data augmentation strategy that generates multiple augmented samples for each original \ac{csi} by adding random noise. By training the \ac{nn} model with these augmented samples, the model is encouraged to learn the mapping that is robust against the variability in the \ac{csi} features, thus preventing it from overfitting to specific realizations corrupted by noise. This approach not only enhances model robustness but also encourages the \ac{nn} model to capture more fundamental localization-related patterns. In particular, we perform a multi-stage training process:
\begin{itemize}
\item In the initialization stage, the \ac{nn} model is pre-trained on the original dataset $\mathcal{S}'$ with the model-based estimates $\hat{\mathbf{p}}_i^{\rm mb}$ as pseudo-labels in a supervised manner, where the loss is $\mathcal{L}_{\rm init} = \frac{1}{N_u} \sum_{i=1}^{N_u} \left\| f_{\Theta} (\mathbf{s}_i) - \hat{\mathbf{p}}_i^{\rm mb} \right\|_2^2$. This pre-training stage helps to provide a good initialization for the \ac{nn} model, which can facilitate the subsequent training stages.
\item In the \ac{cc} training stage (as shown in Fig.~\ref{fig:UNILocPro_training}), given the dataset $\mathcal{S}'$, the \ac{nn} model is trained using the multi-component loss~\eqref{eq:loss_func_UNILocPro} through backpropagation. The goal is to learn a mapping that can provide a good fit, while robustness against noise is not explicitly addressed.
\item In the data augmentation stage, we mainly focus on improving the generalization performance of the \ac{nn} model by performing data augmentation to mitigate the impact of noise. Specifically, for each original \ac{csi} $\mathbf{H}_i$, we generate multiple augmented samples by adding random noise matrix $\mathbf{H}_i^{\mathrm{aug},k} = \mathbf{H}_i + \bar{\mathbf{N}}_i^k$, $k = 1,2,\dots,K$, where $K$ is the number of augmented samples generated for each original \ac{csi}, and $\bar{\mathbf{N}}_i^k$ denotes the artificial noise matrix. Note that $\operatorname{vec} (\bar{\mathbf{N}}_i^k) \sim \mathcal{CN}(0, (\bar{\sigma}_i^k)^2 \mathbf{I}_{M N_c})$, where $(\bar{\sigma}_i^k)^2$ is randomly generated from a predefined range $[\sigma_{\min}^2, \sigma_{\max}^2]$, which is chosen to be the range that the \ac{nn} model is expected to be robust against. Then, given the trained \ac{nn} model from the \ac{cc} training stage, denoted by $f_{\Theta_0}(\cdot)$, we feed $\mathbf{H}_i$ into the \ac{nn} model to obtain the position estimate $\hat{\mathbf{p}}_i^{\mathrm{aug}, k} =f_{\Theta_0}[f_{\rm extr}(\mathbf{H}_i)]$, which is treated as the pseudo-label for the augmented samples. Therefore, the augmented dataset can be constructed as $\{ (\mathbf{H}_i^{\mathrm{aug},k}, \hat{\mathbf{p}}_i^{\mathrm{aug},k}) \}_{i,k}$, which is used to further train the \ac{nn} model with a loss between the \ac{nn} output and the pseudo-label, as given by
\begin{equation}\nonumber
\mathcal{L}_{\rm aug} = \frac{1}{N_u K} \sum_{i=1}^{N_u} \sum_{k=1}^K \left\| f_{\Theta} [ f_{\rm extr} (\mathbf{H}_i^{\mathrm{aug},k}) ] - \hat{\mathbf{p}}_i^{\mathrm{aug},k} \right\|_2^2.
\end{equation}
\end{itemize}

It is noted that while the \ac{cc} model can be directly trained with augmented samples after the initialization stage, the proposed multi-stage training strategy is designed for a more practical and efficient implementation. Instead of constructing pairwise/triplet relationships and computing the Sinkhorn-based \ac{ot} loss over the augmented samples, the proposed strategy allows the augmented samples to inherit pseudo-labels from the \ac{cc} model trained on the original dataset and be directly used for the supervision of regression, which significantly reduces computational complexity and memory consumption while preserving essentially the same localization objective.

\section{Multi-Component \ac{cc} Loss}\label{sec:cc_loss}
In this section, we present each loss term in the overall \ac{cc} loss $\mathcal{L}_{\rm CC}$ in \eqref{eq:loss_func_UNILocPro} and the proposed dissimilarity metrics.

\subsection{Baseline CC Loss}\label{sec:baseline_cc_loss}
In the literature, the baseline \ac{cc} loss functions usually contain the pairwise distance loss $\mathcal{L}_{\rm PW}$ and the triplet loss $\mathcal{L}_{\rm Tri}$, based on which $\mathcal{L}_{\rm B-CC}$ can be written as~\cite{Chan_Char_Stud_2018,Trip_Wire_Ferr_2021}
\begin{equation}
    \mathcal{L}_{\rm B-CC} = \omega_{\rm PW} \mathcal{L}_{\rm PW} + \omega_{\rm Tri} \mathcal{L}_{\rm Tri},
\end{equation}
where $\omega_{\rm PW}$ and $\omega_{\rm Tri}$ are non-negative coefficients. Given the dissimilarity metric $d(\cdot, \cdot)$, $\mathcal{L}_{\rm PW}$ preserves the distance $d_{i,j} = d(\mathbf{H}_i, \mathbf{H}_j)$ in the channel chart, as given by~\cite{Chan_Char_Stud_2018,Angl_Prof_Step_2024}
\begin{equation}\label{eq:loss_func_pairwise}
    \mathcal{L}_{\rm PW} = \frac{1}{N_u^2}\sum_{\mathbf{s}_i \in \mathcal{S}'} \sum_{\mathbf{s}_j \in \mathcal{S}'} \frac{\left( \| f_{\Theta}\left(\mathbf{s}_i\right) - f_{\Theta}\left(\mathbf{s}_j\right) \|_2 - d_{i,j} \right) ^2}{d_{i,j}}.
\end{equation}
Given a triplet $(\mathbf{H}_p, \mathbf{H}_n, \mathbf{H}_r) \in \mathcal{T}$, where $d(\mathbf{H}_r, \mathbf{H}_p) < d(\mathbf{H}_r, \mathbf{H}_n)$ and $\mathcal{T} \subseteq \mathcal{S}^3$ is the triplet set, $\mathcal{L}_{\rm Tri}$~\eqref{eq:triplet_loss} with $\gamma > 0$ being a margin parameter preserves the distance relationship of triplet samples in the channel chart, i.e., $\| \hat{\mathbf{p}}_p^{\rm nb} - \hat{\mathbf{p}}_r^{\rm nb} \|_2 < \| \hat{\mathbf{p}}_n^{\rm nb} - \hat{\mathbf{p}}_r^{\rm nb} \|_2$~\cite{Trip_Wire_Ferr_2021,Chan_Char_Sued_2025}.

\begin{figure*}
    \begin{equation}\label{eq:triplet_loss}
    \mathcal{L}_{\rm Tri} = \frac{1}{|\mathcal{T}|} \sum_{(\mathbf{H}_p, \mathbf{H}_n, \mathbf{H}_r) \in \mathcal{T}} \max\left(0, \| f_{\Theta}\left[f_{\rm extr}(\mathbf{H}_p)\right] - f_{\Theta}\left[f_{\rm extr}(\mathbf{H}_r)\right] \|_2 - \| f_{\Theta}\left[f_{\rm extr}(\mathbf{H}_n)\right] - f_{\Theta}\left[f_{\rm extr}(\mathbf{H}_r)\right] \|_2 + \gamma\right).
\end{equation}
\rule{\textwidth}{0.05pt}
\vspace{-5mm}
\end{figure*}

For localization, it is desired that $d(\mathbf{H}_i, \mathbf{H}_j) = \| \mathbf{p}_i - \mathbf{p}_j \|_2$, which, however, is not directly accessible since supervision is not available. In practice, it is challenging to design, in an unsupervised manner, a metric that is consistent with the physical distance based solely on \ac{csi}, i.e., only relying on manipulations of \ac{csi}, as the physical distance may not be directly reflected in \ac{csi} pairs~\cite{Chan_Char_Tous_2022,Angl_Feat_Uyka_2024}. In this paper, we use two different metrics (as well as their fusion) based on the \ac{gospa} metric and the Wasserstein distance, which could serve as approximations of the physical distance and thus achieve good \ac{cc} and localization performance.

With the estimated channel parameters of the $\hat{L}_i$ recovered paths $\{ \hat{\beta}_l^i, \hat{\theta}_l^i, \hat{\tau}_l^i \}$, $\hat{L}_i$ positions can be generated for $\mathbf{H}_i$ via the channel parameters-to-position function $f_{\rm pe}(\cdot)$, as denoted by $\{ \hat{\mathbf{p}}_{i,l}^{\rm mb} \}_{l=1,\ldots,\hat{L}_i}$ (can be regarded as a signature of $\mathbf{H}_i$), where $\hat{\mathbf{p}}_{i,l}^{\rm mb}$ is the model-based position estimate for the $l$-th channel path of $\mathbf{H}_i$. Here, we treat every path as \ac{los} to compute a coarse estimate, which serves not as an accurate location, but as a geometric surrogate that can be used for dissimilarity computations. Then, the dissimilarity metric $d(\mathbf{H}_i, \mathbf{H}_j)$ can be designed as the distance between the sets of positions $\{\hat{\mathbf{p}}_{i,l}^{\rm mb}\}_{l=1,\ldots,\hat{L}_i}$ and $\{\hat{\mathbf{p}}_{j,l}^{\rm mb}\}_{l=1,\ldots,\hat{L}_j}$, as written by
\begin{equation}\label{eq:dissimilarity_metric}
    d(\mathbf{H}_i, \mathbf{H}_j) = \mathscr{D}\left( \{\hat{\mathbf{p}}_{i,l}^{\rm mb}\}_{l=1,\ldots,\hat{L}_i}, \{\hat{\mathbf{p}}_{j,l}^{\rm mb}\}_{l=1,\ldots,\hat{L}_j} \right),
\end{equation}
where $\mathscr{D}(\cdot, \cdot)$ is a metric that measures the dissimilarity between two sets of positions. In this paper, we use the \ac{gospa} metric and the Wasserstein distance as $\mathscr{D}(\cdot, \cdot)$, both of which are effective for measuring the dissimilarity between sets of points.

The developed dissimilarity metric admits a clear physical interpretation rooted in the geometric structure of wireless propagation. Instead of directly comparing \ac{csi}, it evaluates the distance between users through the similarity of their multipath-induced geometric signatures. Specifically, each \ac{csi} sample is transformed into a set of path-wise position hypotheses using the estimated channel parameters, effectively mapping the \ac{csi} into a cloud of virtual points that represent the underlying propagation environment. In this view, each user is no longer represented by a single location, but by a distribution of geometric hypotheses corresponding to its multipath components. The dissimilarity between two \ac{csi} samples is thus defined as the distance between their corresponding signatures in the induced geometric space, which can be interpreted as approximating spatial proximity through the similarity of induced multipath patterns. This approach leverages the inherent structure of the propagation environments, providing a physically meaningful foundation for geometry preservation in \ac{cc}.

\subsubsection{\ac{gospa}-Based Dissimilarity}

\begin{figure*}[t]
    \centering
    \begin{subfigure}{0.22\textwidth}
        \includegraphics[width=\linewidth]{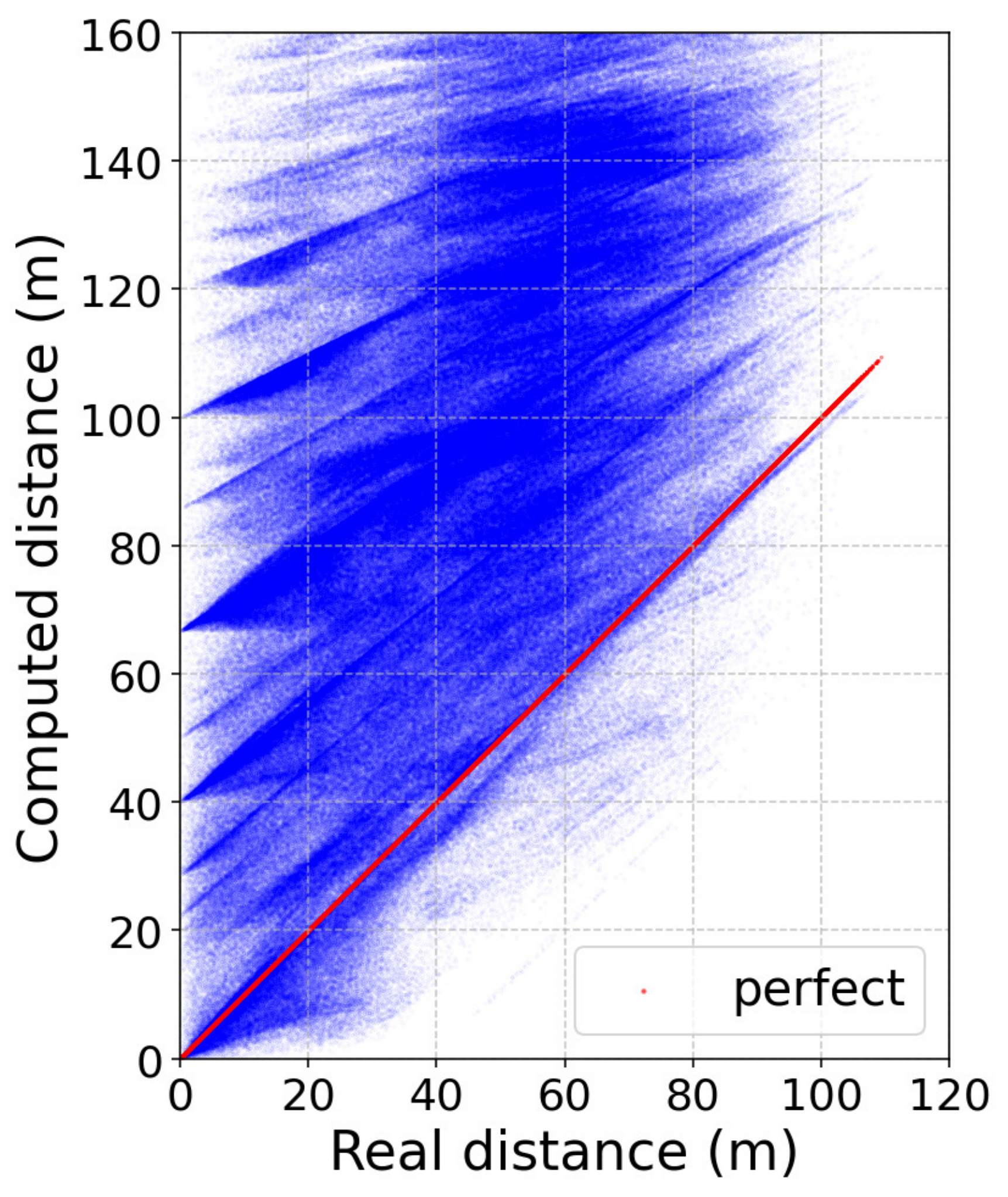}
        \caption{$d_{i,j}^{\mathrm{GOSPA}}$ ($\zeta = 200$)}
    \end{subfigure}
    \begin{subfigure}{0.22\textwidth}
        \includegraphics[width=\linewidth]{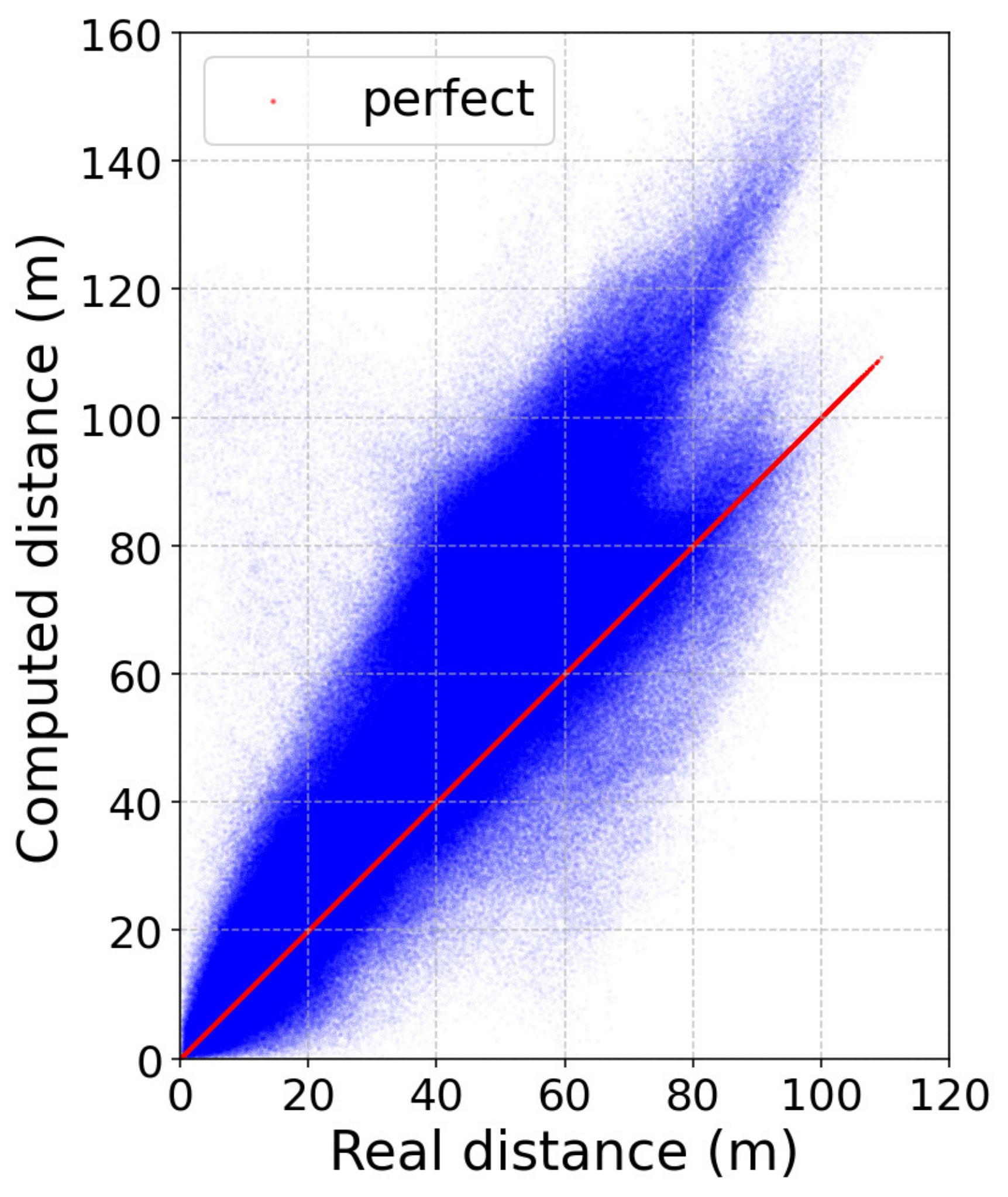}
        \caption{$d_{i,j}^{\mathrm{G-GOSPA}}$ ($\zeta = 20$)}
    \end{subfigure}
    \begin{subfigure}{0.22\textwidth}
        \includegraphics[width=\linewidth]{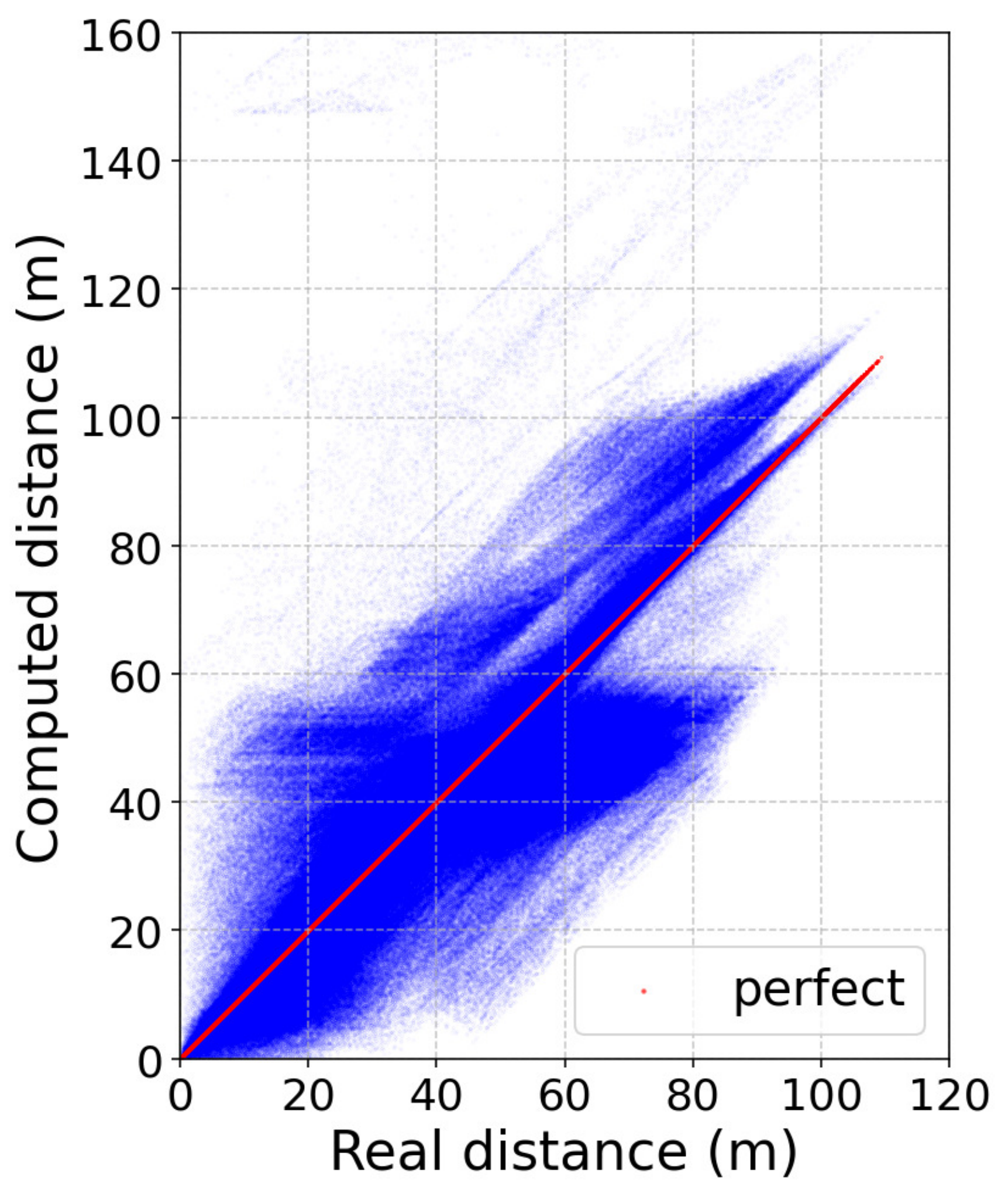}
        \caption{$d_{i,j}^{\mathrm{Wass}}$}
    \end{subfigure}
    \begin{subfigure}{0.22\textwidth}
        \includegraphics[width=\linewidth]{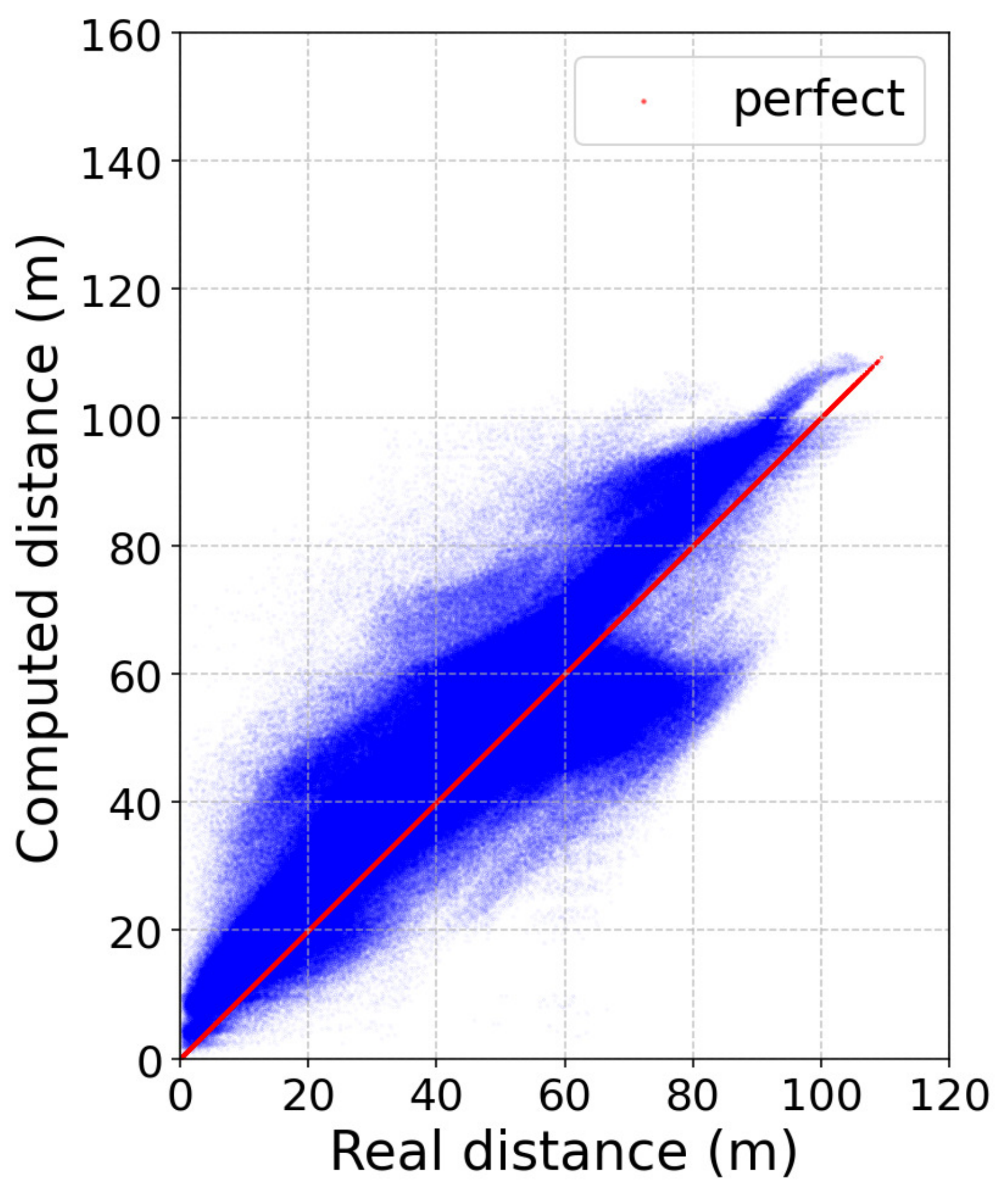}
        \caption{$d_{i,j}^{\mathrm{Fusi}}$}
    \end{subfigure}
    \caption{The computed dissimilarity versus the physical distance between user positions, where the red line represents the real distance (using the same system setting as described in Sec.~\ref{sec:system_setting}).}
    \label{fig:distance_results}
    \vspace{-4mm}
\end{figure*}

As a metric between two sets of points, the \ac{gospa} metric~\eqref{eq:gospa_metric} with $p = 1$, $\varpi = 2$, and $d(\mathbf{x},\mathbf{y}) = \| \mathbf{x} - \mathbf{y} \|_2$ can be used for $d(\mathbf{H}_{i}, \mathbf{H}_{j})$ to indicate the physical distance, i.e.,
\begin{equation}\label{eq:GOSPA_dissimilarity}
    d_{i,j}^{\mathrm{GOSPA}} = \frac{\mathscr{G}_{1}^{(\zeta,2)}(\{\hat{\mathbf{p}}_{i,l}^{\rm mb}\}_{l=1,\ldots,\hat{L}_i}, \{\hat{\mathbf{p}}_{j,l}^{\rm mb}\}_{l=1,\ldots,\hat{L}_j})}{(\hat{L}_i + \hat{L}_j)/2},
\end{equation}
where normalization is applied such that $d_{i,j}^{\mathrm{GOSPA}}$ has the scale of the physical distance between a pair of positions. As shown in Fig.~\ref{fig:distance_results}, directly using $d_{i,j}^{\mathrm{GOSPA}}$ by setting the cut-off value $\zeta$ to a large value is not appropriate since reflection, diffraction, and/or scattering may bias the \ac{gospa} metric. Therefore, in order to have a better representation, we can set $\zeta$ to a relatively small value and use the geodesic dissimilarity~\cite{Indo_Loca_Stah_2023,Angl_Prof_Step_2024} based on $\{ d_{i,j}^{\mathrm{GOSPA}}\}$ to compute the geodesic \ac{gospa} metric as
\begin{equation}\label{eq:gospa_geodesic}
    d_{i,j}^{\mathrm{G-GOSPA}} = \sum_{k \in \mathcal{P}_{i,j}} d(\mathbf{H}_{k}, \mathbf{H}_{k+1}) = \sum_{k \in \mathcal{P}_{i,j}} d_{k,k+1}^{\mathrm{GOSPA}},
\end{equation}
where $k$ and $k+1$ are the indices of all neighboring pairs of points on the shortest path $\mathcal{P}_{i,j}$ between $\mathbf{H}_{i}$ and $\mathbf{H}_{j}$, which can be obtained by utilizing shortest path algorithms, e.g., Dijkstra's algorithm, on the neighbor graphs constructed from $\{ d_{i,j}^{\mathrm{GOSPA}}\}$. More details about the geodesic dissimilarity can be found in~\cite{Indo_Loca_Stah_2023,Angl_Prof_Step_2024}. As shown in Fig.~\ref{fig:distance_results}, the geodesic \ac{gospa} metric $d_{i,j}^{\mathrm{G-GOSPA}}$ is more consistent with the physical distance than the \ac{gospa} metric $d_{i,j}^{\mathrm{GOSPA}}$.

\subsubsection{Wasserstein-Based Dissimilarity}
As the Wasserstein distance is a metric that measures the dissimilarity between two probability distributions, it can also be used as $\mathscr{D}(\cdot, \cdot)$ when probability distributions are defined for the sets of positions $\{\hat{\mathbf{p}}_{i,l}^{\rm mb}\}_{l=1,\ldots,\hat{L}_i}$ and $\{\hat{\mathbf{p}}_{j,l}^{\rm mb}\}_{l=1,\ldots,\hat{L}_j}$. Then, our next task is to construct a spatial probability distribution, denoted by $\mathbf{u}_i$, for each set of positions $\{\hat{\mathbf{p}}_{i,l}^{\rm mb}\}_{l=1,\ldots,\hat{L}_i}$, $\forall i$. Following~\cite{Chan_Char_Tous_2022}, where the probability distribution can be generated from the received path powers, we can assign a probability mass to each recovered position in $\{\hat{\mathbf{p}}_{i,l}^{\rm mb}\}_{l=1,\ldots,\hat{L}_i}$, $\forall i$, based on the reliability of its underlying estimated parameters $\{ \hat{\beta}_l^i, \hat{\theta}_l^i, \hat{\tau}_l^i \}$, which can be written as $[\mathbf{u}_i]_l = p(\hat{\beta}_l^i, \hat{\theta}_l^i, \hat{\tau}_l^i)$, $\forall l$, where $p(\cdot)$ is a probability mass function. In general, stronger paths are typically estimated with higher accuracy and are more likely to represent genuine physical geometry. Furthermore, the \ac{los} path and the earliest \ac{nlos} path are often the most geometrically reliable. Therefore, we aim to assign higher probability to paths that have high energy and arrive early, prioritizing the most stable and geometrically informative components of the channel. As the channel paths that arrive early usually have higher channel gains, being an example implementation, we simply use
\begin{equation}\label{eq:probability_emd}
    p(\hat{\beta}_l^i, \hat{\theta}_l^i, \hat{\tau}_l^i) = \frac{\exp(-\kappa \hat{\tau}_l^i)}{\sum_{l} \exp(-\kappa \hat{\tau}_l^i)}, \quad \forall l,
\end{equation}
where $\kappa>0$ is a parameter. For implementation, we set $\kappa$ to be a relatively large value to emphasize the short paths.

Given $\{\hat{\mathbf{p}}_{i,l}^{\rm mb}\}_{l=1,\ldots,\hat{L}_i}$ and $\{\hat{\mathbf{p}}_{j,l}^{\rm mb}\}_{l=1,\ldots,\hat{L}_j}$, the first-order Wasserstein distance between $\mathbf{u}_i$ and $\mathbf{u}_j$ can be used as the dissimilarity metric $d(\mathbf{H}_i, \mathbf{H}_j)$, which is written as
\begin{equation}
    d_{i,j}^{\mathrm{Wass}} = \min_{\mathbf{\Gamma} \in \mathcal{B}} \quad \langle\mathbf{\Gamma}, \mathbf{C}\rangle_F,
\end{equation}
where $\mathbf{C} \in \mathbb{R}_+^{\hat{L}_i \times \hat{L}_j}$ with $[\mathbf{C}]_{l,l'} = \| \hat{\mathbf{p}}_{i,l}^{\rm mb} - \hat{\mathbf{p}}_{j,l'}^{\rm mb} \|_2$; and $\mathcal{B} = \{ \mathbf{\Gamma} \in \mathbb{R}_+^{\hat{L}_i \times \hat{L}_j} | \mathbf{\Gamma} \cdot \mathbf{1}_{\hat{L}_j} = \mathbf{u}_i, \,\, \mathbf{\Gamma}^{\mathsf{T}} \cdot \mathbf{1}_{\hat{L}_i} = \mathbf{u}_j\}$. Similar to the \ac{gospa} metric $d_{i,j}^{\mathrm{GOSPA}}$, the Wasserstein distance $d_{i,j}^{\mathrm{Wass}}$ can also be used to compute the geodesic dissimilarity~\eqref{eq:gospa_geodesic}. However, as $d_{i,j}^{\mathrm{Wass}}$ with a very large $\kappa$ is used in the later fusion step~\eqref{eq:dissimilarity_fusion} to capture the global geometry, the geodesic dissimilarity is not used for $d_{i,j}^{\mathrm{Wass}}$ in our implementation, which in general is optional. Moreover, as shown in Fig.~\ref{fig:distance_results}, unlike the \ac{gospa} metric $d_{i,j}^{\mathrm{GOSPA}}$, $d_{i,j}^{\mathrm{Wass}}$ can sufficiently represent the physical distance, which validates our choice of treating geodesic computation as optional.

\subsubsection{Dissimilarity Fusion}
\begin{figure}[t]
    \centering
    \begin{tikzpicture}[every node/.style={font=\footnotesize}]
    \node (image) [anchor=south west]{\includegraphics[width=0.95\linewidth]{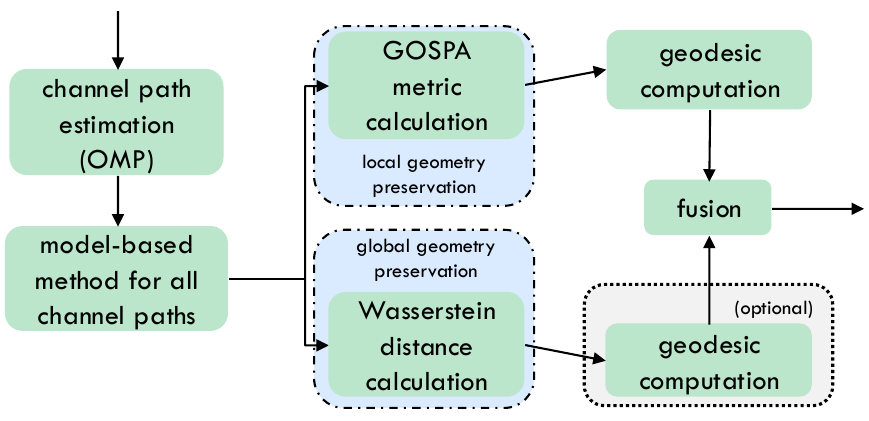}};
    \gettikzxy{(image.north east)}{\ix}{\iy};
    
    \node at (0.1*\ix,0.97*\iy)[rotate=0,anchor=north]{{$\{\mathbf{H}_i\}$}};
    \node at (0.05*\ix,0.6*\iy)[rotate=0,anchor=north]{{$\{ \hat{\beta}_l^i, \hat{\theta}_l^i, \hat{\tau}_l^i \}$}};
    \node at (0.3*\ix,0.48*\iy)[rotate=0,anchor=north]{{$\{\hat{\mathbf{p}}_{i,l}^{\rm mb}\}$}};
    \node at (0.93*\ix,0.635*\iy)[rotate=0,anchor=north]{{$d_{i,j}^{\mathrm{Fusi}}$}};
    
    \end{tikzpicture}
    \caption{The flow graph of computing the fused dissimilarity metric.}
    \label{fig:dissimilarity_metric_flow}
\end{figure}

In addition to the \ac{gospa}-based and Wasserstein-based dissimilarity metrics, we can also fuse these two metrics to achieve a better dissimilarity metric. Before fusion, we first explore these two dissimilarity metrics such that a proper fusion strategy can be designed. As the \ac{gospa}-based metric is sensitive to outliers such that a cut-off value $\zeta$ and geodesic dissimilarity are adopted, it is more effective in capturing local geometry, while by setting a large value for $\kappa$ in \eqref{eq:probability_emd}, the Wasserstein distance can represent the overall shape of global geometry by emphasizing the most reliable paths and being robust to outliers. Therefore, fusing these two metrics can potentially achieve a better representation of the physical distance by capturing both local and global geometry, which is crucial for accurate localization. The computed dissimilarity metrics $d_{i,j}^{\mathrm{G-GOSPA}}$ and $d_{i,j}^{\mathrm{Wass}}$ are shown in Fig.~\ref{fig:distance_results}, where it can be observed that $d_{i,j}^{\mathrm{G-GOSPA}}$ is more consistent when the physical distance is small, while $d_{i,j}^{\mathrm{Wass}}$ is more consistent when the physical distance is large. Therefore, to blend the complementary benefits of the two approaches, we can fuse the \ac{gospa}-based and Wasserstein-based metrics by
\begin{equation}\label{eq:dissimilarity_fusion}
    d_{i,j}^{\mathrm{Fusi}} = \alpha_{i,j} \cdot d_{i,j}^{\mathrm{G-GOSPA}} + (1-\alpha_{i,j}) \cdot d_{i,j}^{\mathrm{Wass}},
\end{equation}
where $\alpha_{i,j} \in [0,1]$ is a coefficient. The overall flow graph of computing the fused dissimilarity metric is shown in Fig.~\ref{fig:dissimilarity_metric_flow}.

To assign a large value to $\alpha_{i,j}$ when $d_{i,j}^{\mathrm{G-GOSPA}}$ is small and a small value when $d_{i,j}^{\mathrm{Wass}}$ is large, we adopt the following method to compute $\alpha_{i,j}$ in our implementation:
\begin{equation}\nonumber
    \alpha_{i,j} = \left\{ \begin{array}{ll}
        1, & {\rm if} \, d_{i,j}^{\mathrm{G-GOSPA}} \leq d_{\rm thre};  \\
        (1 + \vartheta \cdot d_{i,j}^{\mathrm{G-GOSPA}})^{-1}, & {\rm if} \, d_{i,j}^{\mathrm{G-GOSPA}} > d_{\rm thre},
    \end{array} \right.
\end{equation}
where $\vartheta \geq 0$ and $d_{\rm thre} \geq 0$ are prespecified parameters that control the fusion of the two metrics.

\subsubsection{Timestamps}
When timestamps are available for \ac{csi} measurements, they indicate the time difference between \ac{csi} samples, which can be exploited to enhance \ac{cc} and localization performance~\cite{Trip_Wire_Ferr_2021,Chan_Char_Sued_2025,Angl_Prof_Step_2024}. For a trajectory $\mathcal{S}_{\text{tr}}$, each \ac{csi} measurement $\mathbf{H}_i$ is associated with a position $\mathbf{p}_i$ and a timestamp $t_i$, i.e., $\mathcal{S}_{\text{tr}} = \{(\mathbf{H}_i, \mathbf{p}_i, t_i)\}$. Intuitively, the time difference between two \ac{csi} samples can be used as a dissimilarity metric~\cite{Angl_Prof_Step_2024}:
\begin{equation}\label{eq:distance_time_difference}
d_{i,j}^{\mathrm{time}} = |t_i - t_j|, \quad \forall i,j.
\end{equation}
However, $d_{i,j}^{\mathrm{time}}$ becomes unreliable for large time gaps, limiting \ac{cc} performance if directly applied in the pairwise distance loss $\mathcal{L}_{\rm PW}$~\eqref{eq:loss_func_pairwise}. Instead, it can be used to construct triplet sets for training~\cite{Trip_Wire_Ferr_2021,Chan_Char_Sued_2025}. Specifically, based on $\{d_{i,j}^{\mathrm{time}}\}$, the triplet set $\mathcal{T}_{\text{tr}} \subseteq \mathcal{S}_{\text{tr}}^3$ can be defined as
\begin{equation}\label{eq:triplet_set}
\mathcal{T}_{\text{tr}} = \{(\mathbf{H}_p, \mathbf{H}_n, \mathbf{H}_r)\} \subseteq \mathcal{S}_{\text{tr}}^3,
\end{equation}
where $0 \le |t_p - t_r| \le T_{\rm L} < |t_n - t_r| \le T_{\rm U}$ with $T_{\rm L}$ and $T_{\rm U}$ being parameters. Then, the corresponding triplet loss can be obtained by summing over all trajectories as defined in~\eqref{eq:triplet_loss}.

\subsection{LoS-Based Loss}
Since precise position estimates can be obtained by model-based methods for \ac{los} users, a loss term corresponding to the estimation accuracy of the identified \ac{los} user positions, i.e., $\mathcal{L}_{\rm LoS}$ in \eqref{eq:loss_func_UNILocPro}, is designed as
\begin{equation}\label{eq:loss_func_LoS}
    \mathcal{L}_{\rm LoS} = \frac{1}{\sum_i I'(\mathbf{H}_i)} \sum_{\{\mathbf{s}_i | I'(\mathbf{H}_i)=1\}} \| f_{\Theta}\left(\mathbf{s}_i\right) - \hat{\mathbf{p}}_i^{\rm mb} \|^2_2.
\end{equation}
Note that $\mathcal{L}_{\rm LoS}$ is only applied to the identified \ac{los} user positions, i.e., when $I'(\mathbf{H}_i) = 1$, which encourages the \ac{nn} model to learn the mapping from \ac{csi} to the position for \ac{los} users, leveraging the accurate model-based estimates as supervision. This is critical for \ac{cc} to preserve the global geometry, as the model-based estimates for identified \ac{los} users are used as anchor locations, such that the generated channel chart can match the global coordinates during training.

\subsection{OT-based Loss}
For identified \ac{los} user positions, it can be seen that the position estimate $\hat{\mathbf{p}}_i^{\rm nb}$ from the \ac{nn} model is in the \ac{los} region, i.e., $\hat{\mathbf{p}}_i^{\rm nb} \in \mathcal{R}_{\rm LoS}$, since the \ac{los}-based loss $\mathcal{L}_{\rm LoS}$~\eqref{eq:loss_func_LoS} is adopted, where $\hat{\mathbf{p}}_i^{\rm mb} \in \mathcal{R}_{\rm LoS}$ (minimizing $\mathcal{L}_{\rm LoS}$~\eqref{eq:loss_func_LoS} provides $\hat{\mathbf{p}}_i^{\rm nb} = \hat{\mathbf{p}}_i^{\rm mb}$). However, for identified \ac{nlos} user positions, the position estimate $\hat{\mathbf{p}}_i^{\rm nb}$ is not necessarily in the \ac{nlos} region, i.e., $\hat{\mathbf{p}}_i^{\rm nb} \notin \mathcal{R}_{\rm NLoS}$ for some user positions. With the map information, it is known that all \ac{nlos} user positions should be in the \ac{nlos} region, and therefore the \ac{ot}-based loss $\mathcal{L}_{\rm OT}$ can be introduced to ensure that the position estimates $\hat{\mathbf{p}}_i^{\rm nb}$ for identified \ac{nlos} user positions are in the \ac{nlos} region, i.e., $\hat{\mathbf{p}}_i^{\rm nb} \in \mathcal{R}_{\rm NLoS}$ if $I'(\mathbf{H}_i)=0, \forall i$.

In particular, when resorting to \ac{ot}~\eqref{eq:OT_opti_discrete_relaxation}, we sample $N_t$ (proportional to $N_s = N_u-\sum_i I'(\mathbf{H}_i)$) candidate user positions in the \ac{nlos} region according to the \ac{pdf} of user distribution $f_{\mathrm{D}}^{\mathrm{NLoS}}(\cdot)$, yielding a user position set $\{\bar{\mathbf{p}}_i\}_{i = 1,2,\ldots,N_t}$, where each position $\bar{\mathbf{p}}_i$ is drawn i.i.d. from the distribution:
\begin{equation}\label{eq:user_position_generation}
    \bar{\mathbf{p}}_i \overset{\text{i.i.d.}}{\sim} f_{\mathrm{D}}^{\mathrm{NLoS}}(\mathbf{p}), \quad i = 1,\ldots,N_t.
\end{equation}
Then, in the \ac{ot} problem~\eqref{eq:OT_opti_discrete_relaxation}, we can let the source set $\mathcal{X}_s = \{\hat{\mathbf{p}}_i^{\rm nb}|I'(\mathbf{H}_i) = 0\}_{i = 1,2,\ldots,N_u}$, the target set $\mathcal{X}_t = \{\bar{\mathbf{p}}_i\}_{i = 1,2,\ldots,N_t}$, the probability vector $\mathbf{u}_s = \frac{1}{N_s} \cdot \mathbf{1}_{N_s}$ (for user positions in the source set $\mathcal{X}_s$) and the probability vector $\mathbf{u}_t = \frac{1}{N_t} \cdot \mathbf{1}_{N_t}$ (for user positions in the target set $\mathcal{X}_t$). Based on the source set $\mathcal{X}_s$, the target set $\mathcal{X}_t$, and their position probability vectors $\mathbf{u}_s$ and $\mathbf{u}_t$, the \ac{ot}-based loss $\mathcal{L}_{\rm OT}$ can be designed as the first-order Wasserstein distance, which can be written as
\begin{equation}\label{eq:loss_func_OT}
    \mathcal{L}_{\rm OT} = \mathscr{W}_1\left( \mathbf{u}_s, \mathbf{u}_t \right) = \min_{\mathbf{\Gamma} \in \mathcal{B}} \langle\mathbf{\Gamma}, \mathbf{C}\rangle_F = \langle\mathbf{\Gamma}^*, \mathbf{C}\rangle_F,
\end{equation}
where the cost associated with two positions is set as the Euclidean distance, i.e., $[\mathbf{C}]_{i',j} = \|\hat{\mathbf{p}}_{i}^{\rm nb} - \bar{\mathbf{p}}_j \|_2$.\footnote{\label{footnote:index_ordering}The ${i'}$-th data sample in $\{\hat{\mathbf{p}}_i^{\rm nb}|I'(\mathbf{H}_i) = 0\}_{i = 1,2,\ldots,N_u}$ is the $i$-th data sample in $\{\hat{\mathbf{p}}_i^{\rm nb}\}_{i = 1,2,\ldots,N_u}$.}

\subsubsection{Loss Computation}
Therefore, by minimizing the \ac{ot}-based loss $\mathcal{L}_{\rm OT}$ during training, the position estimates $\{\hat{\mathbf{p}}_i^{\rm nb}|I'(\mathbf{H}_i) = 0\}$ can be aligned with the sampled positions $\{\bar{\mathbf{p}}_i\}$, which gives $\hat{\mathbf{p}}_i^{\rm nb} \in \mathcal{R}_{\rm NLoS}$ for $I'(\mathbf{H}_i) = 0$. Note that the estimates $\{\hat{\mathbf{p}}_i^{\rm nb}|I'(\mathbf{H}_i) = 1\}$ for identified \ac{los} user positions are not included in the \ac{ot}~\eqref{eq:loss_func_OT} since they will not be transported and hence have no impact on the \ac{ot}-based loss. It is clear that $\mathcal{L}_{\rm OT}$ involves an \ac{ot} optimization problem, which can be shown to be differentiable with respect to the parameters $\Theta$ of the \ac{nn} model by utilizing convex optimization layers~\cite{Diff_Opti_Bran_2017,Diff_Conv_Agra_2019,Alte_Diff_Haix_2023}. Specifically, we can define a convex optimization layer $f_{\rm CL}(\cdot)$, which takes the outputs of the \ac{nn} model $\{f_{\Theta}(\mathbf{s}_i)|I'(\mathbf{H}_i) = 0\}_{i = 1, 2, \ldots, N_s}$ as its input. The output of $f_{\rm CL}(\cdot)$ is $\mathbf{\Gamma}^* = f_{\rm CL}(\{f_{\Theta}(\mathbf{s}_i)|I'(\mathbf{H}_i) = 0\})$, which is the optimal solution of~\eqref{eq:loss_func_OT} with $[\mathbf{C}]_{i',j} = \|f_{\Theta}(\mathbf{s}_i) - \bar{\mathbf{p}}_j \|_2$. It can be seen that $\langle\mathbf{\Gamma}^*, \mathbf{C}\rangle_F$ is differentiable w.r.t. $\mathbf{\Gamma}^*$ and $\mathbf{C}$; and $\mathbf{C}$ is differentiable w.r.t. $\{f_{\Theta}(\mathbf{s}_i)\}$. Also, it can be proved that the output of convex optimization layers, i.e., the optimal solution of the corresponding optimization problem, is differentiable w.r.t. the input of the layer~\cite{Diff_Opti_Bran_2017,Diff_Conv_Agra_2019} (i.e., $\mathbf{\Gamma}^*$ is differentiable w.r.t. $\{f_{\Theta}(\mathbf{s}_i)|I'(\mathbf{H}_i) = 0\}$), based on which we can conclude that $\mathcal{L}_{\rm OT} = \langle\mathbf{\Gamma}^*, \mathbf{C}\rangle_F$ is differentiable w.r.t. the parameters $\Theta$ since $f_{\Theta}(\mathbf{s}_i)$ is differentiable w.r.t. $\Theta$. Moreover, the forward pass of $f_{\rm CL}(\cdot)$ can be implemented by solving~\eqref{eq:loss_func_OT}, and its gradients w.r.t. $\{f_{\Theta}(\mathbf{s}_i)|I'(\mathbf{H}_i) = 0\}$ can be computed by differentiating its \ac{kkt} conditions according to the implicit function theorem~\cite{Diff_Conv_Agra_2019}, based on which the backpropagation of $\mathcal{L}_{\rm OT}$ can be performed through this layer and the \ac{nn} model w.r.t. $\Theta$. More details about the convex optimization layers can be found in~\cite{Diff_Opti_Bran_2017,Diff_Conv_Agra_2019,Alte_Diff_Haix_2023}.

\subsubsection{Sinkhorn Iteration}
However, training the \ac{nn} model directly using $\mathbf{\Gamma}^* = f_{\rm CL}(\{f_{\Theta}(\mathbf{s}_i)|I'(\mathbf{H}_i) = 0\})$ would require very high computational complexity (prohibitive for large-scale datasets). For example, its training complexity is $\mathcal{O}( N_u\sum_{l=1}^{L_{\rm MLP}} n_{l-1} n_l + N_u^3 N_t^3)$ per epoch for an $L_{\rm MLP}$-layer \ac{mlp}, where $\mathcal{O}(N_u^3 N_t^3)$ is induced by convex optimization layers~\cite{Alte_Diff_Haix_2023}. Instead, we can use the Sinkhorn algorithm to approximately compute $\mathcal{L}_{\rm OT}$ in a more efficient and differentiable manner~\cite{Sink_Dist_Cutu_2013,Lear_with_Frog_2015,Lear_Gene_Gene_2018}. In particular, by adding a regularization term to the \ac{ot} problem, $\mathcal{L}_{\rm OT}$ can be reformulated as
\begin{equation}\label{eq:loss_func_OT_approx}
    \mathcal{L}_{\rm OT} = \min_{\mathbf{\Gamma} \in \mathcal{B}} \quad \langle\mathbf{\Gamma}, \mathbf{C}\rangle_F - \frac{1}{\varepsilon} E(\mathbf{\Gamma}),
\end{equation}
where $E(\mathbf{\Gamma}) = -\sum_{i,j} [\mathbf{\Gamma}]_{i,j} \log([\mathbf{\Gamma}]_{i,j})$ is the entropy term, and $\varepsilon > 0$ is a regularization parameter. It can be proved that~\eqref{eq:loss_func_OT_approx} is strongly convex and admits an optimal solution, as given by~\cite{Sink_Dist_Cutu_2013,Lear_with_Frog_2015}
\begin{equation}
    \mathbf{\Gamma}^* = \mathrm{diag} (\mathbf{a}) \cdot \exp (- \varepsilon \mathbf{C})  \cdot \mathrm{diag} (\mathbf{b}),
\end{equation}
where $\mathbf{a} \in \mathbb{R}_+^{N_u \times 1}$ and $\mathbf{b} \in \mathbb{R}_+^{N_t \times 1}$ are iteratively updated by
\begin{equation}\label{eq:sinkhorn_iter}
    \mathbf{a} \leftarrow \mathbf{u}_s \oslash [\exp (- \varepsilon \mathbf{C}) \cdot \mathbf{b}],\quad \mathbf{b} \leftarrow \mathbf{u}_t \oslash [\exp (- \varepsilon \mathbf{C})^\mathsf{T} \cdot \mathbf{a}],
\end{equation}
until convergence. Therefore, by adopting $I_{\rm Iter}$ iterations in \eqref{eq:sinkhorn_iter}, $\mathcal{L}_{\rm OT}$ can be computed approximately, which is also differentiable with respect to the parameters $\Theta$ since the Sinkhorn update~\eqref{eq:sinkhorn_iter} is differentiable. In this case, the training complexity is $\mathcal{O}( N_u\sum_{l=1}^{L_{\rm MLP}} n_{l-1} n_l + I_{\rm Iter} N_u N_t)$ per epoch for an $L_{\rm MLP}$-layer \ac{mlp}, which is much lower than that of the training using the convex optimization layers.

\section{UNILoc: A Low-complexity Approach}\label{sec:UNILoc}

\begin{figure}[t]
    \centering
    \begin{tikzpicture}[every node/.style={font=\footnotesize}]
    \node (image) [anchor=south west]{\includegraphics[width=0.86\linewidth]{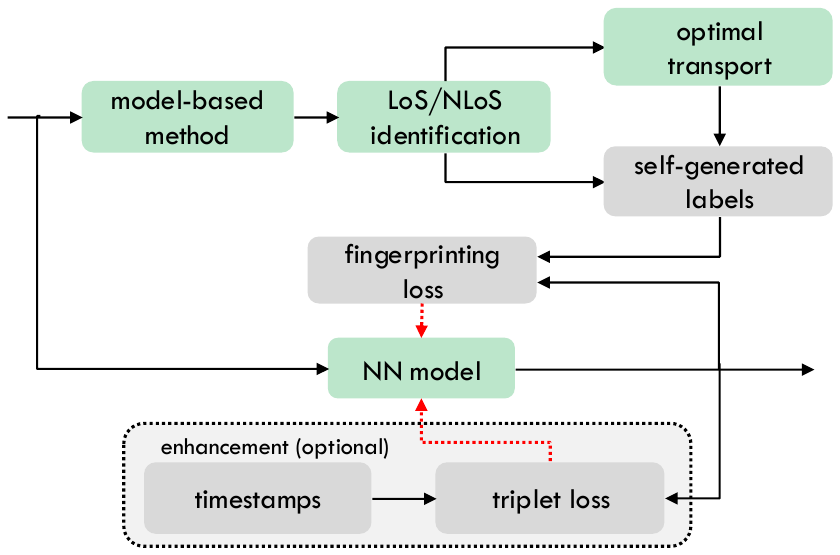}};
    \gettikzxy{(image.north east)}{\ix}{\iy};
    
    \node at (0.06*\ix,0.87*\iy)[rotate=0,anchor=north]{{$\{\mathbf{H}_i\}$}};
    \node at (0.895*\ix,0.435*\iy)[rotate=0,anchor=north]{{$\{\hat{\mathbf{p}}_i^{\rm nb}\}$}};
    \node at (0.61*\ix,0.975*\iy)[rotate=0,anchor=north]{{$I'(\mathbf{H}_i) = 0$}};
    \node at (0.61*\ix,0.68*\iy)[rotate=0,anchor=north]{{$I'(\mathbf{H}_i) = 1$}};
    \node at (0.89*\ix,0.61*\iy)[rotate=0,anchor=north]{{$\{\tilde{\mathbf{p}}_i\}$}};

    \end{tikzpicture}
    \caption{The training process of the \ac{nn} model in UNILoc (the red dashed arrow line indicates the path of gradient backpropagation).}
    \label{fig:UNILoc_training}
\end{figure}

In UNILocPro, the Sinkhorn algorithm needs to be performed to compute the \ac{ot}-based loss $\mathcal{L}_{\rm OT}$ and the backpropagation then traces back through the Sinkhorn iterations in each epoch during training, which could be computationally expensive, especially for very large-scale datasets. In this section, we introduce UNILoc, a low-complexity approach that can achieve performance comparable to UNILocPro with much lower complexity for the \ac{nn} model training. Similarly to UNILocPro, the model-based and \ac{nn}-based methods are integrated as~\eqref{eq:unified_localization}, while the \ac{nn} model is trained using self-generated training labels $\{\tilde{\mathbf{p}}_i\}$ with optional enhancement by timestamps, as shown in Fig.~\ref{fig:UNILoc_training}. Note that the self-generated labels $\{\tilde{\mathbf{p}}_i\}$ are obtained from the model-based estimates $\{\hat{\mathbf{p}}_i^{\rm mb}\}$ by applying the \ac{ot} once, which eliminates the need for the \ac{ot} at each epoch, thus significantly reducing the training complexity.

\subsection{Self-generated Labels}
Given the \ac{csi} dataset $\{\mathbf{H}_i\}_{i = 1,2,\ldots,N_u}$, position estimates $\{\hat{\mathbf{p}}_i^{\rm mb}\}_{i = 1,2,\ldots,N_u}$ can be attained based on the model-based method. For identified \ac{los} user positions, $\{\hat{\mathbf{p}}_i^{\rm mb}\}$ can be used as training labels, i.e., $\tilde{\mathbf{p}}_i = \hat{\mathbf{p}}_i^{\rm mb}$ if $I'(\mathbf{H}_i) = 1$; however, these estimates $\hat{\mathbf{p}}_i^{\rm mb}$ may not be accurate for identified \ac{nlos} user positions and do not match the map. Therefore, we aim to find a transformation $\mathbf{T}^*(\cdot)$ that maps $\{\hat{\mathbf{p}}_i^{\rm mb}|I'(\mathbf{H}_i) = 0\}$ to some other positions that can match the \ac{nlos} region on the map.

To achieve this goal, we resort to the discrete \ac{ot} in \eqref{eq:OT_opti_discrete_relaxation}, where we let the source set $\mathcal{X}_s = \{\hat{\mathbf{p}}_i^{\rm mb}|I'(\mathbf{H}_i) = 0\}_{i = 1,2,\ldots,N_u}$ and the target set $\mathcal{X}_t = \{\bar{\mathbf{p}}_i\}_{i = 1,2,\ldots,N_t}$, where $\{\bar{\mathbf{p}}_i\}_{i = 1,2,\ldots,N_t}$ is a set of candidate user positions in the \ac{nlos} region sampled according to the \ac{pdf} of user distribution $f_{\mathrm{D}}^{\mathrm{NLoS}}(\cdot)$, as given in~\eqref{eq:user_position_generation}. Moreover, by setting $[\mathbf{C}]_{i',j} = \|\hat{\mathbf{p}}_{i}^{\rm mb} - \bar{\mathbf{p}}_j \|_2^2$, $\mathbf{u}_s = \frac{1}{N_s} \cdot \mathbf{1}_{N_s}$, and $\mathbf{u}_t = \frac{1}{N_t} \cdot \mathbf{1}_{N_t}$, the joint probability matrix $\mathbf{\Gamma} \in \mathbb{R}_+^{N_s \times N_t}$ ($N_s = N_u-\sum_i I'(\mathbf{H}_i)$) can be optimized by the \ac{ot} problem~\eqref{eq:OT_opti_discrete_relaxation}. To solve the problem~\eqref{eq:OT_opti_discrete_relaxation}, simplex methods can be used, e.g., the network simplex algorithm with a complexity of $\mathcal{O}( (N_s+N_t)^3 \log (N_s+N_t)^2)$~\cite{A_Poly_Jame_1997,Fast_Robu_Pele_2009}. Moreover, low-complexity algorithms can be adopted to approximately solve~\eqref{eq:OT_opti_discrete_relaxation}, e.g., the Sinkhorn algorithm~\cite{Opti_Tran_Cour_2017,Lear_with_Frog_2015,Sink_Dist_Cutu_2013} with a complexity of $\mathcal{O}( I_{\rm Iter} N_s N_t)$, where $I_{\rm Iter}$ is the number of Sinkhorn iterations.

Once the optimal joint probability matrix $\mathbf{\Gamma}^*$ is obtained, the transformed position for each identified \ac{nlos} user can be written as~\cite{Opti_Tran_Cour_2017}
\begin{equation}\label{eq:OT_label_generation}
    \tilde{\mathbf{{P}}}_{\rm NLoS} = {\rm diag}(\mathbf{\Gamma}^* \cdot \mathbf{1}_{N_t})^{-1} \cdot \mathbf{\Gamma}^* \cdot \bar{\mathbf{P}},
\end{equation}
where $\bar{\mathbf{P}} \in \mathbb{R}^{N_t \times 3}$ is the collection of all $\{\bar{\mathbf{p}}_i\}_{i = 1,2,\ldots,N_t}$; $\tilde{\mathbf{{P}}}_{\rm NLoS} \in \mathbb{R}^{N_s \times 3}$ with $[\tilde{\mathbf{{P}}}_{\rm NLoS}]_{i',:} = \mathbf{T}^*(\hat{\mathbf{p}}_{i}^{\rm mb})$ being the mapped position\footref{footnote:index_ordering}. Then, combining the identified \ac{los} and \ac{nlos} user positions, the self-generated labels $\{\tilde{\mathbf{p}}_i\}_{i = 1,2,\ldots,N_u}$ can be expressed as
\begin{equation}\label{eq:training_label_generation}
    \tilde{\mathbf{p}}_i = \left\{ \begin{array}{ll}
        \hat{\mathbf{p}}_i^{\rm mb}, & {\rm if} \, I'(\mathbf{H}_i) = 1;  \\
        \mathbf{T}^*(\hat{\mathbf{p}}_i^{\rm mb}), & {\rm if} \, I'(\mathbf{H}_i) = 0.
    \end{array} \right.
\end{equation}

\subsection{NN-model Training}
With the self-generated labels $\{\tilde{\mathbf{p}}_i\}$~\eqref{eq:training_label_generation}, a dataset $\mathcal{S}'' = \{(\mathbf{H}_i,\tilde{\mathbf{p}}_i)\}_{i = 1,2,\ldots,N_u}$ can be constructed, based on which the \ac{nn} model can be trained in a supervised manner. In particular, the \ac{nn} model can be trained using a fingerprinting loss and the optional triplet loss (if timestamps are available), as given by
\begin{equation}\label{eq:loss_func_UNILoc}
    \mathcal{L} = \frac{1}{N_u} \sum_{\mathbf{H}_i, \tilde{\mathbf{p}}_i \in \mathcal{S}''} \| f_{\Theta}[f_{\rm extr} (\mathbf{H}_i)] - \tilde{\mathbf{p}}_i \|^2_2 + \omega_{\rm Tri} \cdot \mathcal{L}_{\rm Tri},
\end{equation}
where $\omega_{\rm Tri} \geq 0$ is a non-negative coefficient ($\omega_{\rm Tri} = 0$ and $\mathcal{L}_{\rm Tri}$ is not included if timestamps are not available). With the consideration of imperfect \ac{csi}, the \ac{nn} model of UNILoc can also be enhanced by data augmentation, similar to UNILocPro. Note that the training of the \ac{nn} model in UNILoc does not involve \ac{ot}, and thus, with the self-generated labels $\{\tilde{\mathbf{p}}_i\}$, the parameters $\Theta$ can be updated by backpropagation by only tracing back through the \ac{nn} model. For an $L_{\rm MLP}$-layer \ac{mlp}, the training complexity of UNILoc is $\mathcal{O}(N_u \sum_{l=1}^{L_{\rm MLP}} n_{l-1} n_l)$ per epoch, which is much lower than that of UNILocPro, i.e., $\mathcal{O}( N_u\sum_{l=1}^{L_{\rm MLP}} n_{l-1} n_l + I_{\rm Iter} N_u N_t)$ per epoch, especially for large-scale datasets.

\begin{remark}
    It is noted that UNILoc can be regarded as a specific implementation of UNILocPro by setting $\kappa \rightarrow \infty$ in~\eqref{eq:probability_emd}, $\alpha_{i,j} = 1$ in~\eqref{eq:dissimilarity_fusion}, and detaching the \ac{ot}-based loss $\mathcal{L}_{\rm OT}$ from the iterative training process. Actually, the dissimilarity metric $\{d_{i,j}^{\mathrm{Fusi}}\}$ would be the pairwise distance of $\{\hat{\mathbf{p}}_i^{\rm mb}\}_{i = 1,2,\ldots,N_u}$ when $\alpha_{i,j} = 1$ and $\kappa \rightarrow \infty$ such that the pairwise distance loss $\mathcal{L}_{\rm PW}$ and the \ac{los}-based loss $\mathcal{L}_{\rm LOS}$ would generate a channel chart that is exactly the model-based estimates $\{\hat{\mathbf{p}}_i^{\rm mb}\}_{i = 1,2,\ldots,N_u}$. Then, applying the \ac{ot}-based loss $\mathcal{L}_{\rm OT}$ once would generate the self-generated labels $\{\tilde{\mathbf{p}}_i\}$, which are used as training labels for the \ac{nn} model in UNILoc.
\end{remark}

\section{Numerical Experiments}\label{sec:experiments}

\subsection{System Setting}\label{sec:system_setting}
In this section, numerical experiments in a street canyon scenario, as shown in Fig.~\ref{fig:system_setting}, are carried out. The carrier frequency of the system is $f_c = 10\, {\rm GHz}$, and a bandwidth of $50\, {\rm MHz}$ is used with a subcarrier spacing of $120\, {\rm kHz}$. Being equipped with a \ac{ula} ($M=256$ antennas and the inter-antenna distance set as $d = \lambda/2$), the \ac{bs} is placed at $\mathbf{p}_{\rm BS} = [0,-9,57]^{\mathsf{T}}$, and the height of each user position is set as $1.5\, {\rm m}$, i.e., $\mathbf{p}_i = [ x_i, y_i, 1.5]^\mathsf{T} $. Without other specifications, user positions in both training and testing datasets are generated randomly according to a uniform distribution in the first quadrant of the map. For each user position, the \ac{csi} $\mathbf{H}_i$ is generated synthetically by Sionna RT~\cite{sionna}, where ray-tracing techniques are used to generate the propagation paths (including the direct path, reflection, diffusion, and diffraction), based on which the channel $\mathbf{H}_i$ is computed\footnote{We also conduct additional experiments using the DICHASUS dataset by selecting one antenna array as the \ac{bs}~\cite{A_Dist_Flor_2021}; however, due to the lack of synchronization between the \ac{bs} and users, unknown user-specific clock biases introduce substantial offsets in the estimated \acp{toa}, severely degrading model-based localization. While the ground-truth positions can be used to correct the clock biases (though this is not a practical solution), the performance is still unsatisfactory, especially for \ac{nlos} users. As future work, we will consider the joint estimation of user positions and clock biases to enable the application of our framework to non-synchronized datasets.}. For better comparison and analysis, the channel $\mathbf{H}_i$ is generated in a noise-free manner, except for the discussion on the noise impact in Sec.~\ref{sec:localization_results}.

Around 1800 user positions, consisting of 550 \ac{los} and 1250 \ac{nlos}, are generated independently for both training and testing datasets (as unsupervised learning is considered in this paper, the ground-truth positions $\{\mathbf{p}_i\}$ in the training dataset are only used for validation). Without other specifications, when using \ac{ot} in $\mathcal{L}_{\rm OT}$~\eqref{eq:loss_func_OT} and in label self-generation~\eqref{eq:OT_label_generation}, the target domain $\mathcal{X}_t = \{\bar{\mathbf{p}}_i\}_{i = 1,2,\ldots,N_t}$ is generated by selecting $N_t$ grid points in the \ac{nlos} region, i.e., $\bar{\mathbf{p}}_i \in \mathcal{G}_{\rm NLoS} \subseteq \mathcal{R}_{\rm NLoS}$, where $\mathcal{G}_{\rm NLoS}$ contains all grid points with a spacing distance of $\Delta_d = 0.5\, {\rm m}$ in $\mathcal{R}_{\rm NLoS}$. 

For the training dataset $\{(\mathbf{H}_i,\mathbf{p}_i)\}$, the timestamps $\{t_i\}$ are generated as $t_0 = 0$ and $t_i = t_{i-1} + \| \mathbf{p}_i - \mathbf{p}_{i-1} \|_2/v_{i}$, $i > 1$ along the shortest trajectory; and $v_i$ is the velocity in the $i$-th segment of the trajectory. With consideration of velocity variation, we set $v_k$ as a truncated Gaussian random variable, i.e., $v_k = \max (0, v_k')$, where $v_k' \sim \mathcal{N}(\mu, \sigma_v^2)$ with $\mu$ being the mean and $\sigma_v$ being the standard deviation. Without other specifications, we set $\mu = 10\, {\rm m/s}$ and $\sigma_v = 0\, {\rm m/s}$. When generating the triplet set $\mathcal{T}$, we set $T_{\rm L} = 10/\mu$ and $T_{\rm U} = 50/\mu$ in~\eqref{eq:triplet_set}. Note that the timestamps $\{t_i\}$ are only generated for the training dataset and are not needed for the testing dataset.

For the hyper-parameters involved in this paper, we set the regularization parameter $\varepsilon = 2/3$ in~\eqref{eq:loss_func_OT_approx} when conducting the Sinkhorn iteration; the trust region distance $\zeta = 20$ is adopted for computing the \ac{gospa}-based dissimilarity~\eqref{eq:GOSPA_dissimilarity}; when fusing the \ac{gospa}-based and Wasserstein-based metrics, we use $\vartheta = 0.03$ and $d_{\rm thre} = 10$; when computing the triplet loss~\eqref{eq:triplet_loss}, $\gamma = 0.1$ is used. As all of the loss terms in~\eqref{eq:loss_func_UNILocPro} (resp.~\eqref{eq:loss_func_UNILoc}) have similar magnitudes, we set $\omega_{\rm B-CC} = 1$, $\omega_{\rm LoS} = 1$, $\omega_{\rm OT} = 1$, $\omega_{\rm PW} = 1$, and $\omega_{\rm Tri} = 1$ (resp. $\omega_{\rm Tri} = 1$) when timestamps are available without other specifications ($\omega_{\rm Tri} = 0$ if timestamps are not available).

As for the \ac{nn} model, following the same \ac{nn} architecture and hyper-parameters as in~\cite{Trip_Wire_Ferr_2021,Angl_Prof_Step_2024}, an \ac{mlp} is adopted for $f_{\Theta}(\cdot)$, which consists of 5 hidden layers (each with 1024, 512, 256, 128, 64 neurons, respectively, ReLU activation, and batch normalization) and an output layer (with 2 neurons and linear activation). Moreover, Adam optimizer with a decayed learning rate is employed for training. All the \ac{nn} models are implemented in TensorFlow and the training is performed on a machine with an NVIDIA Tesla T4 GPU (16~GB RAM). Note that if a noise-free scenario is considered and no other specifications are provided, the training of the \ac{nn} model would not include the data augmentation stage.

\subsection{CC Results}\label{sec:cc_results}
\begin{table}[t]
\caption{The \ac{cc} performance for different dissimilarity metrics (the values in parentheses are computed directly from the dissimilarity metrics).}
\label{tab:CC_results}
\centering
\renewcommand{\arraystretch}{1.2} 
\begin{tabular}{|l|c|c|c|}
\hline
Dissimilarity metric & $\mathcal{CT} \uparrow$ & $\mathcal{TW} \uparrow$ & $\mathcal{KS} \downarrow$ \\
\hline
\multicolumn{4}{|l|}{\textbf{without timestamps} ($\mathcal{L}_{\rm PW}$)} \\
\hline
$d_{i,j}^{\mathrm{G-CIRA}}$  & 0.824 (0.840) & 0.827 (0.790) & 0.712 (0.702) \\
$d_{i,j}^{\mathrm{G-Fro}}$ & 0.770 (0.749) & 0.773 (0.721) & 0.540 (0.471) \\
$d_{i,j}^{\mathrm{G-Corr}}$ & 0.852 (0.843) & 0.816 (0.766) & 0.478 (0.434) \\
$d_{i,j}^{\mathrm{G-ADP}}$ & 0.836 (0.833) & 0.768 (0.766) & 0.432 (0.394) \\
$d_{i,j}^{\mathrm{G-GOSPA}}$ & 0.950 (0.983) & 0.960 (0.979) & 0.273 (0.262) \\
$d_{i,j}^{\mathrm{Wass}}$ & 0.971 (0.971) & 0.971 (0.972) & 0.262 (0.262) \\
$d_{i,j}^{\mathrm{Fusi}}$ & \textbf{0.980} (\textbf{0.985}) & \textbf{0.980} (\textbf{0.985}) & \textbf{0.206} (\textbf{0.210}) \\
\hline
\multicolumn{4}{|l|}{\textbf{with timestamps} ($\mathcal{L}_{\rm PW} + \mathcal{L}_{\rm Tri}$)} \\
\hline
$d_{i,j}^{\mathrm{time}}$ ($\mathcal{L}_{\rm PW}$) & 0.763 & 0.709 & 0.593 \\
$\mathcal{L}_{\rm Tri}$ & 0.907 & 0.749 & 0.536 \\
$d_{i,j}^{\mathrm{G-FADP}}$ ($\mathcal{L}_{\rm PW}$) & 0.915 & 0.924 & 0.306 \\
$d_{i,j}^{\mathrm{G-Fro}}$ & 0.928 & 0.831 & 0.467\\
$d_{i,j}^{\mathrm{G-Corr}}$ & 0.933 & 0.866 & 0.461 \\
$d_{i,j}^{\mathrm{G-ADP}}$ & 0.962 & 0.913 & 0.314 \\
$d_{i,j}^{\mathrm{G-GOSPA}}$ & 0.986 & 0.983 & 0.241\\
$d_{i,j}^{\mathrm{Wass}}$ & 0.981 & 0.974 & 0.226 \\
$d_{i,j}^{\mathrm{Fusi}}$ & \textbf{0.988} & \textbf{0.986} & \textbf{0.180}  \\
\hline
\end{tabular}
\end{table}

\begin{figure}[t]
    \centering
    \begin{subfigure}{0.23\textwidth}
        \includegraphics[width=\linewidth]{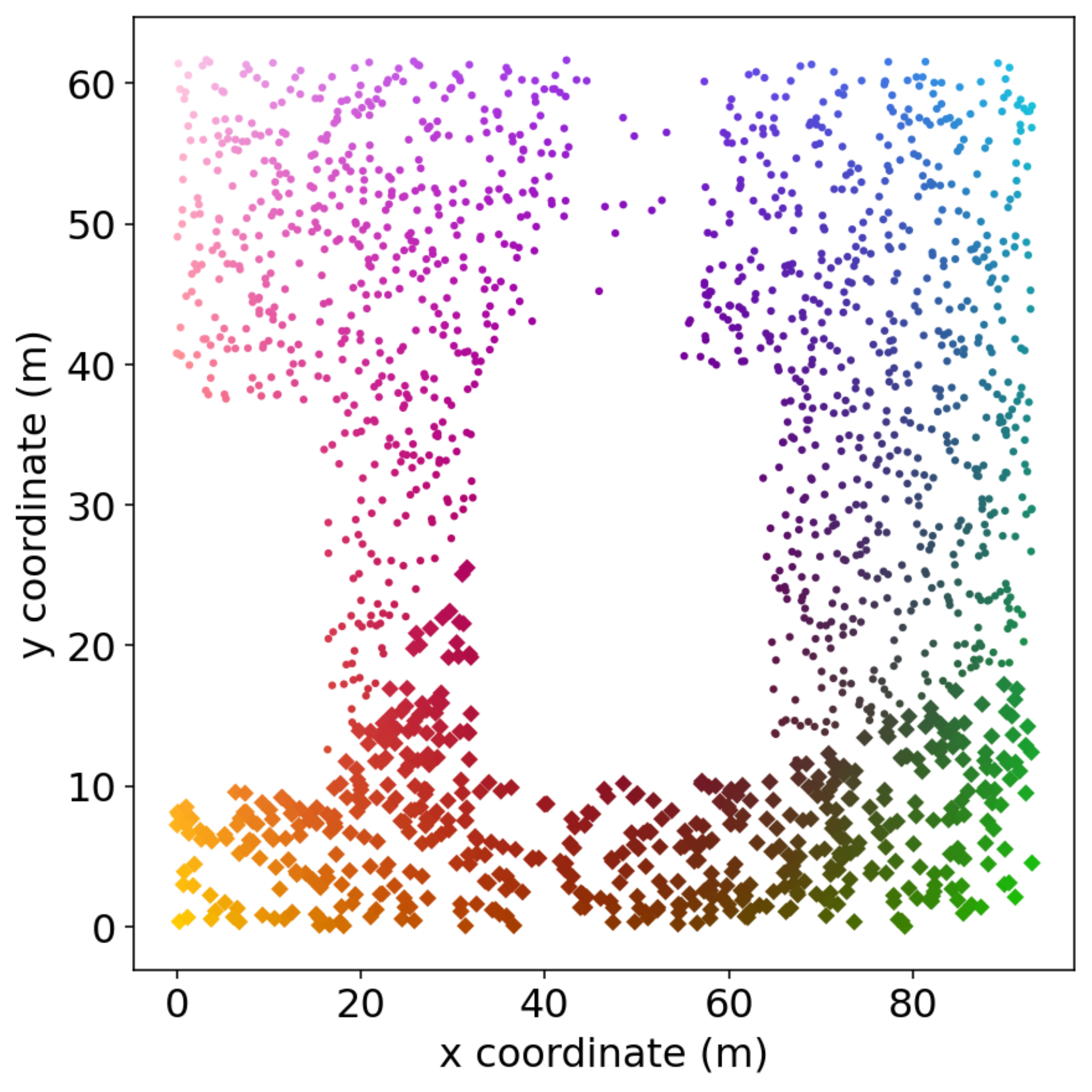}
        \caption{ground-truth positions}
    \end{subfigure}
    \begin{subfigure}{0.23\textwidth}
        \includegraphics[width=\linewidth]{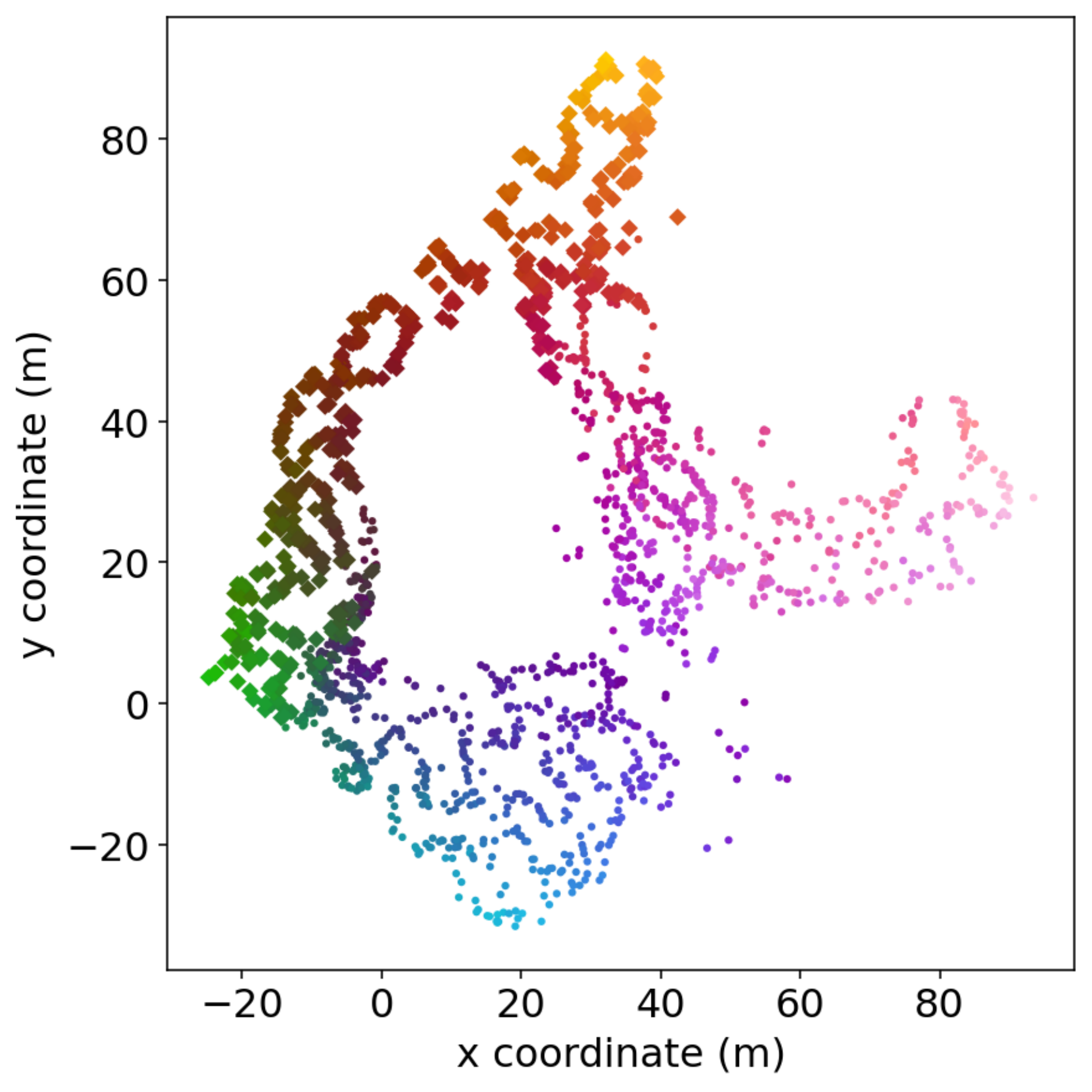}
        \caption{$d_{i,j}^{\mathrm{Fusi}}$ ($\mathcal{L}_{\rm PW} + \mathcal{L}_{\rm Tri}$)}
    \end{subfigure}
    \caption{Channel chart generated by $\{d_{i,j}^{\mathrm{Fusi}}\}$ with timestamps (training dataset; every user is associated with a unique color; dot marks represent \ac{nlos} user positions and diamond marks represent \ac{los} user positions).}
    \label{fig:CC_results}
    \vspace{-2mm}
\end{figure}

We evaluate the \ac{cc} performance of different dissimilarity metrics, including our proposed \ac{gospa}-based metric and its fusion with the Wasserstein-based metric, as well as some dissimilarity metrics used in the literature: 1) a \ac{csi} Frobenius norm-based metric $d_{i,j}^{\mathrm{Fro}}$~\cite[Eq. (1)]{Abso_Posi_Pihl_2020}; 2) a \ac{csi} correlation-based metric $d_{i,j}^{\mathrm{Corr}}$~\cite[Eq. (2)]{Chan_Char_Stud_2018}; 3) a metric based on channel impulse response amplitude $d_{i,j}^{\mathrm{CIRA}}$~\cite[Eq. (15)]{Indo_Loca_Stah_2023}; 4) a dissimilarity metric based on the channel angle-delay profile $d_{i,j}^{\mathrm{ADP}}$~\cite[Eq. (6)]{Angl_Prof_Step_2024}; 5) the time difference $d_{i,j}^{\mathrm{time}}$ in~\eqref{eq:distance_time_difference}; 6) a fusion of $d_{i,j}^{\mathrm{time}}$ and $d_{i,j}^{\mathrm{ADP}}$, as denoted by $d_{i,j}^{\mathrm{FADP}}$~\cite[Eq. (15)]{Angl_Prof_Step_2024}; and 7) the geodesic variants of these metrics (denoted with the prefix ``G-''), including $d_{i,j}^{\mathrm{G-Fro}}$, $d_{i,j}^{\mathrm{G-Corr}}$, $d_{i,j}^{\mathrm{G-CIRA}}$, $d_{i,j}^{\mathrm{G-ADP}}$, and $d_{i,j}^{\mathrm{G-FADP}}$. Without timestamps, only the pairwise distance loss $\mathcal{L}_{\rm PW}$~\eqref{eq:loss_func_pairwise} based on these dissimilarity metrics is used to train the \ac{nn} model $f_{\Theta}(\cdot)$, i.e., $\mathcal{L} = \mathcal{L}_{\rm PW}$. When timestamps are available, the triplet loss $\mathcal{L}_{\rm Tri}$~\eqref{eq:triplet_loss} constructed by $\{ t_i\}$ can be integrated with the pairwise distance loss $\mathcal{L}_{\rm PW}$ for enhancement, i.e., $\mathcal{L} = \mathcal{L}_{\rm PW} + \mathcal{L}_{\rm Tri}$.

In order to quantify the \ac{cc} performance, we use the continuity ($0 \leq \mathcal{CT} \leq 1$), the trustworthiness ($0 \leq \mathcal{TW} \leq 1$), and the Kruskal stress ($0 \leq \mathcal{KS} \leq 1$) to evaluate different dissimilarity metrics~\cite{Chan_Char_Stud_2018,Chan_Char_Sued_2025,Angl_Prof_Step_2024,Trip_Wire_Ferr_2021}. Computed from the generated channel charts, the \ac{cc} performance of different dissimilarity metrics is summarized in Table~\ref{tab:CC_results}. It is observed that the \ac{gospa}-based and Wasserstein-based metrics significantly outperform the state-of-the-art dissimilarity metrics, and their fusion achieves the best performance. It is also seen that timestamps cannot generate a good channel chart directly, e.g., by using $d_{i,j}^{\mathrm{time}}$ or the triplet loss $\mathcal{L}_{\rm Tri}$ alone, but it can improve \ac{cc} performance significantly for existing metrics through adding the triplet loss $\mathcal{L}_{\rm Tri}$. Moreover, as shown in Fig.~\ref{fig:CC_results}, the proposed metrics not only achieve remarkable \ac{cc} performance, but also preserve the overall global spatial structure (up to a rotation and translation ambiguity) as the basic shape of the channel chart closely resembles the ground-truth positions.

\subsection{Localization Results}\label{sec:localization_results}
\begin{figure}[t]
    \centering
    \begin{subfigure}{0.23\textwidth}
        \includegraphics[width=\linewidth]{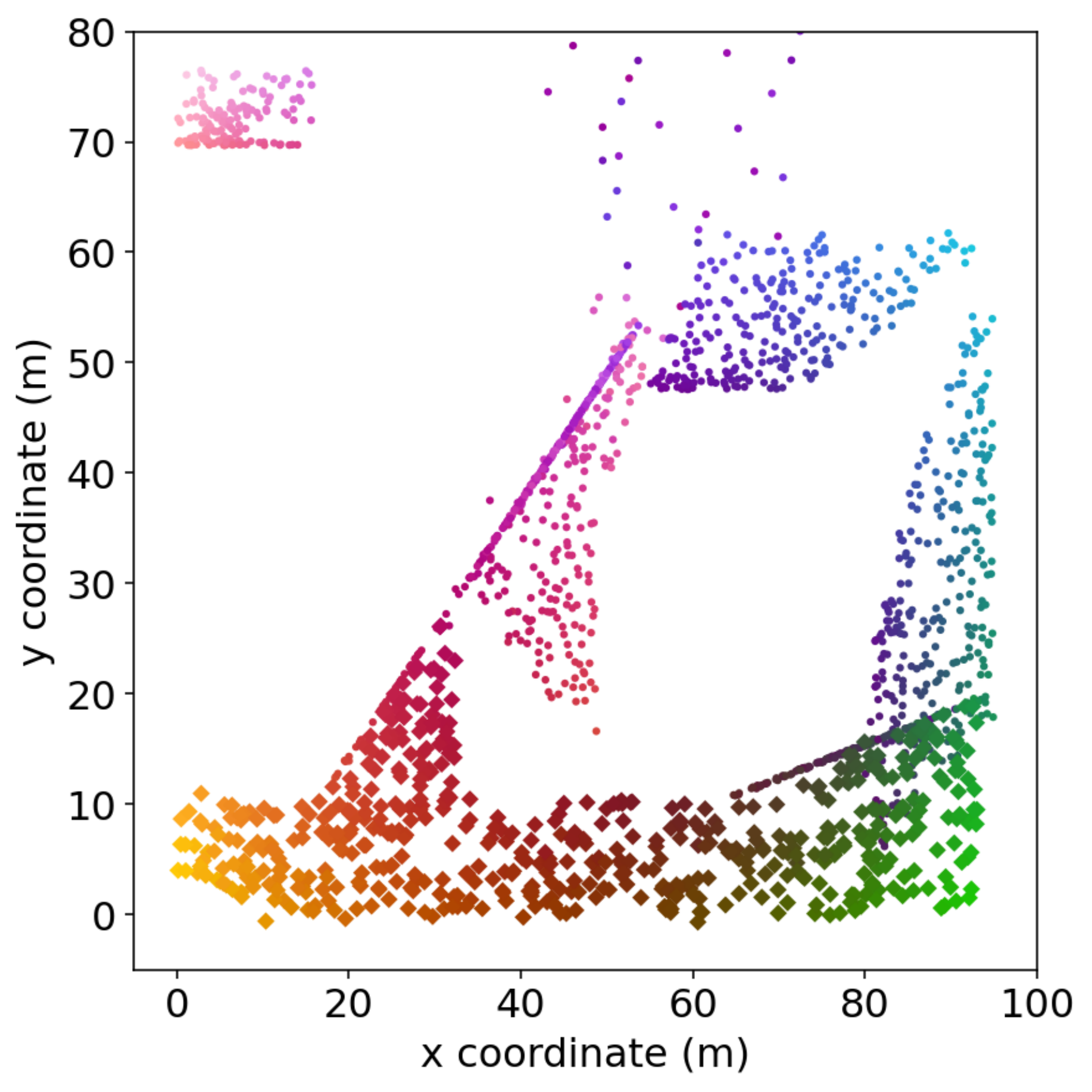}
        \vspace{-6mm}
        \caption{model-based method}
    \end{subfigure}
    \begin{subfigure}{0.23\textwidth}
        \includegraphics[width=\linewidth]{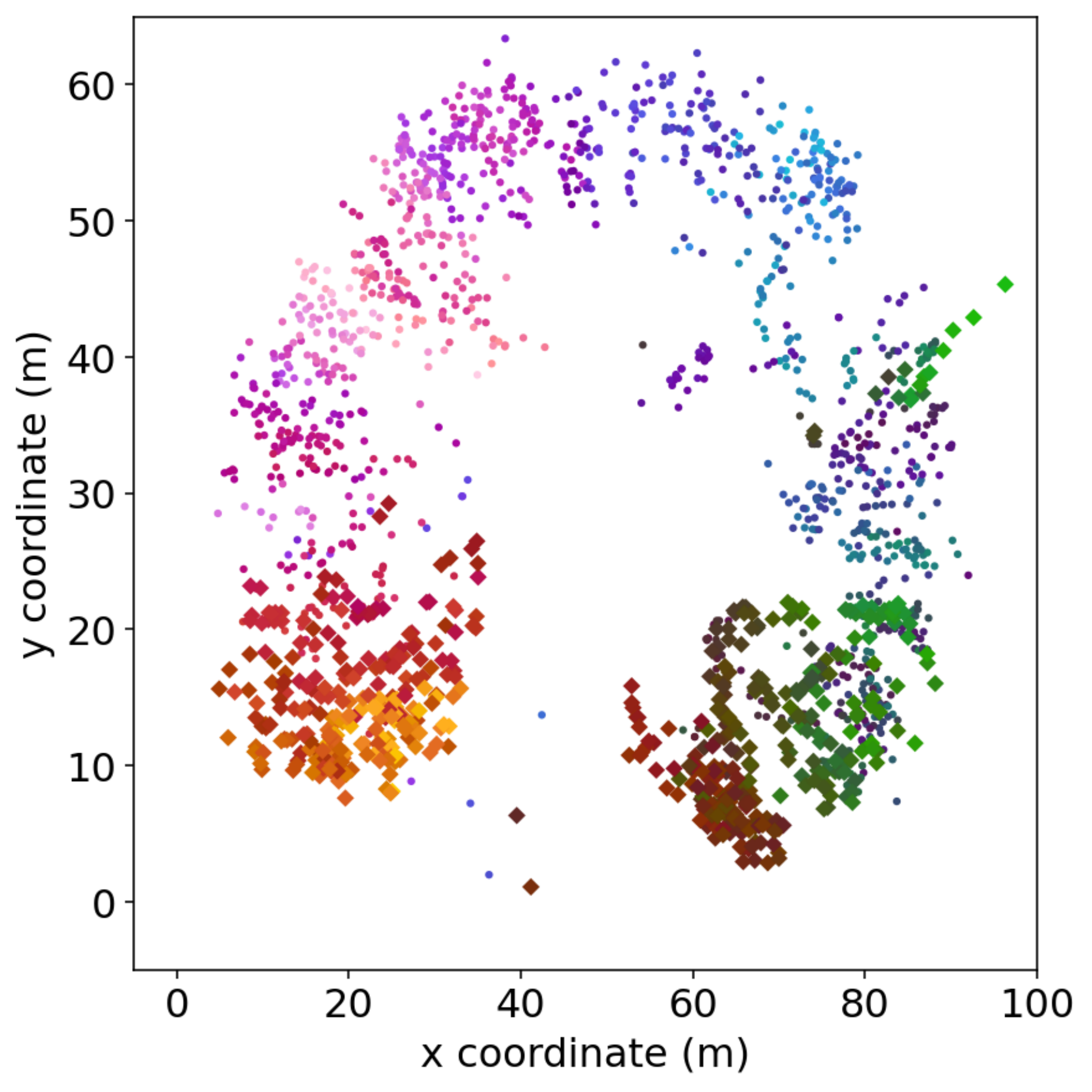}
        \vspace{-6mm}
        \caption{\ac{cc}-based method}
    \end{subfigure}
    \begin{subfigure}{0.23\textwidth}
        \vspace{2mm}
        \includegraphics[width=\linewidth]{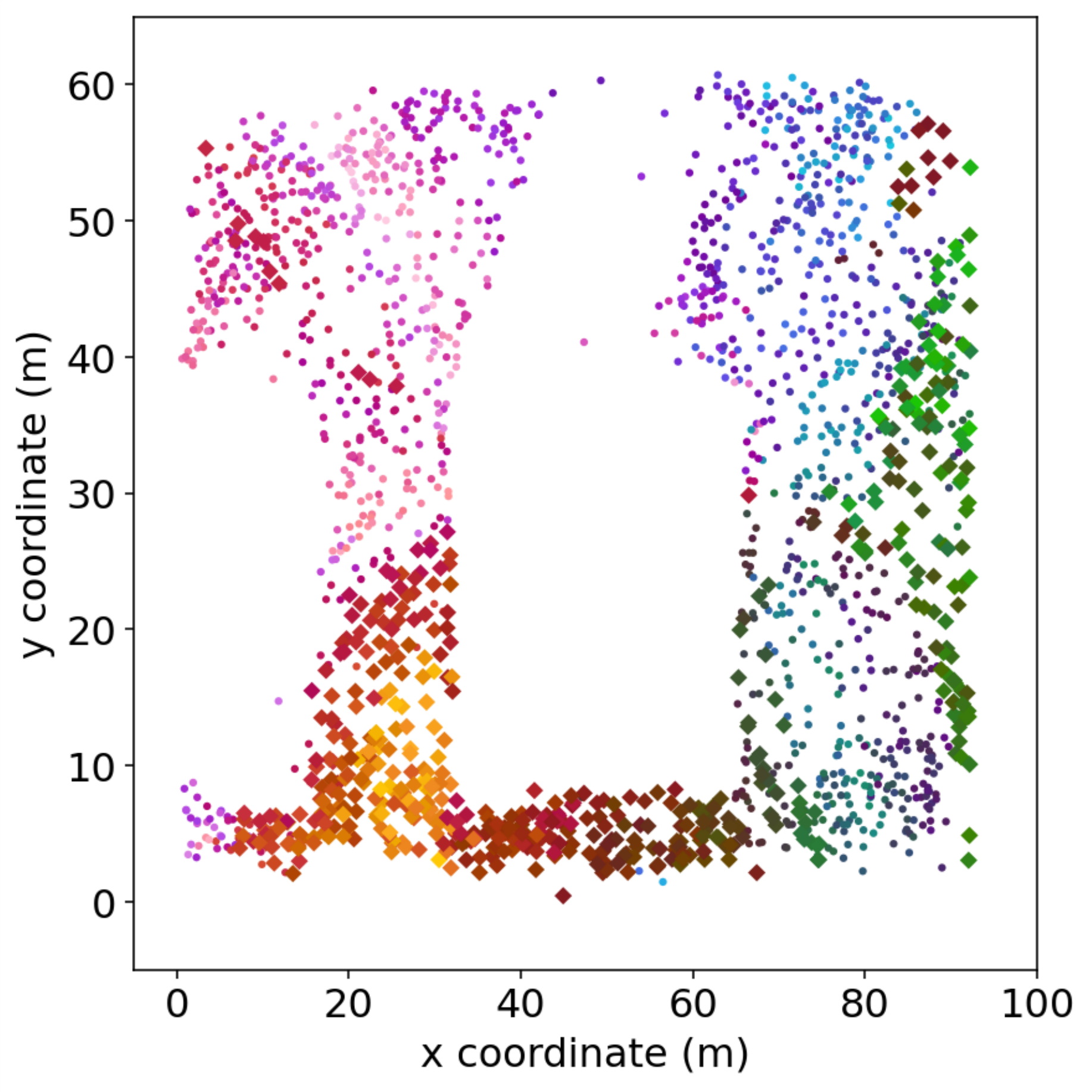}
        \vspace{-6mm}
        \caption{\ac{cc}-based method with \ac{ot}}
    \end{subfigure}
    \begin{subfigure}{0.23\textwidth}
        \vspace{2mm}
        \includegraphics[width=\linewidth]{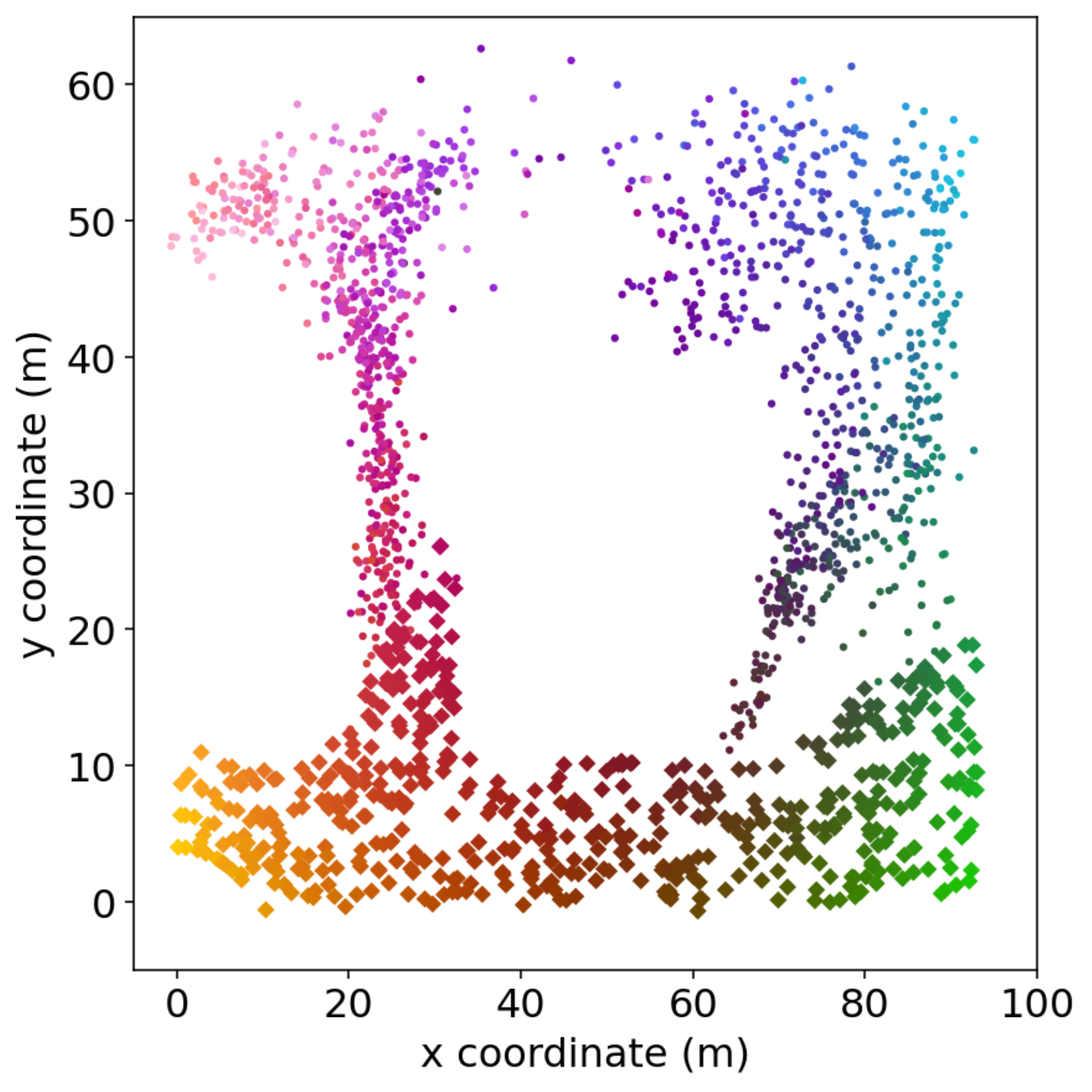}
        \vspace{-6mm}
        \caption{UNILoc with $\{t_i\}$ ($p_I = 1$)}
    \end{subfigure}
    \begin{subfigure}{0.23\textwidth}
        \vspace{2mm}
        \includegraphics[width=\linewidth]{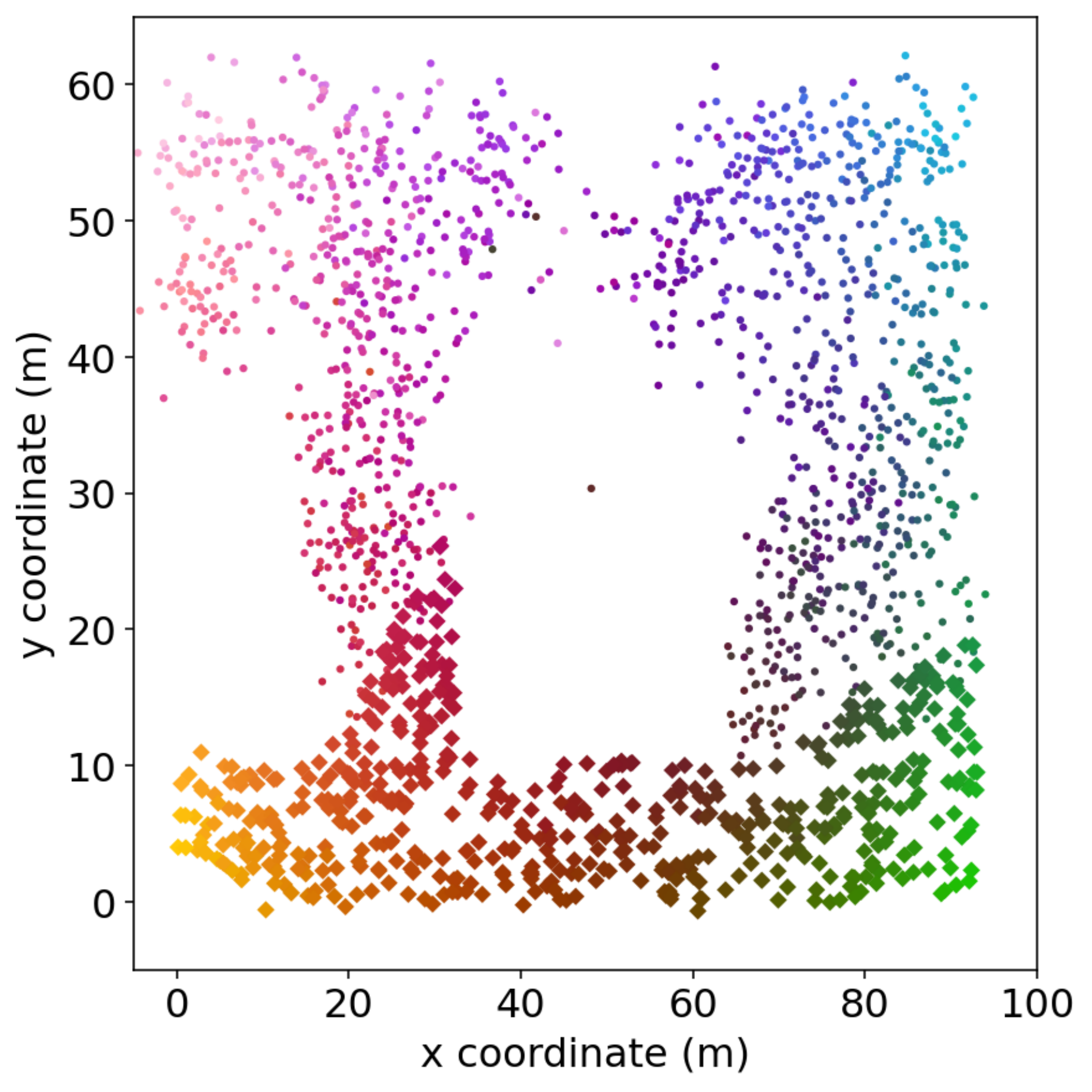}
        \vspace{-6mm}
        \caption{UNILocPro with $\{t_i\}$ ($p_I = 1$)}
    \end{subfigure}
    \begin{subfigure}{0.23\textwidth}
        \vspace{2mm}
        \includegraphics[width=\linewidth]{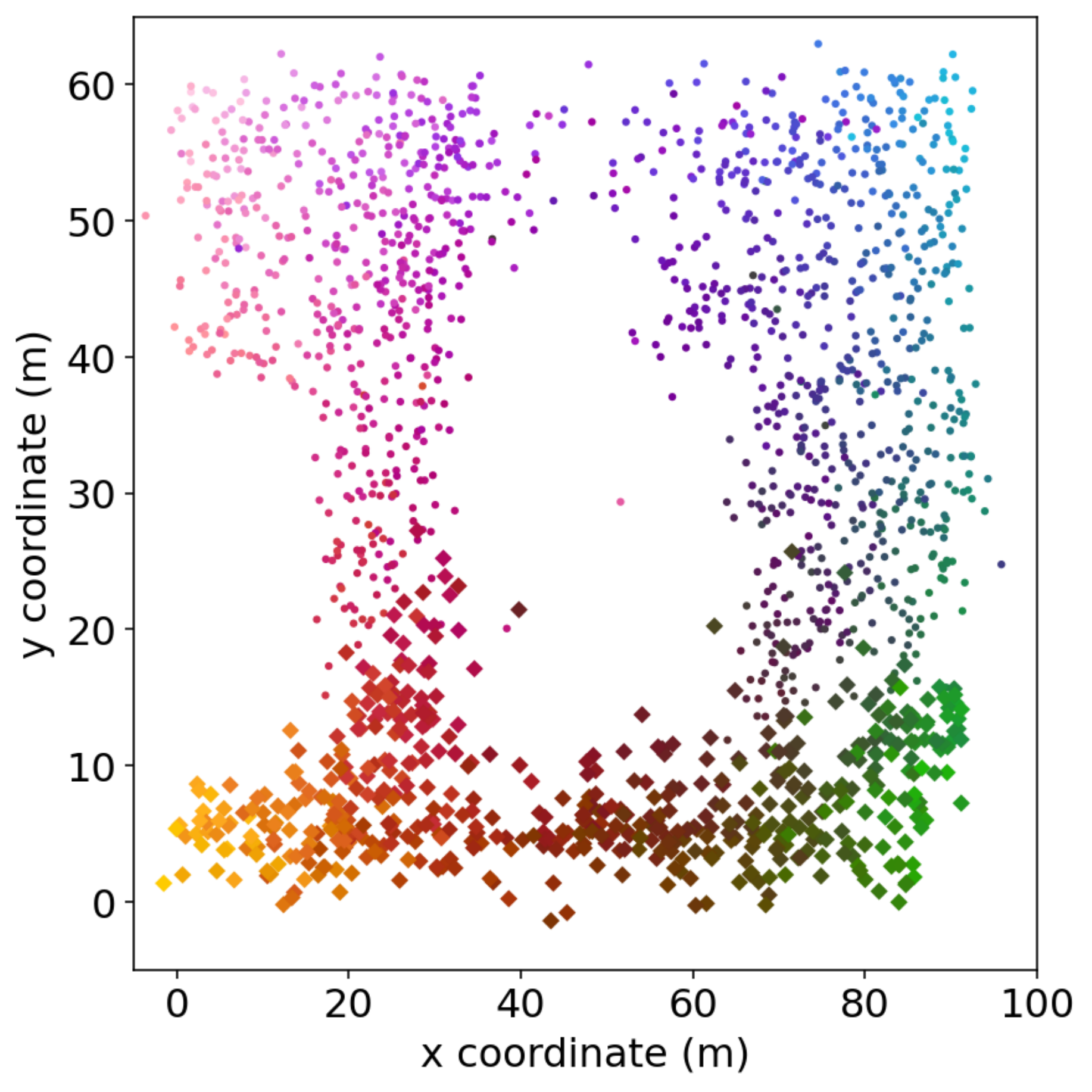}
        \vspace{-6mm}
        \caption{fingerprinting ($\Delta_s = 0.5\, {\rm m}$)}
    \end{subfigure}
    \caption{Position estimates for different localization methods (testing dataset; every user is associated with a unique color; dot marks represent \ac{nlos} user positions and diamond marks represent \ac{los} user positions).}
    \label{fig:localization_results}
\end{figure}

\begin{table}[t]
\caption{The MAE (m) for different localization methods (testing dataset).}
\label{tab:localization_performance}
\centering
\renewcommand{\arraystretch}{1.2} 
\begin{tabular}{|l|ccc|}
\hline
 & \ac{los} & \ac{nlos} & all \\
\hline
\multicolumn{4}{|l|}{\textbf{Unsupervised Methods without Timestamps}} \\
\hline
model-based method & 0.34 & 13.79 & 9.81 \\
UNILocPro ($p_I = 1$) & \textbf{0.34} & \textbf{7.21} & \textbf{5.18} \\
UNILocPro (conservative) & 0.52 & 7.83 & 5.67 \\
UNILoc ($p_I = 1$) & 0.34 & 7.67 & 5.50 \\
UNILoc (conservative) & 0.51 & 8.49 & 6.13 \\
\hline
\multicolumn{4}{|l|}{\textbf{Unsupervised Methods with Timestamps}} \\
\hline
\ac{cc}-based method ($d_{i,j}^{\mathrm{G-FADP}}$) & 12.68 & 14.16 & 13.71 \\
\ac{cc}-based method with \ac{ot} ($d_{i,j}^{\mathrm{G-FADP}}$) & 15.09 & 17.02 & 16.45 \\
UNILocPro ($p_I = 1$) & \textbf{0.34} & \textbf{5.36} & \textbf{3.88} \\
UNILocPro (GMM-based) & 0.45 & 5.64 & 4.11 \\
UNILocPro (conservative) & 0.49 & 6.87 & 4.98 \\
UNILoc ($p_I = 1$) & 0.34 & 6.37 & 4.58 \\
UNILoc (GMM-based) & 0.42 & 6.34 & 4.59 \\
UNILoc (conservative) & 0.42 & 6.97 & 5.03 \\
\hline
\multicolumn{4}{|l|}{\textbf{Supervised Methods}} \\
\hline
fingerprinting ($\Delta_s = 2\, {\rm m}$) & 4.69 & 6.06 & 5.65 \\
fingerprinting ($\Delta_s = 1\, {\rm m}$) & 3.89 & 5.43 & 4.98 \\
fingerprinting ($\Delta_s = 0.5\, {\rm m}$) & 3.95 & \textbf{4.48} & \textbf{4.46} \\
fingerprinting & \textbf{3.88} & 4.72 & 4.47 \\
\hline
\end{tabular}
\end{table}

For evaluation and analysis, \ac{mae} is adopted for localization, i.e., $\mathrm{MAE} = \frac{1}{N_u}\sum_{i=1}^{N_u} \| \hat{\mathbf{p}}_i - \mathbf{p}_i \|_2$. For comparison, model-based, fingerprinting, and \ac{cc}-based localization methods are considered as benchmarks:
\begin{itemize}
    \item \textbf{Model-based method}: position estimates are obtained based on the \ac{omp} algorithm and geometric relationships, i.e., $\hat{\mathbf{p}}_i^{\rm mb} = f_{\rm pe}\left[ f_{\rm ce}(\mathbf{H}_i) \right]$, $\forall i$;
    \item \textbf{Fingerprinting}: position estimates are obtained from \ac{nn}, i.e., $\hat{\mathbf{p}}_i^{\rm nb} = f_{\Theta}\left[f_{\rm extr}(\mathbf{H}_i)\right]$, where $f_{\Theta}(\cdot)$ is trained by assuming ground-truth labels, i.e., $\tilde{\mathbf{p}}_i = \mathbf{p}_i$, $\forall i$. It is a fully-supervised approach;
    \item \textbf{\ac{cc}-based method ($d_{i,j}^{\mathrm{G-FADP}}$)}~\cite{Angl_Prof_Step_2024}: position estimates are obtained as $\hat{\mathbf{p}}_i = \mathbf{T}(f_{\Theta}\left[f_{\rm extr}(\mathbf{H}_i)\right])$, where the \ac{nn} model $f_{\Theta}(\cdot)$ is trained based on the pairwise distance loss $\mathcal{L}_{\rm PW}$~\eqref{eq:loss_func_pairwise} with $d_{i,j} = d_{i,j}^{\mathrm{G-FADP}}$, and $\mathbf{T}(\mathbf{p}) = \mathbf{A}^* \cdot \mathbf{p} + \mathbf{b}^*$ is the optimal affine transformation that solves the following \ac{ls} problem: $\mathbf{A}^*, \mathbf{b}^* = \arg \min_{\mathbf{A}, \mathbf{b}} \, \sum_i \| \mathbf{A} \cdot f_{\Theta}\left[f_{\rm extr}(\mathbf{H}_i)\right] + \mathbf{b} - \mathbf{p}_i \|_2^2$, where $\mathbf{A} \in \mathbb{R}^{3 \times 3}$ is a rotation and scaling matrix, and $\mathbf{b} \in \mathbb{R}^{3 \times 1}$ is a translation vector. It is an unsupervised approach except for the affine transformation $\mathbf{T}(\cdot)$;
    \item \textbf{\ac{cc}-based method with \ac{ot} ($d_{i,j}^{\mathrm{G-FADP}}$)}: position estimates are obtained as $\hat{\mathbf{p}}_i = \mathbf{T}_{\rm OT}\left(\mathbf{T}(f_{\Theta}\left[f_{\rm extr}(\mathbf{H}_i)\right])\right)$, where the \ac{nn} model $f_{\Theta}(\cdot)$ and the affine transformation $\mathbf{T}(\cdot)$ are the same as the \ac{cc}-based method ($d_{i,j}^{\mathrm{G-FADP}}$); and $\mathbf{T}_{\rm OT}(\cdot)$ is a transformation that maps $\{ \mathbf{T}(f_{\Theta}\left[f_{\rm extr}(\mathbf{H}_i)\right]) \}$ to grid positions in $\mathcal{R}$ with a spacing distance of $\Delta_d$ via \ac{ot}. It is an unsupervised approach except for the affine transformation $\mathbf{T}(\cdot)$.
\end{itemize}

In this section, we also implement the conservative variant of UNILoc and UNILocPro, where the conservative \ac{los}/\ac{nlos} identification $I_{\rm conserv}'(\cdot)$~\eqref{eq:LoS_identification_conservative} is adopted instead of $I'(\cdot)$~\eqref{eq:LoS_identification}. For fingerprinting, grid positions on the map with the corresponding \ac{csi} are also used for training. Specifically, the grid points on the map, denoted by $\mathcal{G}$, with a spacing distance of $\Delta_s$, i.e., $\mathbf{p}_i \in \mathcal{G} \subseteq \mathcal{R}$, $\forall i$, are selected as the training dataset. While testing, user positions are placed randomly, which is the same as in the other methods considered.

\subsubsection{Positioning Accuracy}
\begin{figure}[!t]
\centering
\begin{minipage}[t]{0.97\linewidth}
\centering
    \include{Figures/fig_cdf_com.tex}
    \vspace{-1 cm}
\end{minipage}
\caption{The \ac{cdf} w.r.t.~the positioning error for different methods (testing dataset).}
\label{fig:cdf}
\end{figure}

Fig.~\ref{fig:localization_results} shows the position estimates for different localization methods; for a quantitative evaluation, the \ac{cdf} of estimation errors and performance metrics are also shown in Fig.~\ref{fig:cdf} and Table~\ref{tab:localization_performance}, respectively. It is observed that the model-based method achieves high positioning accuracy for \ac{los} users ($\mathrm{MAE} = 0.34\, {\rm m}$), while the estimates for \ac{nlos} users are significantly distorted. The fingerprinting has the lowest estimation error for \ac{nlos} user positions (also preserves the global shape of \ac{nlos} users), while the estimates of \ac{los} users are not as precise as those produced by the model-based method. Being an integrated strategy, UNILoc and UNILocPro (as well as their variants) can not only attain precise estimations for \ac{los} users but also preserve the global position of the \ac{nlos} users, demonstrating the effectiveness of integrating model-based geometry and \ac{cc}, the new dissimilarity metrics, and the \ac{ot}-based loss. This indicates that the model-based methods could provide valuable information for unsupervised learning approaches to improve estimation accuracy. Moreover, UNILocPro is able to provide better localization compared with UNILoc, as UNILocPro incorporates \ac{ot} loss during training rather than a single-shot \ac{ot} before training, which offers a global geometry that is close to the ground-truth positions.

It can be seen that although the local geometry, i.e., the neighboring relationship between user positions, is preserved in general, the \ac{cc}-based method ($d_{i,j}^{\mathrm{G-FADP}}$) cannot capture the global geometry due to the lack of physical distance information in the dissimilarity metric, resulting in large position estimation errors. Note that adding the \ac{ot} directly to the \ac{cc}-based method ($d_{i,j}^{\mathrm{G-FADP}}$) could not improve the performance (on the contrary, it would deteriorate positioning accuracy), which highlights that \ac{ot} can only improve localization performance when the generated channel chart is sufficiently good, i.e., the local geometry is preserved and the global geometry is not very far from the ground-truth positions. This is because the \ac{ot}, which performs a transformation based solely on transportation cost minimization (this process is purely metric-related and lacks higher-level geometric constraints), can only refine a localization estimate but cannot directly recover the underlying global geometry autonomously. In other words, the \ac{ot} cannot correct large errors in the position estimates.

It is intuitive that introducing timestamps would further improve the performance of UNILoc and UNILocPro. Notably, UNILocPro with timestamps achieves performance very close to that of fully-supervised fingerprinting. With the lightweight \ac{gmm}-based \ac{los}/\ac{nlos} identification, as described in Appendix~\ref{app:LoS_NLoS_identification}, the performance is very close to the case with perfect \ac{los}/\ac{nlos} identification. It is also noted that without external \ac{los}/\ac{nlos} identification, the conservative variant of UNILocPro can still achieve satisfactory performance (close to the fully-supervised fingerprinting with $\Delta_s = 2\, {\rm m}$). This indicates that the proposed unified localization frameworks do not rely on external \ac{los}/\ac{nlos} identification and are also robust to the identification accuracy, should one be available.

\subsubsection{Discussion on LoS/NLoS Identification}

\begin{figure}[t]
\centering
\begin{minipage}[t]{0.97\linewidth}
\centering
%
%
\begin{tikzpicture}

\begin{axis}[%
width=7.2cm,
height=4cm,
at={(0in,0in)},
scale only axis,
xmin=0.5,
xmax=1,
xlabel style={font=\color{white!15!black},font=\footnotesize, yshift=0.2cm},
xlabel={$p_I$},
ymin=3,
ymax=11,
yminorticks=false,
ylabel style={font=\color{white!15!black},font=\footnotesize, yshift=-0.7cm},
ylabel={MAE (m)},
axis background/.style={fill=white},
xmajorgrids,
ymajorgrids,
xticklabel style={font=\footnotesize},
yticklabel style={font=\footnotesize},
legend style={fill opacity=0.5, draw opacity=1, text opacity=1, at={(0.98,0.98)},anchor=north east,legend cell align=left, align=left, font=\scriptsize, legend columns=1, draw=white!15!black}
]

\addplot [color=black, line width=1pt, mark size=2 pt, mark=x, mark options={solid, black}]
  table[row sep=crcr]{%
0.5 	10.404203\\
0.55	9.973633\\
0.6 	9.648494\\
0.65	8.597662\\
0.7 	7.9189487\\
0.75	7.203807\\
0.8 	6.4084787\\
0.85	5.777803\\
0.9 	5.3356576\\
0.95	4.6166296\\
1   	3.88\\
};
\addlegendentry{$I(\cdot)$}

\addplot [color=red, line width=1pt, mark size=1.5 pt, mark=o, mark options={solid, red}]
  table[row sep=crcr]{%
0.5 	6.404031\\
0.55	6.1410837\\
0.6 	5.885548\\
0.65	5.5324397\\
0.7 	5.249051\\
0.75	5.0035024\\
0.8 	4.8539863\\
0.85	4.5713\\
0.9 	4.438627\\
0.95	4.1807707\\
1   	3.88\\
};
\addlegendentry{$I'(\cdot)$ Eq.~\eqref{eq:LoS_identification}}

\addplot [color=blue, line width=1pt]
  table[row sep=crcr]{%
0.5 	4.98\\
0.55	4.98\\
0.6 	4.98\\
0.65	4.98\\
0.7 	4.98\\
0.75	4.98\\
0.8 	4.98\\
0.85	4.98\\
0.9 	4.98\\
0.95	4.98\\
1   	4.98\\
};
\addlegendentry{$I_{\rm conserv}'(\cdot)$ Eq.~\eqref{eq:LoS_identification_conservative}}

\addplot [color=green!80!black, dashed, line width=1.5pt]
  table[row sep=crcr]{%
0.5 	4.11\\
0.55	4.11\\
0.6 	4.11\\
0.65	4.11\\
0.7 	4.11\\
0.75	4.11\\
0.8 	4.11\\
0.85	4.11\\
0.9 	4.11\\
0.95	4.11\\
1   	4.11\\
};
\addlegendentry{the GMM-based $I'(\cdot)$}

\end{axis}
\end{tikzpicture}%
    \vspace{-1 cm}
\end{minipage}
\caption{The \ac{mae} of UNILocPro with timestamps for different $p_I$ (testing dataset).}
\label{fig:p_I_UNILocPro}
\end{figure}

In order to explore the impact of the \ac{los}/\ac{nlos} identification accuracy, Fig.~\ref{fig:p_I_UNILocPro} shows the \ac{mae} of UNILocPro with timestamps for different $p_I$. It is clear (also intuitive) that the performance of UNILocPro (the same for UNILoc) depends on the identification accuracy $p_I$, and its estimation error would increase as $p_I$ decreases; however, using the model-based estimations to improve the \ac{los}/\ac{nlos} identification~\eqref{eq:LoS_identification} can significantly reduce the positioning error compared to relying only on the identification mechanism $I(\cdot)$. This verifies that the model-based methods can provide information to improve the \ac{los}/\ac{nlos} identification or even replace the identification mechanism $I(\cdot)$ (as the conservative UNILocPro performs quite well), which is one of the advantages of the proposed frameworks. It can also be observed that for a reasonable $p_I \in [0.93, 1]$ (from the existing identification methods in the literature), UNILocPro (UNILoc) performs better than its conservative variant overall. Specifically, the \ac{gmm}-based identification method (as described in Appendix~\ref{app:LoS_NLoS_identification}) can achieve an accuracy of $p_I = 0.940$ for $I(\cdot)$ (the identification accuracy can be further improved to $0.986$ by using $I'(\cdot)$ in~\eqref{eq:LoS_identification}), based on which UNILocPro (UNILoc) can achieve performance very close to the case with perfect \ac{los}/\ac{nlos} identification. This indicates that the proposed frameworks are robust to the identification accuracy and can be effectively integrated with practical \ac{los}/\ac{nlos} identification methods.

\subsubsection{Discussion on Noise Impact}
In the presence of channel estimation noise, the localization performance would degrade due to the imperfect \ac{csi}. To assess this degradation, we examine the \ac{mae} of UNILocPro and some baselines with timestamps under different noise levels $\sigma^2$, which are characterized by the \ac{snr} of the estimated channel at a reference position, denoted as $\mathrm{SNR}_\mathrm{R} = 10 \log_{10} \left( \frac{\|\mathbf{H}_\mathrm{R}\|_F^2}{M N_c \sigma^2} \right)$, where $\mathbf{H}_\mathrm{R}$ is the estimated \ac{csi} at the reference position. All methods are implemented independently for each noise level. In this work, we choose the reference position as the user position that is closest to the \ac{bs} in the training dataset, which indicates that $\mathrm{SNR}_\mathrm{R}$ is approximately the highest \ac{snr} in the system.

\begin{figure}[t]
\centering
\begin{minipage}[t]{0.97\linewidth}
\centering
    \begin{tikzpicture}
\begin{axis}[
    width=9cm,
    height=5cm,
    ybar=1pt,
    bar width=6pt,
    xlabel={$\mathrm{SNR}_\mathrm{R}$ (dB)},
    xlabel style={font=\footnotesize},
    ylabel={MAE (m)},
    ylabel style={font=\footnotesize,yshift=-15pt}, 
    xtick={1,2,3,4,5,6},
    xticklabels={Perfect, 35, 25, 15, 5, 0},
    xticklabel style={rotate=0, anchor=north, font=\footnotesize},
    yticklabel style={font=\footnotesize},
    legend style={at={(0.3,0.97)}, anchor=north, legend columns=1, font=\scriptsize, cells={anchor=west}},
    ymin=0,
    ymax=45,
    grid=major,
    ymajorgrids=true,
]
\addplot[fill=black!70, draw=black!70, opacity=0.9] coordinates {
    (1,9.84) (2,10.04) (3,10.57) (4,16.53) (5,32.83) (6,42.07)
};
\addplot[fill=green!80!black, draw=green!80!black, opacity=0.9] coordinates {
    (1,11.94) (2,14.40) (3,15.66) (4,17.05) (5,24.51) (6,27.24)
};
\addplot[fill=blue!80, draw=blue!80, opacity=0.9] coordinates {
    (1,3.65) (2,5.58) (3,8.25) (4,12.46) (5,19.13) (6,23.22)
};
\addplot[fill=red!80, draw=red!80, opacity=0.9] coordinates {
    (1,3.61) (2,6.31) (3,7.27) (4,10.57) (5,16.46) (6,21.54)
};
\legend{model-based method, CC-based (data augmentation), UNILocPro (data augmentation), fingerprinting (data augmentation)}
\end{axis}
\end{tikzpicture}
    \vspace{-1 cm}
\end{minipage}
\caption{The MAE for different methods with timestamps, where $\mathrm{SNR}_\mathrm{R}$ is the \ac{snr} at the reference position, i.e., the user position closest to the \ac{bs} in the training dataset (testing dataset).}
\label{fig:loc_performance_with_noise}
\end{figure}

\begin{table}[!t]
\caption{The \ac{cc} performance for $d_{i,j}^{\mathrm{Fusi}}$ ($\mathcal{L}_{\rm PW}$) without timestamps, where $\mathrm{SNR}_\mathrm{R}$ is the \ac{snr} at the reference position, i.e., the user position closest to the \ac{bs} in the training dataset (testing dataset).}
\label{tab:CC_results_SNR}
\centering
\renewcommand{\arraystretch}{1.2} 
\begin{tabular}{|l|c|c|c|}
\hline
$\mathrm{SNR}_\mathrm{R}$ & $\mathcal{CT} \uparrow$ & $\mathcal{TW} \uparrow$ & $\mathcal{KS} \downarrow$ \\
\hline
$0\, \mathrm{dB}$ & 0.759 & 0.777 & 0.543 \\
$5\, \mathrm{dB}$ & 0.785 & 0.836 & 0.511 \\
$15\, \mathrm{dB}$ & 0.914 & 0.934 & 0.480 \\
$25\, \mathrm{dB}$ & 0.961 & 0.963 & 0.255 \\
$35\, \mathrm{dB}$ & 0.976 & 0.974 & 0.224 \\
\hline
\end{tabular}
\end{table}

The results are shown in Fig.~\ref{fig:loc_performance_with_noise}, where it is expected that all methods experience performance degradation as $\mathrm{SNR}_\mathrm{R}$ decreases due to the increasing channel estimation errors. However, it can be seen that UNILocPro still outperforms the model-based and the \ac{cc}-based method under all $\mathrm{SNR}_\mathrm{R}$ values, and performs comparably (even better w.r.t. overall performance) to the fingerprinting unless the $\mathrm{SNR}_\mathrm{R}$ is low. It is noted that the reason why the performance degradation of UNILocPro against fingerprinting becomes more severe when $\mathrm{SNR}_\mathrm{R}$ is low is that UNILocPro relies on dissimilarity metrics that are computed based on model-based estimation of the channel parameters (e.g., \ac{aoa} and \ac{toa}), which are affected by the estimation errors. As shown in Table~\ref{tab:CC_results_SNR}, the dissimilarity metrics become less reliable under low $\mathrm{SNR}_\mathrm{R}$, which would lead to a severe degradation of UNILocPro. That being said, UNILocPro relies on the robustness against noise of the model-based methods, which can also be indicated by the performance of the model-based method (as shown in Fig.~\ref{fig:loc_performance_with_noise}, its accuracy remains stable when $\mathrm{SNR}_\mathrm{R}$ is above $25\, \mathrm{dB}$, but degrades significantly when $\mathrm{SNR}_\mathrm{R}$ becomes very low).

Moreover, it is intuitive that \ac{nlos} users are more affected by estimation errors than \ac{los} users since \ac{nlos} users typically have weaker signal strength, which is more vulnerable to noise. Also, it is important to note that data augmentation can improve positioning performance even under perfect \ac{csi}, where the \ac{mae} of UNILocPro (resp. fingerprinting) improves from $3.88\, \mathrm{m}$ to $3.65\, \mathrm{m}$ (resp. from $4.46\, \mathrm{m}$ to $3.61\, \mathrm{m}$). By introducing variability in the training samples, data augmentation can help mitigate overfitting and learn a more robust mapping that is less sensitive to specific realizations of the training data.

\subsubsection{Discussion on Non-uniform User Distribution}
In practice, the user distribution may not be uniform due to various factors such as population density, urban structure, and user behavior patterns. To investigate the impact of non-uniformity in the user distribution, three different distributions are considered:
\begin{itemize}
  \item Case 1: users are randomly distributed across the map with a density that decreases as $y$ increases (the probability of having a user at a given location decreases linearly from 1 at $y=0$ to 0 at $y=63$). This setup represents scenarios in which users are more concentrated in certain areas (e.g., main streets) and sparser in others (e.g., auxiliary lanes and alleys).
  \item Case 2: users are uniformly distributed throughout the entire area, except within the region $y \in [30, 40]$, where no users are present. This setup represents scenarios in which certain regions have very few or no users.
  \item Case 3: users are uniformly distributed throughout the entire area, except within the region $y \in [30, 50]$, where no users are present. This setup is similar to case 2 but represents a more extreme case of non-uniformity.
\end{itemize}

\begin{table}[!tbp]
\caption{The \ac{mae} (m) for UNILocPro with timestamps ($p_I = 1$) under non-uniform user distribution, which is used simultaneously for training and testing datasets generation (performance is on testing dataset).}
\label{tab:localization_performance_non_uniform_distribution}
\centering
\renewcommand{\arraystretch}{1.2} 

\begin{tabular}{|l|ccc|}
\hline
 & LoS & NLoS & all \\
\hline
\multicolumn{4}{|l|}{\textbf{distribution case 1}} \\
\hline
distribution aligned sampling & 0.34 & 5.92 & 3.68 \\
\hline
distribution mismatched sampling & 0.34 & 7.49 & 4.61 \\
\hline
\multicolumn{4}{|l|}{\textbf{distribution case 2}} \\
\hline
distribution aligned sampling & 0.34 & 6.06 & 4.13 \\
\hline
distribution mismatched sampling & 0.34 & 6.89 & 4.54 \\
\hline
\multicolumn{4}{|l|}{\textbf{distribution case 3}} \\
\hline
distribution aligned sampling & 0.34 & 4.94 & 2.95 \\
\hline
distribution mismatched sampling & 0.34 & 7.75 & 4.68  \\
\hline
\end{tabular}
\end{table}

When sampling the target positions for UNILocPro, we can either use the true distribution (distribution aligned) or the uniform distribution (distribution mismatched). As shown in Table~\ref{tab:localization_performance_non_uniform_distribution}, it is observed that UNILocPro can achieve non-decreasing performance when the sampling and true distributions are aligned, which demonstrates the effectiveness of the proposed framework under non-uniform distribution if the real distribution is known in prior or accurately estimated. Moreover, it is also seen that UNILocPro would achieve a much higher \ac{mae} when using the uniform distribution for sampling, which indicates that the distribution mismatch would lead to performance degradation. This is because the \ac{ot}-based loss would try to match the distribution of the generated channel chart to the uniform distribution, which would distort the geometry and lead to worse performance. It is noted that the performance degradation is more severe when the true distribution is more different from the uniform distribution (e.g., case 3 has a more severe degradation than case 2), which implies that the degree of degradation depends on the difference between the true distribution and the sampling distribution.

\subsubsection{Discussion on Velocity Variation}
Then, we investigate the impact of velocity variation. The channel chart generated by the triplet loss $\mathcal{L}_{\rm Tri}$ for $\sigma_v = 0.5 \mu$ (resp. $\sigma_v = \mu$) achieves $\mathcal{CT} = 0.856$, $\mathcal{TW} = 0.720$, and $\mathcal{KS} = 0.544$ (resp. $\mathcal{CT} = 0.733$, $\mathcal{TW} = 0.664$, and $\mathcal{KS} = 0.594$), which shows that the channel chart is significantly distorted when $\sigma_v$ increases. As shown in Table~\ref{tab:std_velocity}, it is intuitive that the increase of $\sigma_v$ would deteriorate localization performance; however, UNILoc and UNILocPro are quite robust to the velocity variation, and the \ac{mae} of UNILocPro is $4.90\, {\rm m}$ when $\sigma_v = \mu$, which is still better than the case without timestamps, and around $1\, {\rm m}$ larger than that when $\sigma_v = 0$ ($\mathrm{MAE} = 3.88\, {\rm m}$). This indicates that the proposed frameworks, by integrating the model-based pairwise distance loss $\mathcal{L}_{\rm PW}$ and the \ac{ot}-based loss $\mathcal{L}_{\rm OT}$, can effectively mitigate the impact of velocity variation on localization performance. In other words, the timestamps, even if not accurate, can still help to improve localization in our proposed frameworks by providing additional information about user positions.

\begin{table}[!tbp]
\caption{The MAE (m) of UNILoc and UNILocPro with timestamps for different $\sigma_v$ used in training (testing dataset).}
\label{tab:std_velocity}
\centering
\renewcommand{\arraystretch}{1.1} 
\begin{tabular}{|l|ccccc|}
\hline
$\sigma_v (\mu)$ & 0.2 & 0.4 & 0.6 & 0.8 & 1 \\
\hline
UNILocPro & 4.38 & 4.43 & 4.49 & 4.58 & 4.90 \\
UNILoc & 4.74 & 4.75 & 4.89 & 5.10 & 5.27 \\
\hline
\end{tabular}
\end{table}

\subsubsection{Discussion on Residual Clock Bias}
\begin{table}[!bp]
\caption{The MAE (m) for different methods under imperfect synchronization (testing dataset).}
\label{tab:localization_performance_bias}
\centering
\renewcommand{\arraystretch}{1.1} 
\begin{tabular}{|l|c|c|c|c|}
\hline
$\epsilon_{\max}\, (\mathrm{ns})$ & 1 & 10 & 50 & 100 \\
\hline
model-based method & 9.93 & 11.27 & 18.80 & 23.06 \\
UNILocPro & 3.94 & 5.81 & 10.84 & 14.99 \\
fingerprinting & 4.42 & 4.90 & 9.19 & 13.28 \\
\hline
UNILocPro (aug) & 3.46 & 4.88 & 8.50 & 12.32 \\
fingerprinting (aug) & 2.90 & 3.15 & 4.23 & 4.76 \\
\hline
\end{tabular}
\end{table}

We consider imperfect synchronization scenarios in which the clock bias is modeled as $\epsilon_i \sim \mathcal{U}[-\epsilon_{\max},\epsilon_{\max}]$~\cite{Requ_For_3GPP_2025}. Like the data augmentation for noise in Sec.~\ref{sec:CC_method}, we adopt a similar strategy that generates multiple augmented samples for each original \ac{csi} by adding random clock bias perturbations. As shown in Table~\ref{tab:localization_performance_bias}, the localization performance of all methods degrades when clock bias is introduced. Data augmentation benefits both UNILocPro and fingerprinting, while the improvement is limited for UNILocPro since its learned geometry is distorted by the clock bias, which cannot be fully corrected by data augmentation. Moreover, the performance degradation is more pronounced for larger $\epsilon_{\max}$: with tight synchronization (e.g., $\epsilon_{\max} = 1\, {\rm ns}$), UNILocPro approaches fingerprinting; it performs acceptably for small clock bias (e.g., $\epsilon_{\max} = 10\, {\rm ns}$); while with loose synchronization (e.g.,  $\epsilon_{\max} = 100\, {\rm ns}$), the performance gap between UNILocPro and fingerprinting can become non-negligible. This is because UNILocPro (also the model-based method) relies on the geometric estimates derived from the \acp{toa}, which are corrupted by clock bias (however, compared with the model-based method, the degradation is less severe for UNILocPro since it involves other geometric constraints, i.e., the triplet loss and the \ac{ot}-based loss, which help to improve positioning accuracy). In contrast, fingerprinting is less sensitive to clock bias since it does not explicitly rely on the absolute \ac{toa} values; however, it requires ground-truth labels for training. The proposed frameworks can tolerate clock biases up to a certain level, but can be sensitive when the clock bias is large and perform inadequately under loose synchronization.

\subsection{Results on Training Strategies}\label{sec:strategies_results}

\subsubsection{Discussion on Loss Terms}
\begin{figure}[t]
    \centering
    \begin{subfigure}{0.23\textwidth}
        \includegraphics[width=\linewidth]{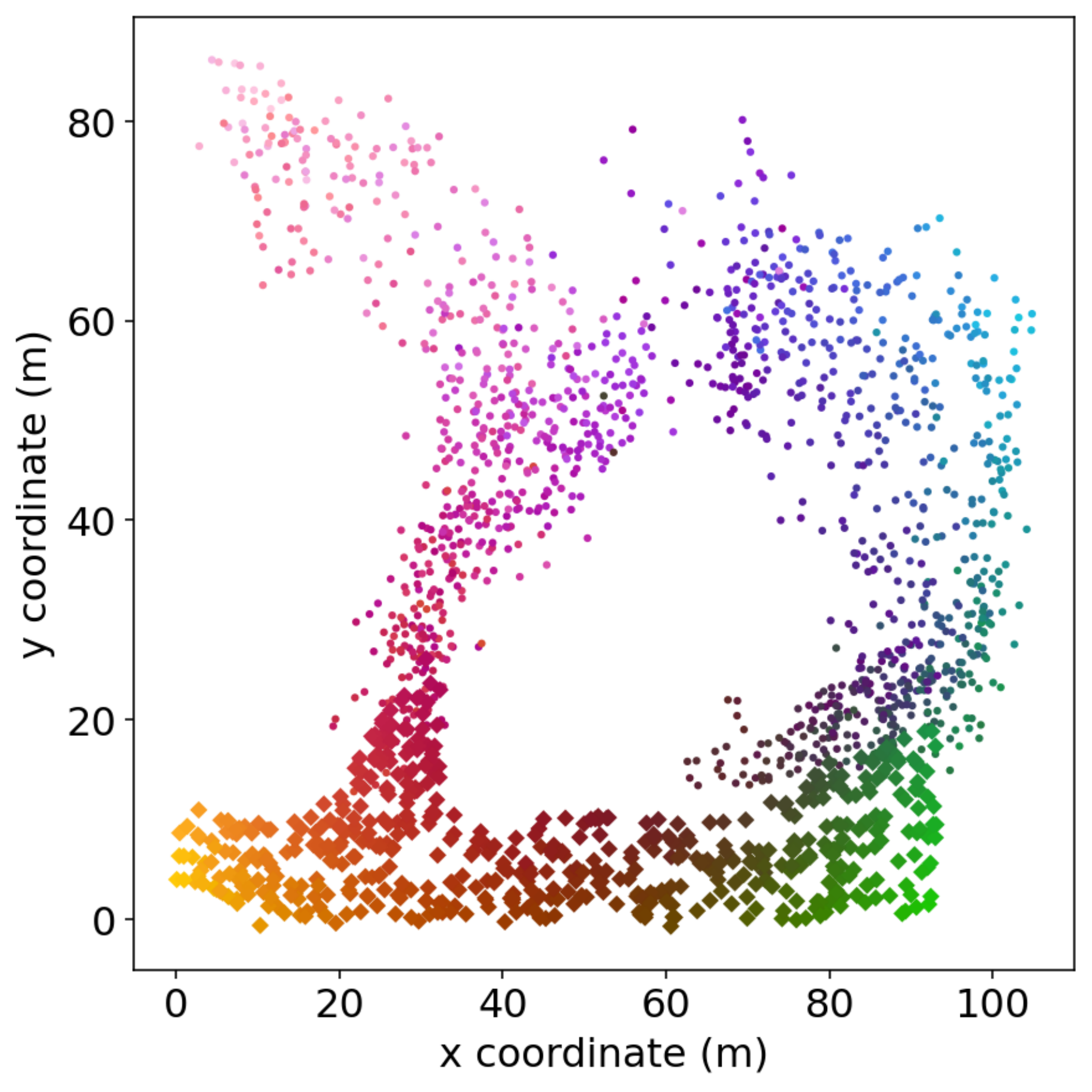}
        \caption{$\mathcal{L}_{\rm CC} = \mathcal{L}_{\rm B-CC} + \mathcal{L}_{\rm LoS}$}
    \end{subfigure}
    \begin{subfigure}{0.23\textwidth}
        \includegraphics[width=\linewidth]{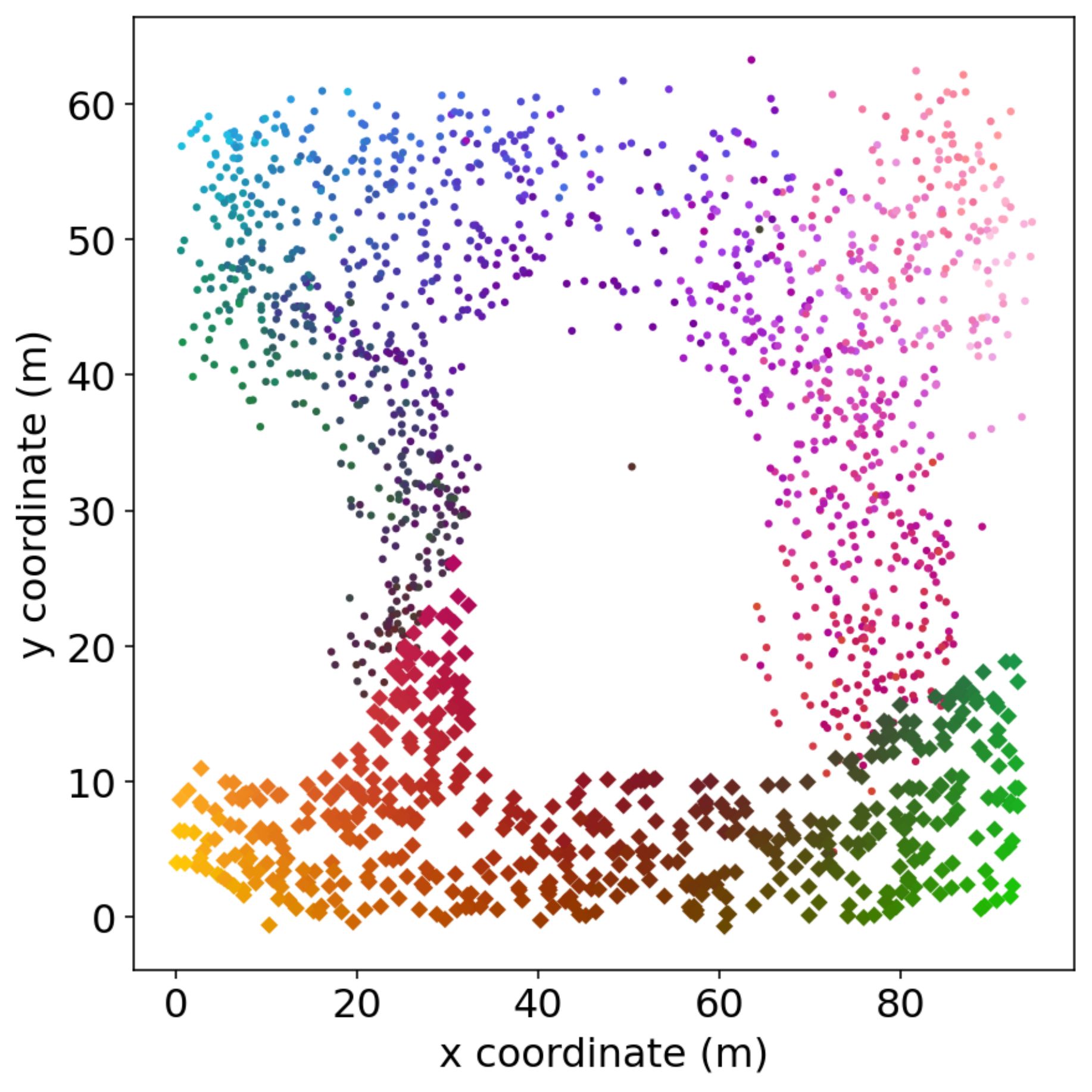}
        \caption{$\mathcal{L}_{\rm CC} = \mathcal{L}_{\rm B-CC} + \mathcal{L}_{\rm OT}$}
    \end{subfigure}
    \caption{Position estimates for UNILocPro with different loss functions (testing dataset).}
    \label{fig:UNILocPro_results_OT_LOS_loss_ablation}
    \vspace{-2mm}
\end{figure}

Apart from $\mathcal{L}_{\rm B-CC}$, which is essential for learning local geometry, the other two loss terms, i.e., $\mathcal{L}_{\rm LoS}$ and $\mathcal{L}_{\rm OT}$, are designed to capture global geometry. The position estimates for UNILocPro with different loss functions are shown in Fig.~\ref{fig:UNILocPro_results_OT_LOS_loss_ablation}, and the corresponding errors are shown in Table~\ref{tab:localization_performance_OT_LOS_loss_ablation}. It can be observed that, when $\mathcal{L}_{\rm OT}$ is disabled, the position estimates of \ac{nlos} users are not aligned with the map, which leads to distorted global geometry and thus performance degradation; whereas when disabling $\mathcal{L}_{\rm LoS}$, \ac{nlos} users are aligned with the map, but they are located entirely in incorrect positions (the generated channel chart for \ac{nlos} users is left-right reversed) as no information is provided to anchor the map, which leads to an even more severe performance degradation. Since $\mathcal{L}_{\rm LoS}$ aims to anchor the generated channel chart to global coordinates and $\mathcal{L}_{\rm OT}$ aims to refine the channel chart by aligning it with the map, both are necessary for achieving accurate localization, while $\mathcal{L}_{\rm LoS}$ (the basis for global geometry) is more crucial than $\mathcal{L}_{\rm OT}$ (refinement for global geometry).

\begin{table}[!htbp]
\caption{The MAE (m) for UNILocPro with timestamps ($p_I = 1$) for different loss functions (testing dataset).}
\label{tab:localization_performance_OT_LOS_loss_ablation}
\centering
\renewcommand{\arraystretch}{1.1} 
\begin{tabular}{|l|ccc|}
\hline
 & LoS & NLoS & all \\
 \hline
$\mathcal{L}_{\rm CC} = \mathcal{L}_{\rm B-CC} + \mathcal{L}_{\rm LoS} + \mathcal{L}_{\rm OT}$ & 0.34 & 5.36 & 3.88 \\
\hline
$\mathcal{L}_{\rm CC} = \mathcal{L}_{\rm B-CC} + \mathcal{L}_{\rm LoS}$ & 0.34 & 15.22 & 10.81 \\
\hline
$\mathcal{L}_{\rm CC} = \mathcal{L}_{\rm B-CC} + \mathcal{L}_{\rm OT}$ & 0.34 & 54.31 & 38.34 \\
\hline
\end{tabular}
\end{table}

\subsubsection{Discussion on Hyper-parameters for Dissimilarity Metrics}
In general, the choice of the hyper-parameters for the dissimilarity metrics, e.g., $\kappa$, $\vartheta$, and $d_{\rm thre}$, can depend on the specific scenario, and it may require some tuning in practice to achieve the best performance. However, we numerically verify that the \ac{cc} and localization performance is robust to the choice of these parameters. For example, the Kruskal stress of \ac{cc} for $d_{i,j}^{\mathrm{Fusi}}$ with different values of $\vartheta$ and $d_{\rm thre}$ is shown in Table~\ref{tab:CC_results_fusion_KS}, where it can be observed that the \ac{cc} performance is not very sensitive to $\vartheta$ and $d_{\rm thre}$, and even $d_{i,j}^{\mathrm{Wass}}$ and $d_{i,j}^{\mathrm{G-GOSPA}}$ can achieve acceptable performance. This demonstrates the robustness of UNILocPro to the choice of these parameters.

\begin{table}[!htbp]
\caption{The $\mathcal{KS} \downarrow$ of \ac{cc} for $d_{i,j}^{\mathrm{Fusi}}$ with different values of $\vartheta$ and $d_{\rm thre}$.}
\label{tab:CC_results_fusion_KS}
\centering
\renewcommand{\arraystretch}{1.1} 
\begin{tabular}{|c|c|c|c|c|}
\hline
 & $d_{\rm thre} = 0$ & $d_{\rm thre} = 10$ & $d_{\rm thre} = 20$ & $d_{\rm thre} = 30$ \\
\hline
$\vartheta = 0$ &  \multicolumn{4}{|c|}{0.267}  \\
\hline
$\vartheta = 0.001$ & 0.257 & 0.254 & 0.268 & 0.251 \\
\hline
$\vartheta = 0.01$ & 0.204 & 0.204 & 0.204 & 0.204 \\
\hline
$\vartheta = 0.1$ & 0.234 & 0.234 & 0.232 & 0.226 \\
\hline
$\vartheta = 10$ & 0.263 & 0.261 & 0.258 & 0.249 \\
\hline
$\vartheta = \infty$ & 0.262 & 0.262 & 0.258 & 0.250 \\
\hline
\end{tabular}
\end{table}

\subsubsection{Discussion on NN Initialization}
In order to investigate the sensitivity of UNILocPro to the initialization of the \ac{nn} parameters, we conduct 10 experiments with randomly initialized \ac{nn} parameters, both with and without pre-training, and the averaged results are shown in Table~\ref{tab:CC_Loc_results_initialization_sensitivity}. It can be observed that $\mathcal{CT}$, $\mathcal{TW}$, and \ac{mae} are stable across different initializations. Moreover, it can also be seen that the pre-training step, where model-based estimates are used as pseudo-labels to pre-train the \ac{nn} for a few epochs, can consistently improve the positioning performance compared to random initialization.

\begin{table}[!htbp]
\caption{The performance of UNILocPro with timestamps ($p_I = 1$) under random initialized \ac{nn} parameters (testing dataset).}
\label{tab:CC_Loc_results_initialization_sensitivity}
\centering
\renewcommand{\arraystretch}{1.1} 
\begin{tabular}{|l|c|c|c|}
\hline
training & $\mathcal{CT} \uparrow$ & $\mathcal{TW} \uparrow$ & MAE (m) \\
\hline
w/o pre-training & 0.977 $\pm$ 0.0015 & 0.976 $\pm$ 0.0016 & 4.77 $\pm$ 0.26\\
with pre-training & 0.979 $\pm$ 0.0008 & 0.978 $\pm$ 0.0011 & 4.40 $\pm$ 0.25 \\
\hline
\end{tabular}
\end{table}

\subsubsection{Discussion on Training Complexity}
\begin{figure}[t]
\centering
\begin{minipage}[t]{0.97\linewidth}
\centering
    \include{Figures/fig_complexity_v3.tex}
    \vspace{-1 cm}
\end{minipage}
\caption{The \ac{mae} of UNILoc and UNILocPro with timestamps during training.}
\label{fig:complexity}
\end{figure}

Finally, we compare the training complexity of UNILoc and UNILocPro by examining the \ac{mae} during training, as shown in Fig.~\ref{fig:complexity}. Without the computation and backpropagation of the iterative Sinkhorn algorithm, the training of UNILoc is much faster than that of UNILocPro (around 4 to 5 times faster in our experiments), which implies that the training complexity of UNILoc is much lower than that of UNILocPro. It also takes a much longer time for UNILocPro before its testing performance becomes stable. Note that with reduced training complexity, UNILoc only leads to an acceptable degradation in performance (within $0.7\, {\rm m}$ in our experiments) compared to UNILocPro. This suggests that the simplified UNILoc is more efficient and suitable for large-scale datasets, while UNILocPro can be utilized when training complexity is not a concern.

\section{Conclusion and Future Work}\label{sec:conclusion}
In this paper, we propose unified localization frameworks by combining model-based geometry and \ac{cc} for single-\ac{bs} massive \ac{mimo} in mixed \ac{los}/\ac{nlos} scenarios. To achieve unsupervised \ac{cc}, we first develop a fused dissimilarity metric, which is shown not only to achieve good \ac{cc} performance but also to preserve the basic global geometry, and then the \ac{cc} model of UNILocPro is trained by integrating multiple losses. Moreover, a low-complexity variant, i.e., UNILoc, is also proposed by simplifying the \ac{ot}-based loss. Neither of the proposed methods requires ground-truth labels, and they can also work without timestamps and external \ac{los}/\ac{nlos} identification. Ray-tracing simulations are carried out to show that the proposed frameworks significantly outperform the model-based and \ac{cc}-based methods in terms of both \ac{cc} and localization performance. Notably, UNILocPro with timestamps can achieve positioning performance that is close to fully-supervised fingerprinting. It is also shown that the proposed frameworks are robust to \ac{los}/\ac{nlos} identification errors, channel estimation noise, and velocity variation, and UNILoc can significantly reduce the training complexity with marginal performance degradation compared to UNILocPro. However, while being able to tolerate small clock bias, the proposed frameworks are sensitive and perform inadequately when the clock bias is large. Therefore, as future work, we will explore their extension to scenarios with loose synchronization. Moreover, in addition to synthetic simulations, we will collect measured data to evaluate the performance of the proposed frameworks in practical environments.

\appendices
\section{GOSPA Metric}
\ac{gospa} is a metric for measuring the difference between two sets of points, widely used in multi-target tracking and localization~\cite{Mode_End_Jose_2025,An_Uncer_Pint_2021,Dyna_Spec_Wang_2024}. Consider two finite sets of points $\mathcal{X}_s = \{\mathbf{x}_s^1, \mathbf{x}_s^2, \ldots, \mathbf{x}_s^{|\mathcal{X}_s|}\}$ and $\mathcal{X}_t = \{\mathbf{x}_t^1, \mathbf{x}_t^2, \ldots, \mathbf{x}_t^{|\mathcal{X}_t|}\}$, where $|\mathcal{X}_s|$ and $|\mathcal{X}_t|$ are the cardinalities of the sets. Let $d(\mathbf{x}_s,\mathbf{x}_t)$ denote a distance for any $\mathbf{x}_s, \mathbf{x}_t \in \mathbb{R}^{N \times 1}$, and its cut-off version is defined as $d^{(\zeta)}(\mathbf{x}_s,\mathbf{x}_t) = \min(d(\mathbf{x}_s,\mathbf{x}_t),\zeta)$, where $\zeta>0$ is a cut-off value that limits the maximum distance between two points, mitigating the influence of outliers and preventing extreme values from distorting the metric. Then, for $|\mathcal{X}_s| \leq |\mathcal{X}_t|$, the \ac{gospa} metric is defined as~\cite{Gene_Opti_Rahm_2017}
\begin{equation}\label{eq:gospa_metric}
    \begin{aligned}
        & \mathscr{G}_{p}^{(\zeta,\varpi)} (\mathcal{X}_s,\mathcal{X}_t) \\
         & = \Big( \min_{\pi \in \Pi_{|\mathcal{X}_t|}} \sum_{i=1}^{|\mathcal{X}_s|} d^{(\zeta)}(\mathbf{x}_s^i, \mathbf{x}_t^{\pi(i)})^p + \frac{\zeta^p}{\varpi} (|\mathcal{X}_t| - |\mathcal{X}_s|) \Big)^{\frac{1}{p}},
    \end{aligned}
\end{equation}
where the exponent $1 \leq p< \infty$ controls the penalization for inaccurate estimations; the parameter $0 < \varpi \leq 2$, together with $\zeta$, controls the penalization for cardinality mismatch; $\Pi_{|\mathcal{X}_t|}$ is the set of all permutations of $\{1,\ldots,|\mathcal{X}_t|\}$, with elements $\pi \in \Pi_{|\mathcal{X}_t|}$ being a sequence of indices $(\pi(1), \ldots, \pi(|\mathcal{X}_t|))$. Note that if $|\mathcal{X}_s| > |\mathcal{X}_t|$, the \ac{gospa} metric is $\mathscr{G}_{p}^{(\zeta,\varpi)}(\mathcal{X}_s,\mathcal{X}_t) = \mathscr{G}_{p}^{(\zeta,\varpi)}(\mathcal{X}_t,\mathcal{X}_s)$.

\section{Discrete OT and Wasserstein Distance}
In this work, we adopt discrete \ac{ot} and more details on general \ac{ot} can be found in~\cite{Opti_Tran_Cour_2017,Comp_Opti_Peyr_2019}. Given the source set $\mathcal{X}_s$ and the target set $\mathcal{X}_t$ with $N_s = |\mathcal{X}_s|$ and $N_t = |\mathcal{X}_t|$, we have two defined probability vectors $\mathbf{u}_s \in \mathbb{R}_+^{N_s \times 1}$ and $\mathbf{u}_t \in \mathbb{R}_+^{N_t \times 1}$ for $\mathcal{X}_s$ and $\mathcal{X}_t$, respectively, where $[\mathbf{u}_s]_i$ (resp. $[\mathbf{u}_t]_i$) is the probability associated with data sample $\mathbf{x}_s^i$ (resp. $\mathbf{x}_t^i$). Then, the \ac{ot} aims to find a transformation from $\mathcal{X}_s$ to $\mathcal{X}_t$, as denoted by $\mathbf{T}^*: \mathcal{X}_s \rightarrow \mathcal{X}_t$, that minimizes the transportation cost on the condition that the transformation $\mathbf{T}^*$ would push the probability vector $\mathbf{u}_s$ in the source set $\mathcal{X}_s$ toward the probability vector $\mathbf{u}_t$ in the target set $\mathcal{X}_t$. In particular, the discrete \ac{ot} problem can be written as~\cite{Comp_Opti_Peyr_2019}
\begin{equation}\label{eq:OT_opti_discrete}
\begin{aligned}
    \mathbf{T}^* =  \arg & \min_{\mathbf{T} : \mathcal{X}_s \rightarrow \mathcal{X}_t} \quad \sum_{\mathbf{x}_s^i \in \mathcal{X}_s} c\left(\mathbf{x}_s^i,\mathbf{T}(\mathbf{x}_s^i)\right), \\
    & {\rm s.t.} \quad [\mathbf{u}_t]_j = \sum_{i:\mathbf{T}(\mathbf{x}_s^i) = \mathbf{x}_t^j} [\mathbf{u}_s]_i,\, \forall \mathbf{x}_t^j \in \mathcal{X}_t,
\end{aligned}
\end{equation}
where $c\left(\mathbf{x}_s,\mathbf{T}(\mathbf{x}_s)\right)$ is the cost of transforming $\mathbf{x}_s$ to $\mathbf{T}(\mathbf{x}_s)$, which can be defined differently for specific applications.

The Kantorovich relaxation of the above problem~\eqref{eq:OT_opti_discrete} can be formulated by finding a joint probability matrix $\mathbf{\Gamma} \in \mathbb{R}_+^{N_s \times N_t}$ ($[\mathbf{\Gamma}]_{i,j}$ is the probability of $(\mathbf{x}_s^i,\mathbf{x}_t^j)$) with marginals $\mathbf{u}_s$ and $\mathbf{u}_t$ to minimize $\sum_{\mathbf{x}_s^i \in \mathcal{X}_s, \mathbf{x}_t^j \in \mathcal{X}_t} c\left(\mathbf{x}_s^i,\mathbf{x}_t^j\right) \cdot [\mathbf{\Gamma}]_{i,j}$, i.e.,
\begin{equation}\label{eq:OT_opti_discrete_relaxation}
    \mathbf{\Gamma}^* = \arg \min_{\mathbf{\Gamma} \in \mathcal{B}} \quad \langle\mathbf{\Gamma}, \mathbf{C}\rangle_F,
\end{equation}
where $\mathbf{C} \in \mathbb{R}_+^{N_s \times N_t}$ is the transportation cost matrix with $[\mathbf{C}]_{i,j} = c(\mathbf{x}_s^i,\mathbf{x}_t^j)$, and $\mathcal{B} = \{ \mathbf{\Gamma} \in \mathbb{R}_+^{N_s \times N_t} | \mathbf{\Gamma} \cdot \mathbf{1}_{N_t} = \mathbf{u}_s, \,\, \mathbf{\Gamma}^{\mathsf{T}} \cdot \mathbf{1}_{N_s} = \mathbf{u}_t\}$. Then, based on the \ac{ot} problem~\eqref{eq:OT_opti_discrete_relaxation}, the Wasserstein distance of order $p$ ($p \geq 1$) between $\mathbf{u}_s$ in $\mathcal{X}_s$ and $\mathbf{u}_t$ in $\mathcal{X}_t$ can be defined as $\mathscr{W}_p(\mathbf{u}_s, \mathbf{u}_t) = \langle\mathbf{\Gamma}^*, \mathbf{C}\rangle_F^{\frac{1}{p}}$, where $c(\mathbf{x}_s^i,\mathbf{x}_t^j) = [d(\mathbf{x}_s^i,\mathbf{x}_t^j)]^p$ with $d(\mathbf{x}_s^i,\mathbf{x}_t^j)$ being a distance~\cite{Comp_Opti_Peyr_2019,Opti_Tran_Cour_2017,Opti_Mass_Kolo_2017}. It is observed that~\eqref{eq:OT_opti_discrete_relaxation} is a linear programming problem and can be solved by, e.g., simplex methods~\cite{Opti_Tran_Cour_2017}, which usually incur polynomial time complexity.

\section{An Example of CSI-based LoS/NLoS Identification}\label{app:LoS_NLoS_identification}
In this work, we adopt a lightweight \ac{csi}-based \ac{los}/\ac{nlos} identification method, which is based on the fact that the \ac{csi} of \ac{los} users typically has a stronger signal strength and a more dominant direct path component compared with that of \ac{nlos} users. Specifically, given $\{ \hat{\beta}_l^i, \hat{\theta}_l^i, \hat{\tau}_l^i \}_{l}$, we extract two statistical channel features, i.e., the peak path power $P_{\mathrm{peak}}^i$ and the confidence metric $C_i$~\cite{An_Intr_Zeka_2012}, based on which a two-component \ac{gmm} clusters the users into \ac{los} and \ac{nlos} clusters~\cite{Patt_Reco_Bish_2006}. The probability density of the feature vector $\mathbf{x}_i = [P_{\mathrm{peak}}^i, C_i]^{\mathsf T}$ is modeled as $p(\mathbf{x}_i) = \sum_{k=1}^{2} \pi_k \mathcal{N} \left(\mathbf{x}_i; \boldsymbol{\mu}_k, \boldsymbol{\Sigma}_k\right)$, $\forall i$, where $\pi_k$, $\boldsymbol{\mu}_k$, and $\boldsymbol{\Sigma}_k$ denote the mixture weight, mean vector, and covariance matrix of the $k$-th Gaussian component, and $\sum_{k=1}^{2}\pi_k=1$. The parameters of the \ac{gmm} can be estimated using the \ac{em} algorithm, which iteratively updates the responsibilities and the model parameters until convergence. Consequently, each user is assigned to the Gaussian component with the highest responsibility~\cite{Patt_Reco_Bish_2006}
\begin{equation}
  \gamma_i^k = \frac{\hat{\pi}_k \mathcal{N}(\mathbf{x}_i;\hat{\boldsymbol{\mu}}_k,\hat{\mathbf{\Sigma}}_k)} {\sum_{j=1}^{2} \hat{\pi}_j \mathcal{N}(\mathbf{x}_i;\hat{\boldsymbol{\mu}}_j,\hat{\mathbf{\Sigma}}_j)}, \quad k=1,2,
\end{equation}
where $\hat{\pi}_k$, $\hat{\boldsymbol{\mu}}_k$, and $\hat{\mathbf{\Sigma}}_k$ are the estimated parameters. Since the \ac{los} users are expected to exhibit larger values of both $P_{\mathrm{peak}}^i$ and $C_i$, the Gaussian component with a larger $\hat{\boldsymbol{\mu}}_k$ is assigned to the \ac{los} cluster, while the other component is assigned to the \ac{nlos} cluster. Accordingly, the predicted propagation type of user $i$ can be obtained as $\hat{k}_i = \arg\max_{k\in\{1,2\}} \gamma_i^k$ followed by the above physical interpretation of the two Gaussian components. Without requiring any labeled data, the \ac{gmm} can be trained in an unsupervised manner using the extracted features from the training dataset. This procedure provides an unsupervised \ac{los}/\ac{nlos} identification mechanism $I(\cdot)$ that can be directly integrated into UNILocPro, where $I(\mathbf{H}_i)$ outputs 1 if the user is identified as \ac{los} and 0 if identified as \ac{nlos}.

\bibliographystyle{IEEEtran}
\bibliography{main}

\vfill

\end{document}